\documentclass[11pt, tightenlines, aps, prd, onecolumn, groupedaddress, superscriptaddress]{revtex4-2}
\usepackage{graphicx}
\usepackage{float}
\usepackage{amssymb}
\usepackage{amsmath}
\usepackage{array}
\usepackage{lineno}
\usepackage{scrextend}
\usepackage{textcomp}
\usepackage{gensymb}
\usepackage{xspace}
\usepackage{subcaption}
\usepackage{hyperref}
\hypersetup{
    colorlinks=true,
    citecolor=blue,
    linkcolor=blue,
    filecolor=magenta,      
    urlcolor=blue,
    pdftitle={Sharelatex Example},
    }
\def\eV{{\ensuremath{\mathrm{\,e\kern -0.1em V}}\xspace}}
\def\keV{{\ensuremath{\mathrm{\,ke\kern -0.1em V}}\xspace}}
\def\MeV{{\ensuremath{\mathrm{\,Me\kern -0.1em V}}\xspace}}
\def\GeV{{\ensuremath{\mathrm{\,Ge\kern -0.1em V}}\xspace}}
\def\TeV{{\ensuremath{\mathrm{\,Te\kern -0.1em V}}\xspace}}
\def\meg{\mu^+ \rightarrow e^+ \gamma}

\def\meee{\mu^+ \rightarrow e^+ e^- e^+}
\def\mec{\mu^-N\rightarrow e^-N}

\begin{document}
\begin{flushright}
FERMILAB-CONF-23-464-PPD\\
CALT-TH-2023-036
\end{flushright}
\vspace{0.5cm}

\title{Workshop on a future muon program at FNAL}


\newif\ifp
\ptrue

\author{S. Corrodi}
\author{Y.~Oksuzian}
\affiliation{Argonne National Laboratory, Lemont, Illinois 60439, USA}

\author{A.~Edmonds}
\affiliation{Boston University, Boston, Massachusetts 02215, USA}
\author{J.~Miller}
\ifp\author{H.~N.~Tran}\fi
\affiliation{Boston University, Boston, Massachusetts 02215, USA}

\author{R.~Bonventre}
\author{D.~N.~Brown}
\affiliation{Lawrence Berkeley National Laboratory, Berkeley, California 94720, USA}

\author{F.~M\'eot}
\affiliation{Brookhaven National Laboratory, Upton, New York 11973, USA}

\ifp\author{V.~Singh}	
\affiliation{University of California, Berkeley, California 94720, USA}\fi

\ifp\author{Y.~Kolomensky}
\affiliation{University of California, Berkeley, California 94720, USA; Lawrence Berkeley National Laboratory, Berkeley, California 94720, USA}\fi

\ifp\author{S.~Tripathy}	
\affiliation{University of California, Davis, California 95616, USA}\fi

\author{L.~Borrel}
\ifp\author{M.~Bub}\fi
\affiliation{California Institute of Technology, Pasadena, California 91125, USA}

\author{B.~Echenard}
\affiliation{California Institute of Technology, Pasadena, California 91125, USA}
\email[corresponding authors: ]{echenard@caltech.edu}

\author{D.G.~Hitlin}
\affiliation{California Institute of Technology, Pasadena, California 91125, USA}

\ifp\author{H.~Jafree}\fi
\author{S. Middleton}
\affiliation{California Institute of Technology, Pasadena, California 91125, USA}

\author{R.~Plestid}
\affiliation{California Institute of Technology, Pasadena, California 91125, USA}

\author{F.~C.~Porter}
\affiliation{California Institute of Technology, Pasadena, California 91125, USA}

\author{R.~Y.~Zhu}
\affiliation{California Institute of Technology, Pasadena, California 91125, USA}

\author{L.~Bottura}
\affiliation{European Organization for Nuclear Research (CERN), Geneva, Switzerland}

\ifp\author{E.~Pinsard}
\author{A.~M.~Teixeira}
\affiliation{Laboratoire de Physique de Clermont (LPC), 63178 Aubière, France}\fi

\author{C.~Carelli}
\affiliation{Italian National Agency for New Technologies (ENEA), 76 - 00196 Rome, Italy}

\author{D.~Ambrose}
\affiliation{Fermi National Accelerator Laboratory, P.O.\  Box 500, Batavia, Illinois 60510, USA}

\author{K.~Badgley}
\ifp\author{G.~D.~Bautista}\fi
\affiliation{Fermi National Accelerator Laboratory, P.O.\  Box 500, Batavia, Illinois 60510, USA}

\author{R.H.~Bernstein}
\affiliation{Fermi National Accelerator Laboratory, P.O.\  Box 500, Batavia, Illinois 60510, USA}

\ifp\author{S.~Boi}\fi
\ifp\author{J.~Crnkovic}\fi
\author{J. Eldred}
\affiliation{Fermi National Accelerator Laboratory, P.O.\  Box 500, Batavia, Illinois 60510, USA}

\author{A.~Gaponenko}
\affiliation{Fermi National Accelerator Laboratory, P.O.\  Box 500, Batavia, Illinois 60510, USA}

\author{C.~Johnstone}
\affiliation{Fermi National Accelerator Laboratory, P.O.\  Box 500, Batavia, Illinois 60510, USA}

\ifp\author{B.~Kiburg}\fi
\ifp\author{R.~Kutschke}\fi
\author{K.~Lynch}
\ifp\author{A.~Mukherjee}\fi
\author{D.\ Neuffer}
\affiliation{Fermi National Accelerator Laboratory, P.O.\  Box 500, Batavia, Illinois 60510, USA}

\author{F.~Pellemoine}
\author{V.\ Pronskikh}
\ifp\author{G.~Rakness}\fi
\ifp\author{J.~Tang}\fi
\ifp\author{R.~Tschirhart}\fi
\author{M.~Yucel}
\ifp\author{J.~Zettlemoyer}\fi
\affiliation{Fermi National Accelerator Laboratory, P.O.\  Box 500, Batavia, Illinois 60510, USA}

\ifp\author{B.~Simons}
\affiliation{Fermi National Accelerator Laboratory, P.O.\  Box 500, Batavia, Illinois 60510, USA; Northern Illinois University, DeKalb, Illinois 60115, USA}\fi

\author{D.~Redigolo}
\affiliation{INFN, Sezione di Firenze, 50019 Sesto Fiorentino, Italy}

\ifp\author{E.~Diociaiuti}\fi
\ifp\author{S.~Giovannella}\fi
\ifp\author{S.~Miscetti}\fi
\author{I.~Sarra}
\affiliation{Laboratori Nazionali di Frascati dell'INFN, I-00044 Frascati, Italy}

\author{ S.~E.~M\"uller}
\affiliation{Helmholtz-Zentrum Dresden-Rossendorf, Dresden 01328, Germany}

\author{W.~Ootani}
\affiliation{International Center for Elementary Particle Physics (ICEPP), The University of Tokyo, Tokyo 113-0033, Japan}

\ifp\author{E.~B.~Yucel}
\affiliation{University of Illinois at Urbana-Champaign, Urbana, Illinois 61801, USA}\fi
 
\author{D.\ M.\ Kaplan}
\author{T.~J.~Phillips}
\affiliation{Illinois Institute of Technology, 10 West 35th Street
Chicago, IL 60616, USA}

\author{J.~Pasternak}
\affiliation{Imperial College London, London, UK}

\author{J.~Zettlemoyer}
\affiliation{Indiana University, Bloomington, Indiana 47405, USA}

\author{D.~Palo}
\affiliation{University of California, Irvine, California 92697, USA}

\author{Y.~Davydov}
\affiliation{Joint Institute for Nuclear Research, 141980, Dubna, Russia}

\ifp\author{D.~Brown}	
\affiliation{Western Kentucky University, Bowling Green, Kentucky 42101, USA}\fi

\ifp\author{S.~Banerjee}	
\affiliation{University of Louisville, Louisville, Kentucky 40292, USA}\fi

\author{D.~Kawall}
\affiliation{University of Massachusetts, Amherst, Massachusetts 01003, USA}

\author{Z.~Hartwig}
\affiliation{Massachusetts Institute of Technology, Cambridge, Massachussetts 02139, USA}

\author{S.~Davidson}
\affiliation{LUPM, CNRS, Universit\'e Montpellier, Montpellier Cedex 5, France}

\ifp\author{R.~Abrams}
\affiliation{Muons, Inc., Batavia, Illinois 60510, USA}\fi

\author{C. Kampa}
\affiliation{Northwestern University, Evanston, Illinois 60208, USA}

\author{M.~Mackenzie}
\ifp\author{M.~Schmitt}\fi
\affiliation
{Northwestern University, Evanston, Illinois 60208, USA}

\ifp\author{P.~Piot}\fi
\author{B. Simons}
\affiliation{Northern Illinois University, DeKalb, Illinois 60115, USA}

\author{Y.~J.~Lee}
\author{V.~Morozov}
\affiliation{Oak Ridge National Laboratory, Oak Ridge, Tennessee 37831, USA}

\author{A.~Sato}
\affiliation{Osaka University, Toyonaka, Osaka 564-0043, Japan}

\ifp\author{S.~Di~Falco}\fi
\author{A.~Gioiosa}
\author{L.~Morescalchi}
\affiliation{INFN Sezione di Pisa, I-56127 Pisa, Italy}

\author{A.~Papa}
\affiliation{INFN, Sezione di Pisa, I-56127 Pisa, Italy;
Paul Scherrer Institut, 5232 Villigen, Switzerland}

\author{M.~T.~Hedges}
\affiliation{Purdue University, West Lafayette, Indiana 47907, USA}

\author{F.~Renga}
\affiliation{INFN Sezione di Roma, P.le A. Moro 2, 00185 Roma, Italy}

\author{J.-B.~Lagrange}
\ifp\author{C.~Rogers}\fi
\author{D.~Wilcox}
\affiliation{STFC Rutherford Appleton Laboratory, Didcot
OX11 0QX, UK}

\author{A.~Petrov}
\affiliation{University of South Carolina, Columbia, South Carolina 29208, USA}

\ifp\author{J.~Tang}\fi
\author{S.~Zhao}
\affiliation{Sun Yat-sen University, Guangzhou, Guangdong 510275, China}

\author{E.~C.~Dukes}
\author{R.~Erlich}
\affiliation{University of Virginia, Charlottesville, Virginia 22904, USA}

\author{C. Group}
\ifp\author{J.~Heeck}\fi
\affiliation{University of Virginia, Charlottesville, Virginia 22904, USA}

\author{G.~Pezzullo}
\affiliation{Yale University, New Haven, Connecticut 06520, USA}

\ifp\author{T.~Nguyen}\fi
\author{J.~L.~Popp}
\affiliation{York College and Graduate Center, The City University of New York, Nedw York, New York 11451, USA}

\date{\today}
\begin{abstract}
The Snowmass report on rare processes and precision measurements recommended Mu2e-II and a next generation muon facility at Fermilab (Advanced Muon Facility) as priorities for the frontier. The {\it Workshop on a future muon program at FNAL} was held in March 2023 to discuss design studies for Mu2e-II, organizing efforts for the next generation muon facility, and identify synergies with other efforts (e.g., muon collider). Topics included high-power targetry, status of R\&D for Mu2e-II, development of compressor rings, FFA and concepts for muon experiments (conversion, decays, muonium and other opportunities) at AMF. This document summarizes the workshop discussions with a focus on future R\&D tasks needed to realize these concepts. 
\end{abstract}

\maketitle

\newpage
\tableofcontents

\def\thefootnote{\fnsymbol{footnote}}
\setcounter{footnote}{0}

\newpage 
\section{Introduction}
\label{sec:introduction}
The concepts of flavor and generations have played a central role in the development of the Standard Model, but the fundamental symmetries underlying the observed structures remain to be discovered. While flavor violation in quark and neutral lepton transitions have already shed some light on this question, charged lepton lepton flavor violation (CLFV) has yet to be seen. An observation would be a clear sign of New Physics (NP), and provide unique insights about the mechanism generating flavor. Furthermore, CLFV is closely linked to the physics of neutrino masses, and these processes can strongly constrain neutrino mass models and open a portal into GUT-scale physics. 

Thanks to the availability of intense sources and their relatively long lifetime, muons offer a promising avenue to search for charged lepton flavor violation. A global experimental program of muon CLFV searches is underway in the US, Europe and Asia. Impressive sensitivity gains are expected in this decade, with up to four orders of magnitude improvements in the rate of $\mec$ conversion and $\meee$ decay searches. Upgrades to the beam lines at PSI, Fermilab, and J-PARC would further extend the discovery potential by orders of magnitude. With the goal of exploiting the full potential of PIP-II, a staged program of next-generation experiments and facilities has been proposed at FNAL. Mu2e-II is a near-term evolution of the Mu2e experiment, proposing to improve Mu2e sensitivity by an order of magnitude. The construction is planned to start before the end of the decade by leveraging existing infrastructure. The Advanced Muon Facility is a more ambitious proposal for a new high-intensity muon science complex, delivering the world's most intense positive and negative muon beams. This facility would enable broad muon science with unprecedented sensitivity, including a suite of CLFV experiments that could improve the sensitivity of planned experiments by orders of magnitude, and constrain the type of operators contributing to NP in case of an observation.

\begin{figure}[ht!]
\centering
\includegraphics[width=\textwidth]{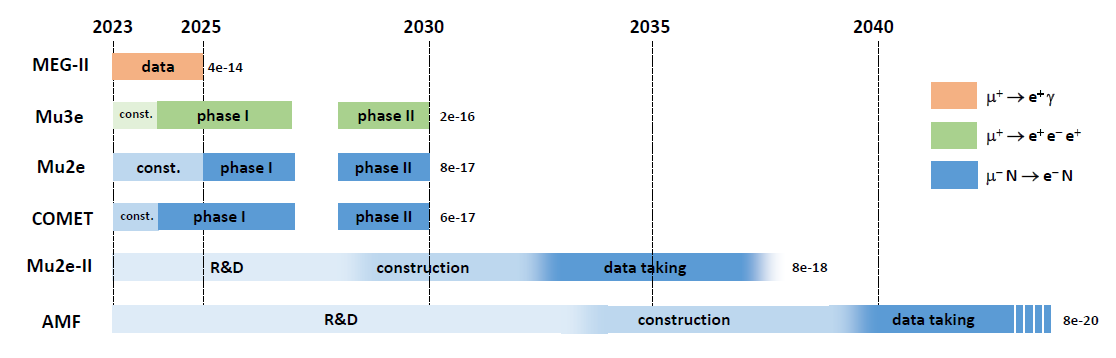}
\caption{Timeline for muon-based charged lepton flavor violation experiments outlined in the Snowmass report on rare processes and precision measurements~\cite{Artuso:2022ouk}. Approximate expected
dates are shown. The number to the right of the timeline is the expected final 90\% CL}
\label{fig:timelineProg}
\end{figure}

The Snowmass report on rare processes and precision measurements~\cite{Artuso:2022ouk} recommended Mu2e-II and a next generation muon facility at Fermilab (Advanced Muon Facility) as priorities for the frontier. The timeline of a muon program outlined in this report is shown in Fig.~\ref{fig:timelineProg}, with Mu2e-II starting shortly after the completion of the Mu2e experiment, followed by the construction and operations of the Advanced Muon Facility. This program would leverage the full power of PIP-II at FNAL, and potential upgrades of the accelerator complex in the future. A strong R\&D program should begin immediately to realize these opportunities.

The {\it Workshop on a future muon program at FNAL} (https://indico.fnal.gov/event/57834/) was held in March 2023 to pursue design studies for Mu2e-II, to organize efforts for the next generation muon facility, and to identify synergies with other R\&D efforts. The workshop comprised plenary and parallel sessions discussing technical aspects and physics capabilities; the workshop program is available in Appendix~\ref{program}. This document provides a summary of the workshop discussion, together with the list of prioritized R\&D tasks and avenues for future investigation.

\newpage
\section{Theory} 
\label{IntroTheory}

The non-zero neutrino masses and mixing angles induce  CLFV via loops. If neutrino masses are generated from Yukawa interactions with the Higgs boson, the CLFV rates are GIM-suppressed by factors of $\sum_{i,j}(\Delta m_{i,j}^2 / m_W^2)^2$, where $\Delta m^2$ denotes the mass-squared difference between the $i$-th and $j$-th neutrino mass eigenstates. The resulting branching fraction for $\mu \rightarrow e \gamma$ is at the level of $10^{-54}$~\cite{Petcov:1976ff}, well below any conceivable experimental sensitivity. However, new sources of CLFV are introduced in many BSM scenarios, leading to rates that are potentially accessible to future experiments (see e.g. Ref.~\cite{Calibbi:2017uvl, Cei:2014jtm}). 

Under the assumption that new particles responsible for CLFV are heavy, Effective Field Theory (EFT) offers a powerful framework to assess the reach and complementarity of CLFV searches. Restricting the discussion to the muon sector, the $\meee$ decay, $\meg$ decay, and (Spin Independent) $\mec$ conversion can be parameterized by the following Lagrangian at the experimental scale ($\sim m_\mu$): 
\begin{align}
&\mathcal{L}_{\mu e} = +\frac{4 G_F}{\sqrt{2}} \sum_{X = L,R}\Big[ m_\mu C_{D,X}\, \overline{e} \sigma^{\alpha\beta} P_X \mu \,F_{\alpha\beta} + C_{S,XX}\, \overline{e} P_X \mu\, \overline{e} P_X e \label{eq:lagr} \\
 & +
 \sum_{Y\in L,R} C_{V,XY} \,\overline{e}\gamma^\alpha P_X \mu \, \overline{e} \gamma_\alpha P_Y e\
 +\sum_{N=p,n}\left( C_{S,X}^{N}\, \overline{e} P_X \mu \, \overline{N} N
+ C_{V,X}^{N}\, \overline{e} \gamma^\alpha P_X \mu \, \overline{N} \gamma_\alpha N \right)\Big]
\nonumber
\end{align}
where $P_{L,R}$ are chiral projection operators, $C_a$ are dimensionless Wilson coefficients, and Spin Dependent conversion is neglected, occurring at a relatively suppressed rate compared to spin-independent conversion~\cite{Cirigliano:2017azj, Davidson:2017nrp, Hoferichter:2022mna}. This Lagrangian provides a model-independent description of CLFV interactions at leading order in $\chi$PT with electrons, muons and nucleons as long as the NP scale is much larger than the $\GeV$ scale. If the underlying model is specified, the Wilson coefficients can be calculated in terms of the model parameters. A similar Lagrangian with more operators could be constructed to describe $\tau$ decays. It is worth noting that low-energy muon CLFV reactions are sensitive to a much larger number of operators than those given in the above Lagrangian, as loop effects ensure that almost every operator with four or less legs contributes to $\meg$, $\meee$ and $\mec$ amplitudes with a suppression factor at most $\sim 10^{-3}$~\cite{Davidson:2020hkf}.

\begin{figure}[htb]
\begin{center}
 \includegraphics[width=0.45\textwidth]{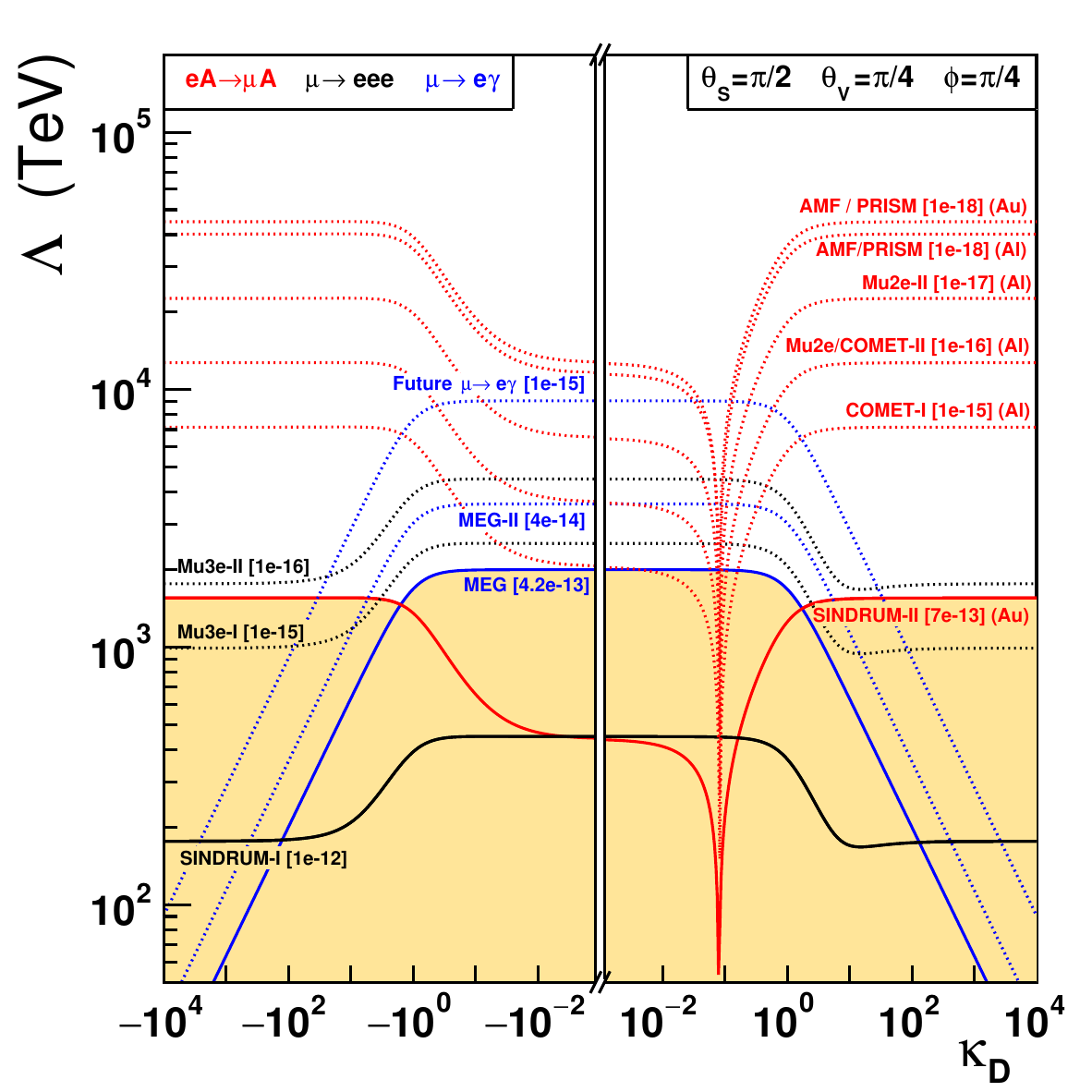}
 \includegraphics[width=0.45\textwidth]{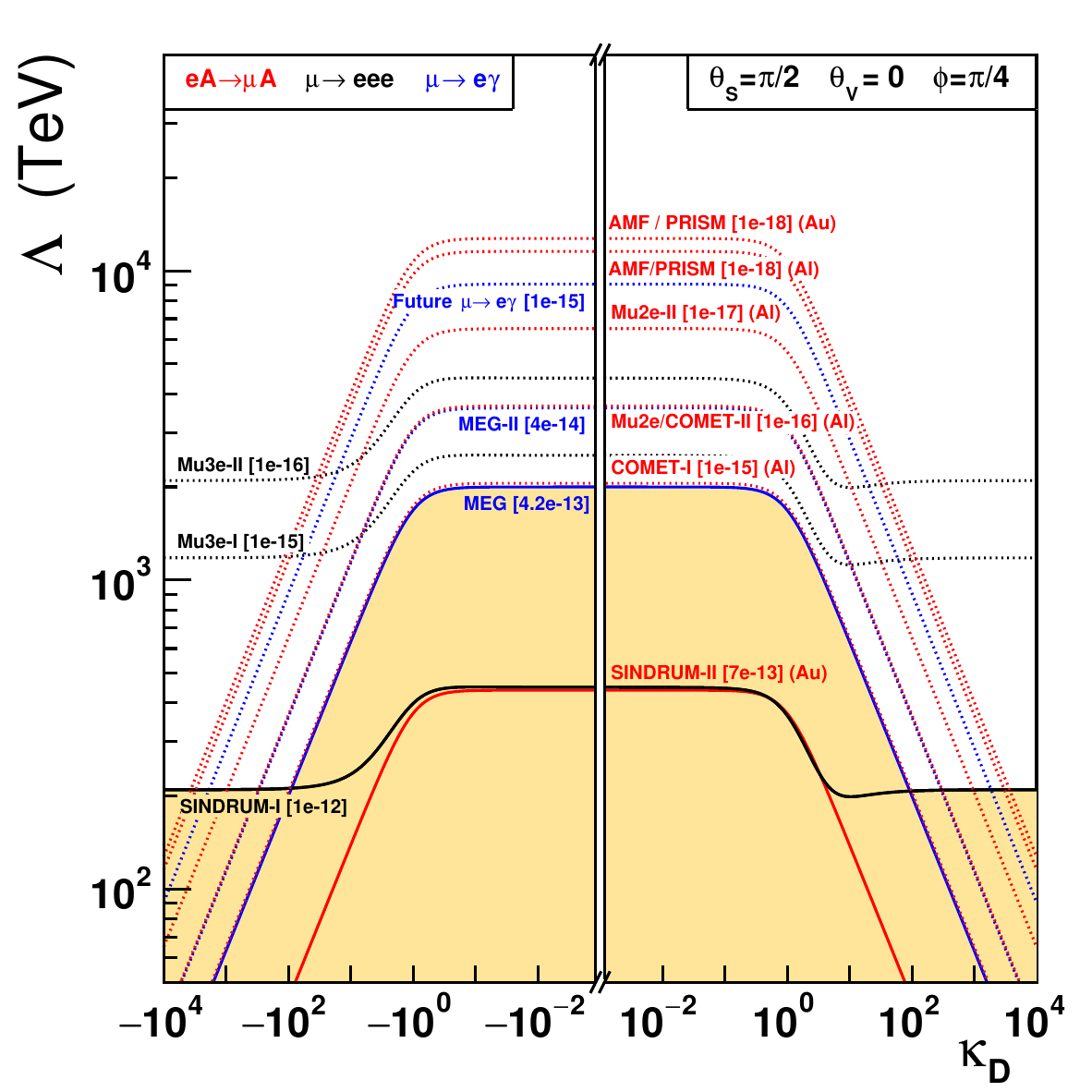}
\end{center}
\caption{Reach in NP scale, $\Lambda$, of past and upcoming muon CLFV searches. The solid region is currently excluded. The parameter $\kappa_D$ describes the relative contribution of the dipole and the four-fermion contact operators. The dipole operator dominates for $|\kappa_D| \ll 1$, while the four-fermion operators dominate for $|\kappa_D| \gg 1$. The remaining parameters are fixed to typical values (see~\cite{Davidson:2022nnl} for details).
\label{fig:NPreach} }
\end{figure}

The reach and complementarity of muon CLFV reactions is shown in Fig.~\ref{fig:NPreach}, expressing the coefficient appearing in the Lagrangian in spherical coordinates~\cite{Davidson:2022nnl}. The variable $\kappa_D$ describes the relative contribution of the dipole and selected four-fermion operators: the dipole dominates for $|\kappa_D| \ll 1$, while the four-fermion operators dominate for $|\kappa_D| \gg 1$. The variable $\theta_V$ describes the angle between  specific four-fermion operators on leptons or quarks, representing the relative rate between $\meee$ and $\mec$ at large $|\kappa_D|$, and $\phi$ distinguishes coefficients probed by $\mu-e$ conversion on light and heavy nuclei.

\begin{figure}[hb]
  \centering
  \includegraphics[width=0.55\textwidth]{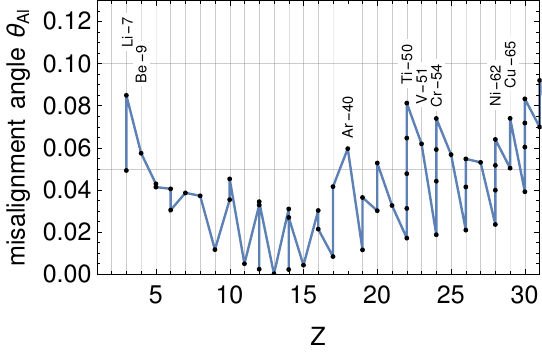}
  \includegraphics[width=0.4\textwidth]{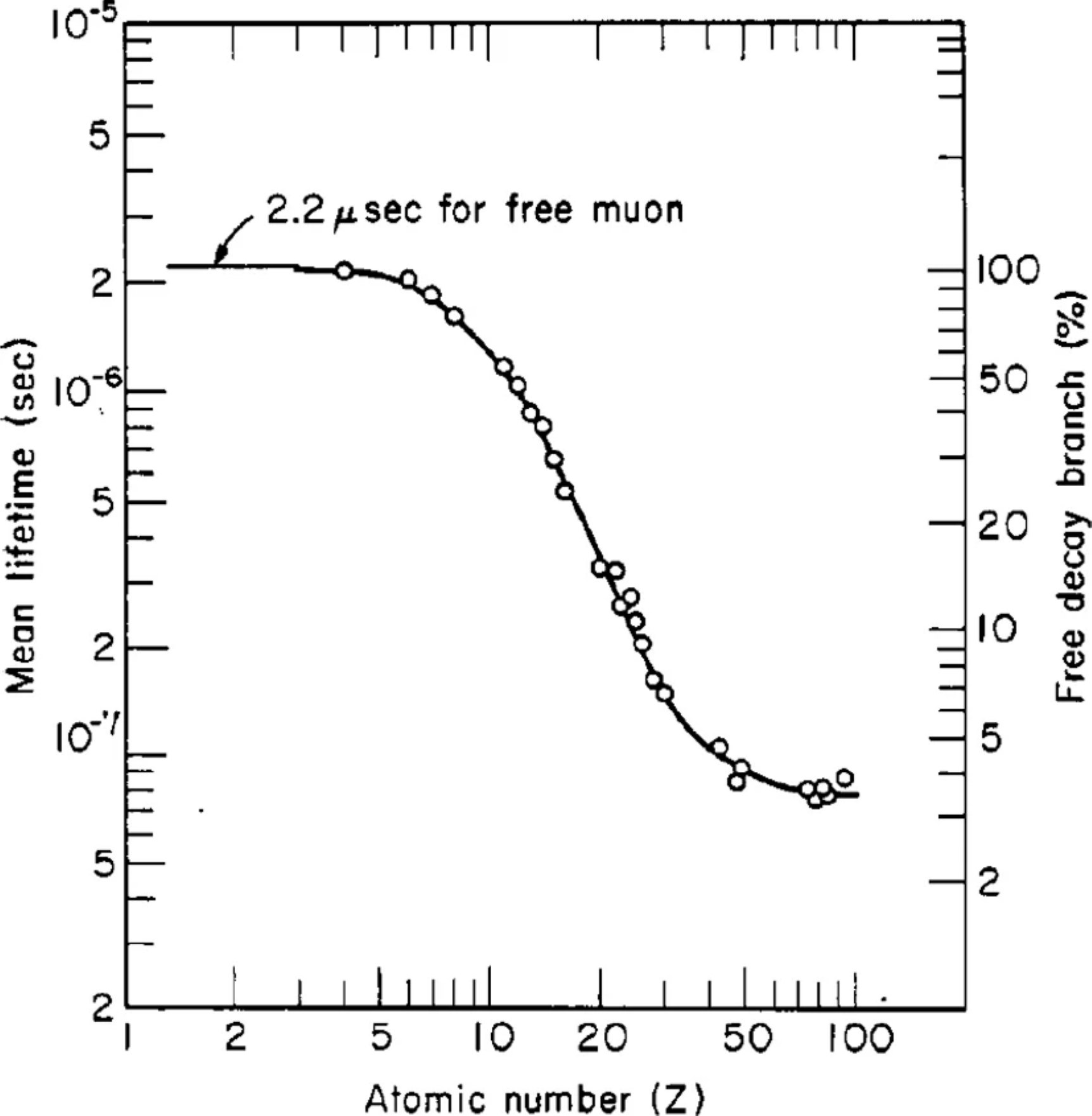}
\caption{Left: misalignment angle with Al, taken from~\cite{Heeck:2022wer}. The misalignment angle increases with the number of neutrons in isotopes. Right: mean muonic atom lifetime as a function of the atomic number. Adapted from Ref.~\protect{\cite{Knecht:2020}}.}
\label{fig:sens-and-lifetime-vs-z}
\end{figure}

Additional observables can be used to further study the underlying NP. For example, the dipole and four-lepton operators in $\meee$ decays can be distinguished by analyzing the final state angular distributions with polarized muons~\cite{Okada:1999zk}. In Spin-Independent $\mec$ conversion, all operators add coherently at the amplitude level, weighted by nucleus-dependent overlap integrals~\cite{Kitano:2002mt,Heeck:2022wer}. A different nuclear target probes a different combinations of coefficients, and multiple measurements can be used to disentangle their contributions. Representing overlap integrals as vectors in the space of operator coefficients, the complementarity can be expressed as a misalignment angle~\cite{Davidson:2018kud, Mu2e-II:2022blh}, as shown in Fig.~\ref{fig:sens-and-lifetime-vs-z}. As pointed out by previous studies~\cite{Cirigliano:2009bz}, light and heavy targets are good complements. However, the shorter muon lifetime in heavier elements presents an experimental challenge that must be addressed by future concepts.

The EFT approach has the advantage of being model agnostic, but the large number of operators is  daunting (although this issue can be circumvented using an observable-motivated operator basis~\cite{Davidson:2020hkf,Davidson:2022nnl}).  The study of simpler models, in particular neutrino-mass scenarios, provides an alternative approach. In the type-I seesaw mechanism~\cite{Minkowski:1977sc,Yanagida:1979as,Gell-Mann:1979vob,Mohapatra:1979ia}, heavy right-handed neutrinos generate a Majorana neutrino mass matrix inducing among others  $\ell_\alpha \to \ell_\beta \gamma$ decays. CLFV processes can be sizable~\cite{Antusch:2006vwa, Abada:2007ux, Coy:2018bxr} and provide information about the seesaw mechanism~\cite{Broncano:2003fq}. In the type-II seesaw mechanism~\cite{Konetschny:1977bn, Magg:1980ut, Schechter:1980gr, Cheng:1980qt, Mohapatra:1980yp}, CLFV processes induced by a $SU(2)_L$-triplet with a flavor structure are directly linked to the neutrino masses and oscillation angles. The observation of CLFV reactions could provide information about neutrino parameters difficult to access otherwise. Similarly, models generating neutrino masses via loops rather naturally require lower new-physics scales and thus enhanced CLFV rates~\cite{Cai:2017jrq}. Models involving light particles also need dedicated analysis as they cannot be described by the SMEFT~\cite{LoI080}. A non-exhaustive list of candidates includes the majoron $J$~\cite{Chikashige:1980ui,Schechter:1981cv}, pseudoscalars (axions, famuilons,...)~\cite{Kim:1986ax, Calibbi:2020jvd}, or $Z'$ gauge bosons~\cite{Foot:1994vd, Heeck:2016xkh}. These scenarios predict a large variety of CLFV signatures, including invisible and displaced decays.  

In summary, muon CLFV reactions are excellent probes of NP, closely linked to the mechanisms generating flavor and neutrino masses. Current measurements of $\meg$, $\meee$ and $\mec$ already set constraints on the NP mass scale at the level of $10^3 \TeV$ for some EFT operators, and future initiatives are poised to improve these bounds by one or more orders of magnitude. Prospects for distinguishing between models can conveniently be explored in EFT. Should a signal be observed, the Z-dependence of the conversion rate will provide critical information about the NP structure. If not, higher intensity muon beams will be required to further improve the sensitivity, probing higher mass scales and constraining models. The physics case for a next generation of experiments is well motivated in both cases. 

\clearpage

\newpage

\section{Mu2e-II}
The Mu2e experiment is designed to improve sensitivity to muon to electron conversion by four orders of magnitude. With the advent of PIP-II, it was recognized that this sensitivity could be improved upon still further using the Mu2e facility as a base. This possibility is referred to as Mu2e-II, with a goal of improving $\mu\to e$ sensitivity by at least an order of magnitude beyond Mu2e's capabilities. Mu2e-II uses the more powerful beam available from PIP-II, including a higher duty cycle, to enable this sensitivity, while re-using a substantial portion of the Mu2e infrastructure. The higher power and higher rates imply challenges, and some aspects of Mu2e will require modification to address these.

Mu2e-II provides a natural evolutionary step in the muon physics program at FNAL. It can follow Mu2e fairly quickly, keeping the muon physics program active. Mu2e-II can further inform as well as fill the gap towards a more ambitious program such as AMF. The R\&D for Mu2e-II, as well as experience from Mu2e-II, is synergistic with R\&D needed for other efforts such as AMF and the muon collider. The following sections describe the status of the thinking about Mu2e-II, including existing and needed R\&D.
\subsection{Accelerator} 
The primary beam for Mu2e-II will be the 800 MeV proton beam from the new PIP-II linac
under construction at Fermilab. The PIP-II linac will be capable of accelerating up to 2 ma
of H$^-$ beam to 800 MeV in CW operation (1.6 MW). Mu2e-II will use a fraction of that potential, with pulses of 100 ns at a 1.7 $\mu$s period. The plan is to use an intensity of ~100 kW, more than 10 times greater than Mu2e.

The initial implementation of PIP-II only includes pulsed beam for the Fermilab Booster. Mu2e-II requires that the PIP-II Linac be upgraded to include CW operation, which will require some power supply and chopper upgrades. It also requires construction of a new beam line from PIP-II to the Mu2e experimental 
hall. An initial design of that beam line exists; construction is a moderate but substantial expense.

The PIP-II linac could be extended to 1 GeV with modest upgrades.
It could also be extended to 2 GeV, with a substantially more expensive upgrade, which may be needed for future Booster replacement.  These extensions could modify the implementation of Mu2e-II.
\subsection{Magnets}

The Mu2e experiment includes three large solenoids, the production solenoid (PS), the transport solenoid (TS) and the detector solenoid (DS).
The Mu2e-II experiment plans to use the same configuration, reusing as much of that infrastructure as possible. 

Some modifications will be required. 800 MeV proton beam trajectories 
would not fit within the mu2e HRS (heat and radiation shield), and the HRS is insufficient to protect the PS from the higher radiation and heat load of the Mu2e-II beam. The HRS should be replaced, and the PS magnet modified to handle the larger heat loads. Mu2e operation may iaradiate the PS magnet to a level that it cannot be modified. It is likely that the most cost-effective solution would be to replace the entire PS assembly. A redesign based on Mu2e-II parameters combined with lessons learned from Mu2e experience should be developed. 

At this workshop, we heard two talks about high temperature superconductivity (HTS), from Luca Bottura (CERN, working on the muon collider), and from Zachary Hartwig (MIT). The 2022 version of the muon collider capture solenoid is a HTS magnet with a field of 20 T running at 20 K. The cost of HTS, thanks to interest and investment from the fusion community, has been coming down dramatically, approaching that of NB$_3$Sn. It no longer seems that cost is a prohibitive factor for a possible HTS replacement Mu2e-II PS. Besides cost, conductor such as REBCO (Rare Earth Barium Copper) is now available in quantity.

A new Python tool, called SolCalc, with a GUI, assists in the design of solenoids to meet magnetic field specifications. Calculations were shown for several choices of conductor towards deign of a Mu2e-II production solenoid, including low temperature superconductor, HTS (VIPER REBCO), and resistive coils. Several choices can potentially meet field specification. An issue for further investigation is whether it is possible to keep the radius of the solenoid from growing compared with the Mu2e magnet.
\subsection{Targetry}
\subsubsection{Possible approach to and challenges for the Mu2e-II target}

The forthcoming research program at Fermilab necessitates the development of innovative, high-power targets that can endure high-energy and high-intensity proton beams throughout their operational lifespan. The Mu2e project has proposed a solution anchored on a radiatively-cooled bicycle wheel structure made of chopped tungsten, while LBNF is strategizing to employ a long graphite target, designed to sustain a 1.2 (2.4) MW, 120-GeV beam. For the Mu2e-II project, however, only conceptual designs and early-stage prototypes exist, as will be discussed in the subsequent subsection. At present, there are no formulated strategies on how to construct the target for the AMF experiment.

The operational regimes anticipated for these future facilities present challenging conditions for materials, exhibiting parallels with those anticipated at the Muon Collider. Consequently, it is imperative to explore synergies and facilitate knowledge transfer with the Muon Collider program, as well as with other target facilities, in order to foster innovation and overcome these challenges.

Among the rapidly advancing initiatives in the field of nuclear physics is the Second Target Station (STS) at Oak Ridge National Laboratory (ORNL), a program predicated on the use of rotating tungsten target technology. The STS is slated to operate a 1.3-GeV 700-kW proton beam, with anticipated radiation damage approximating 1 DPA/yr as noted in reference~\cite{YJL1}. The STS target is designed as a `lasagna'-type tungsten target featuring a copper thermal interface, an Inconel water-cooled vessel, and a protective tantalum cladding to safeguard against in-beam erosion.

STS researchers have underscored~\cite{YJL2} that programs utilizing tungsten targets are currently hampered by an insufficiency of data regarding tungsten's embrittlement, hardening, and diffusivity characteristics. In response to this challenge, the STS is actively conducting irradiation experiments at Los Alamos National Laboratory (LANL), which include fatigue and oxidation tests. Establishing open communication lines or active collaboration with these programs could yield significant benefits for future muon programs.

A highly promising concept for muon-production targets involves the use of fluidized tungsten powder as cited in reference~\cite{Dan}. The strengths of these targets include their ability to endure exceptionally high energy densities and the fact that the technology for handling fluidized powder is already well-established in the industry. They have a lower eruption velocity than mercury and do not suffer from cavitation damage. Nevertheless, there are certain challenges associated with this type of target. Long-term operation could lead to erosion, necessitating further research and development to mitigate this issue. The density of tungsten is higher than that of other industrial materials, which demands enhancements in theoretical understanding and plant design. Additionally, diagnostic tools and process controls for reliable long-term operation are yet to be fully developed. To tackle these challenges, developers are conducting off-line testing. For instance, tests were executed at the HiRadMat facility in 2012 and 2015 to evaluate and improve upon the performance and resilience of these targets.

The development of new targetry technologies necessitates a synergistic approach with nuclear physics programs, which are formulating strategies towards a Muon Collider. Among these strategies, the use of liquid heavy metal targets, as discussed in reference~\cite{Carelli}, is prominent. Three such targets are currently under consideration: a liquid flow lead target, where liquid lead will circulate within a double-wall container, encased by a superconducting solenoid; a lead curtain design target; and a liquid mercury jet target, a design akin to that of the Spallation Neutron Source (SNS). Several issues are expected to arise with these targets, including cavitation, fatigue, shockwave, magneto-hydrodynamics, and stability, all of which demand a substantial amount of research and development. Additionally, the differences in pion spectra between lead and mercury compared to tungsten, the increased production of secondary neutrons from lead compared to tungsten (which requires shielding), and the generation of mixed wastes need to be taken into consideration.

The dearth of information regarding the radiation stability of materials, including tungsten, underscores the need for radiation tests of targets, including the Mu2e baseline one as suggested in reference~\cite{Hedges}. Proposals are being reviewed to conduct tests of the Mu2e "Hayman" tungsten target using an 8-GeV proton beam at AP0. Despite certain limitations, such as the absence of resonant extraction and a smaller beam spot size, such tests can be conducted post completion of the g-2 experiment. The parameters that can be assessed include thermal stresses, oxidation, and creep.

In an effort to fully leverage the experience and successful methodologies developed at other centers, Fermilab is contemplating several strategies, as outlined in reference~\cite{Frederique}. Firstly, the establishment of Post-Irradiation Examination (PIE) facilities, incorporating hot cells and specific characterization equipment, is under consideration. The laboratory generates a substantial volume of highly-activated materials, the examination of which necessitates transportation to collaborating centers equipped with suitable facilities. Having in-house PIE facilities would enable more efficient use of time and resources. Secondly, there is a need for dedicated facilities for radiation tests, including those using low-energy beams, which would facilitate radiation damage and thermal shock measurements. Thirdly, Fermilab currently lacks the capabilities to perform ab initio and molecular dynamics calculations and modeling. Development of these skills is vital for predicting the fundamental response of various classes of materials to irradiation, thereby guiding material selection and experimental designs for future irradiation studies. This includes modeling scenarios such as helium gas bubbles in beryllium and the radiation behavior of novel materials such as High-Entropy Alloys.

\subsubsection{Conveyor target design and particle production rates}
\label{sec:target_particle_yields}

\begin{figure}
    \centering
\includegraphics[angle=90,origin=lB,
width=0.32\linewidth]{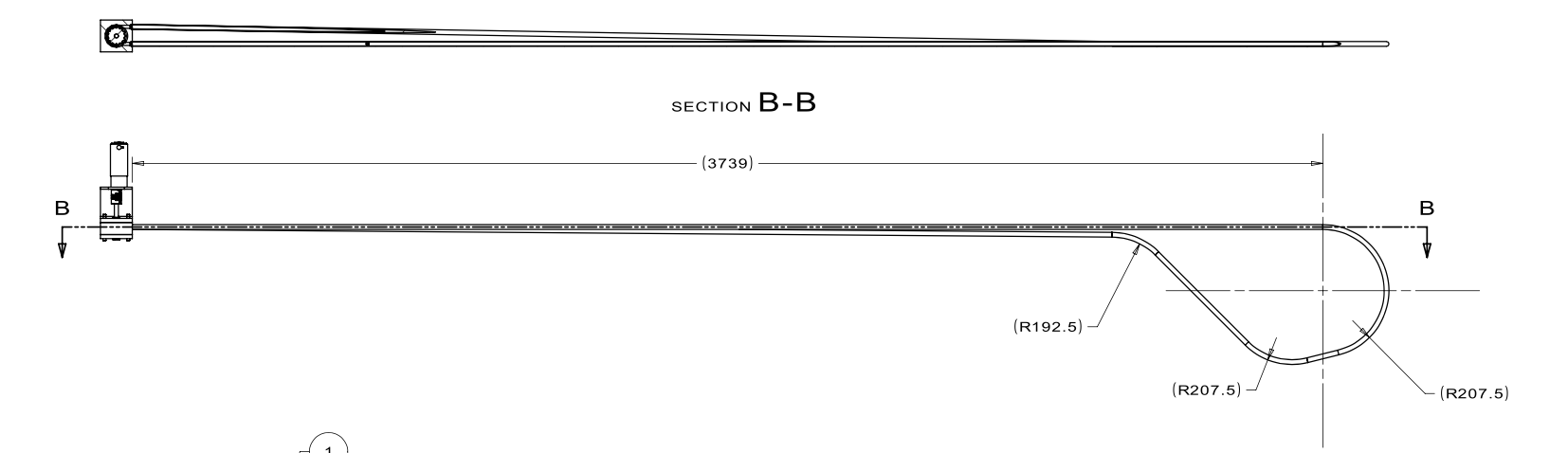}
    \raisebox{1cm}{\includegraphics[width=0.49\linewidth]{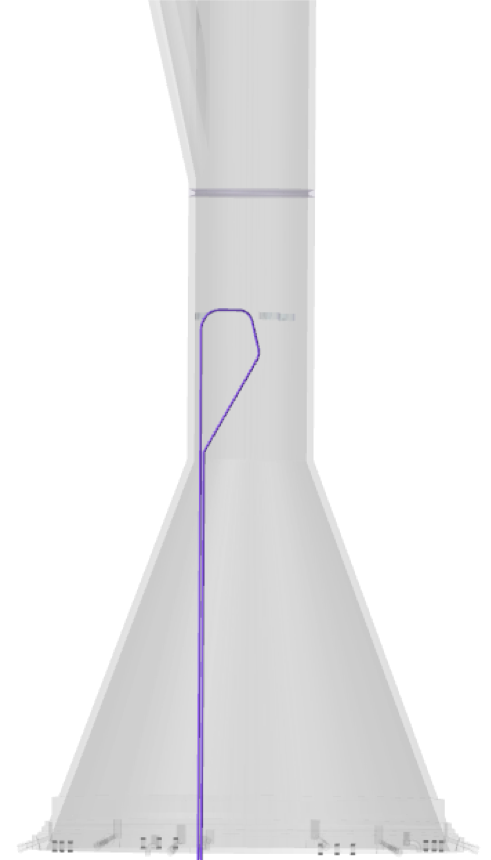}}
    \caption{Conveyor target drawing (left) and conveyor target placement in the HRS (right)}
    \label{fig:conveyor-target}
\end{figure}

Within the recently completed Laboratory Directed Research and Development (LDRD) project, a leading Mu2e-II pion-production target design has been developed, the ``conveyor'' target. This utilizes target spheres that are cycled in and out of the target region, distributing the damage among many spheres. Two prototypes of the conveyor Mu2e-II pion-production target have been developed, a carbon and a tungsten sphere design. The carbon target requires more spheres due to the lower density of carbon. The target design drawing and its deployment in Heat and Radiation Shield (HRS) are shown in Fig.~\ref{fig:conveyor-target}, where only the upper right straight section would be in the beam. While these prototypes show promise, they have not yet reached the final design stage and require additional research, development, and simulation studies.

A critical factor in the design of the Mu2e-II production target is the rate of low momentum pion production to produce a high intensity, low momentum muon beam. The transport solenoid is designed to accept low momentum muons, which have a higher probability of stopping in the muon stopping target in the detector solenoid, while suppressing backgrounds from high energy electrons produced in the production solenoid. The number of conveyor target balls in the target region of the conveyor target is optimized to maximize the number of stopped muons in the muon stopping target using MARS simulations.

The Mu2e-II Offline uses GEANT4~\cite{GEANT4_2003, GEANT4_2006, GEANT4_2016} for particle interaction and transport simulations to model the expected detector pileup environment as well as the signal and dominant background contributions. To validate the simulated particle production rates, Mu2e-II utilizes MARS15, GEANT4, and FLUKA to simulate the primary proton interactions with the production target candidates. The particle yields per proton on target are studied at the entrance to the transport solenoid, focusing on these particles that can contribute to the experiment's sensitivity.

\begin{figure}
    \centering
    \includegraphics[width=0.49\linewidth]{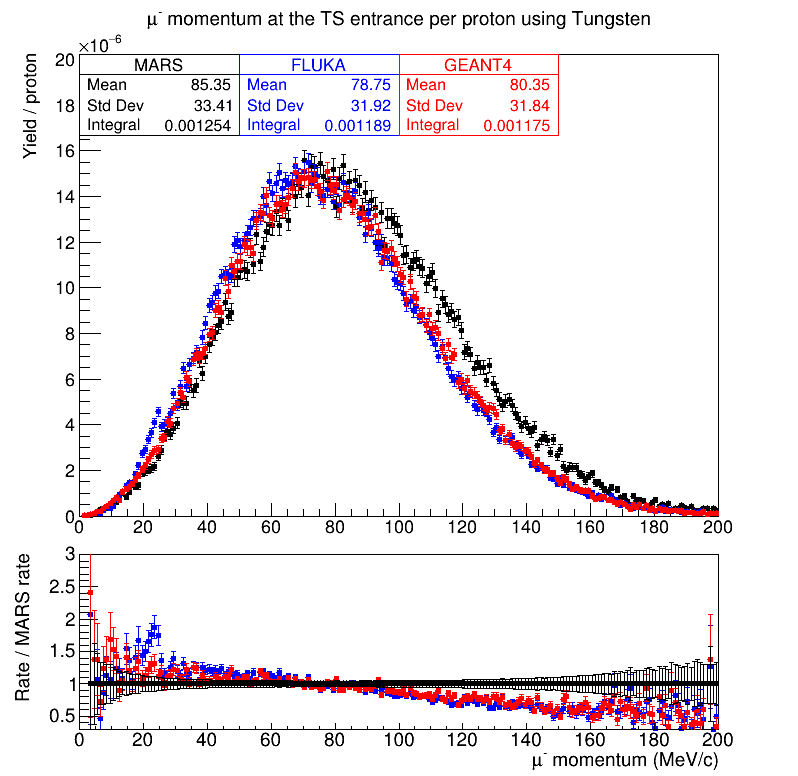}
    \includegraphics[width=0.49\linewidth]{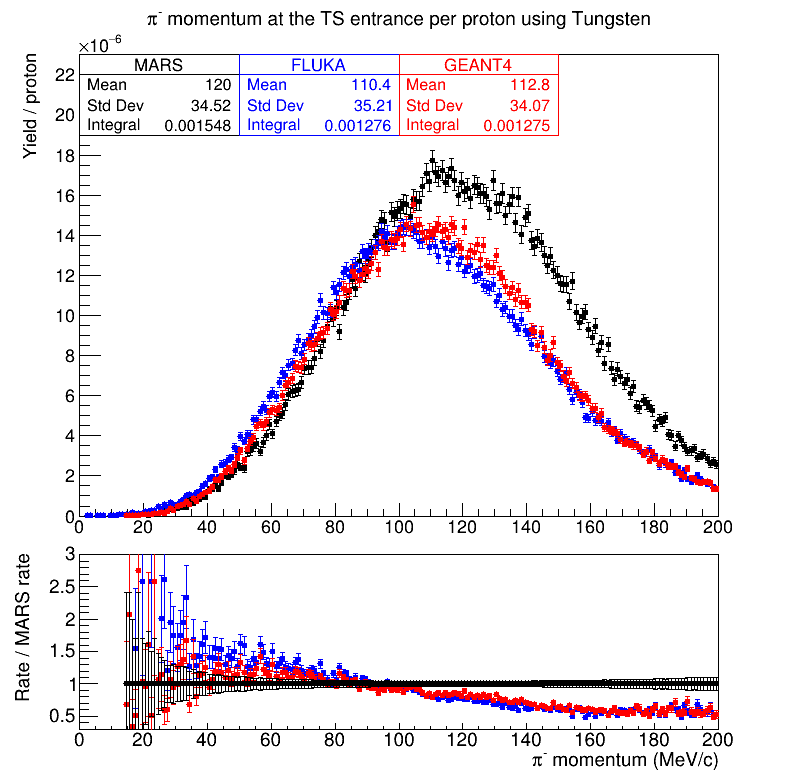}
    \caption{Muon (left) and pion (right) momentum spectrum at the transport solenoid entrance per 800 MeV primary proton on the tungsten conveyor target.}
    \label{fig:target_yield_tungsten}
\end{figure}

Figure \ref{fig:target_yield_tungsten} shows the negative muon and pion momentum spectrum per primary proton on target for the tungsten conveyor target design using the MARS, GEANT4, and FLUKA MC. The three MC agree well, with the muon (pion) yield per primary proton agreeing within 10\% (20\%). The transport solenoid acceptance is very low for particles above 100 MeV/c, so the disagreement in the high momentum region is less important for Mu2e-II. 

\begin{figure}
    \centering
    \includegraphics[width=0.49\linewidth]{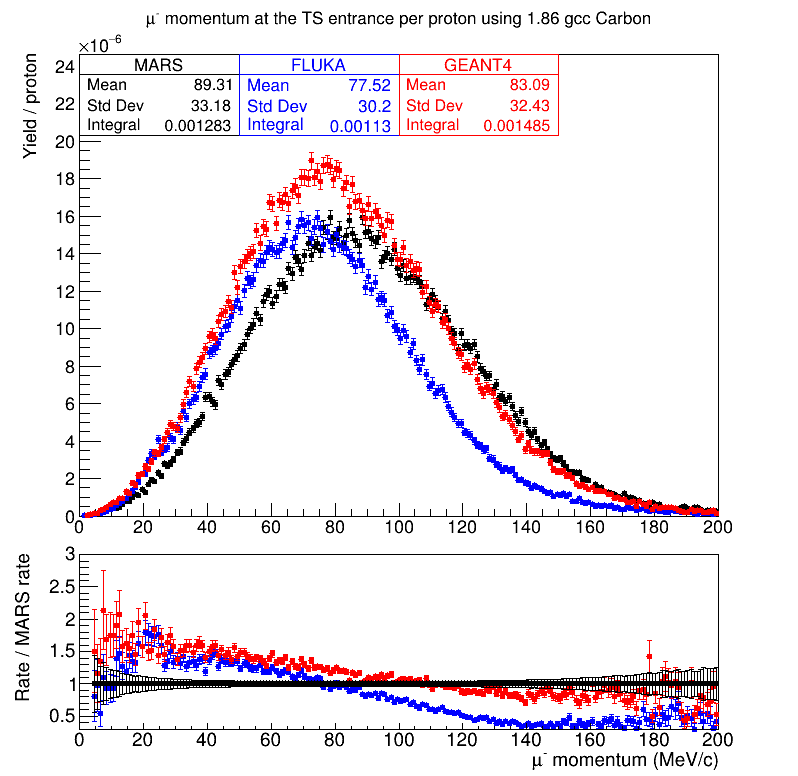}
    \includegraphics[width=0.49\linewidth]{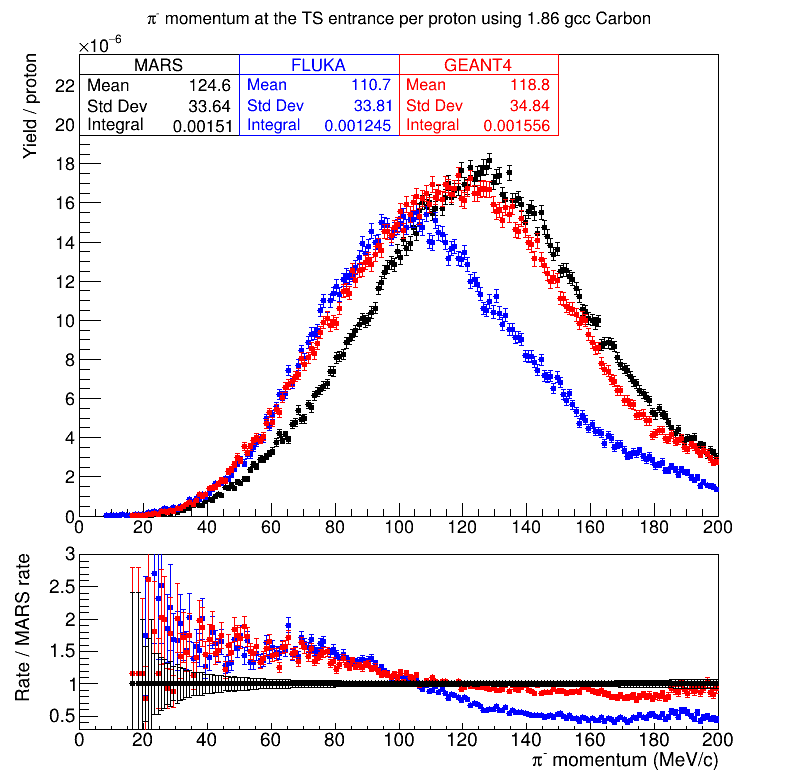}
    \caption{Muon (left) and pion (right) momentum spectrum at the transport solenoid entrance per 800 MeV primary proton on the carbon conveyor target.}
    \label{fig:target_yield_carbon}
\end{figure}

Figure \ref{fig:target_yield_carbon} shows the negative muon and pion momentum spectrum per primary proton on target for the carbon conveyor design using the MARS, GEANT4, and FLUKA MC. Unlike in the tungsten conveyor target case, the MC have large disagreements, most notably GEANT4 which predicts much higher muon and pion yields. GEANT4 predicts higher muon (pion) yields at the transport solenoid entrance using the carbon conveyor target than using the tungsten conveyor target by 26\% (22\%). MARS and FLUKA predict a 2\% (-2\%) and -5\% (-2\%) change in the rates using the carbon conveyor target, respectively, making the large increase predicted using GEANT4 an outlier. The shape of the momentum distributions disagree between the MC using the carbon conveyor target, including between FLUKA and MARS. These differences between the MC are not yet understood, where studies to compare the differential pion production cross section on both carbon and tungsten using 800 MeV protons in the three MC are underway. 

\subsection{Tracker}\label{sec:tracker}
The Mu2e-II tracker discussion focused on the challenges the Mu2e-II environment will create and possible changes and improvements to the current Mu2e straw tube tracker design. Alternative technologies and geometries were also considered.

\subsubsection{Improvements to the current design}\label{sec:trackerimprov}

By the time Mu2e-II is running, the existing Mu2e tracker will have been exposed to significant radiation and time. This causes concerns about aging effects in addition to performance metrics such as leaks, straw sag and radiation damage. A new tracker needs to be built according to the Mu2e-II beam and sensitivity requirements as indicated in the recent Snowmass paper~\cite{Mu2e-II:2022blh}. A straightforward approach is to improve the current design by conducting R\&D and using the latest advancements in technology to update or replace tracker elements. This approach has also the benefit of using the experience gained throughout the life cycle of the Mu2e tracker construction and operations for the design of the new tracker.

\begin{figure}[H]
\begin{center}
     \begin{subfigure}[b]{0.45\textwidth}
         \centering
         \includegraphics[width=\textwidth]{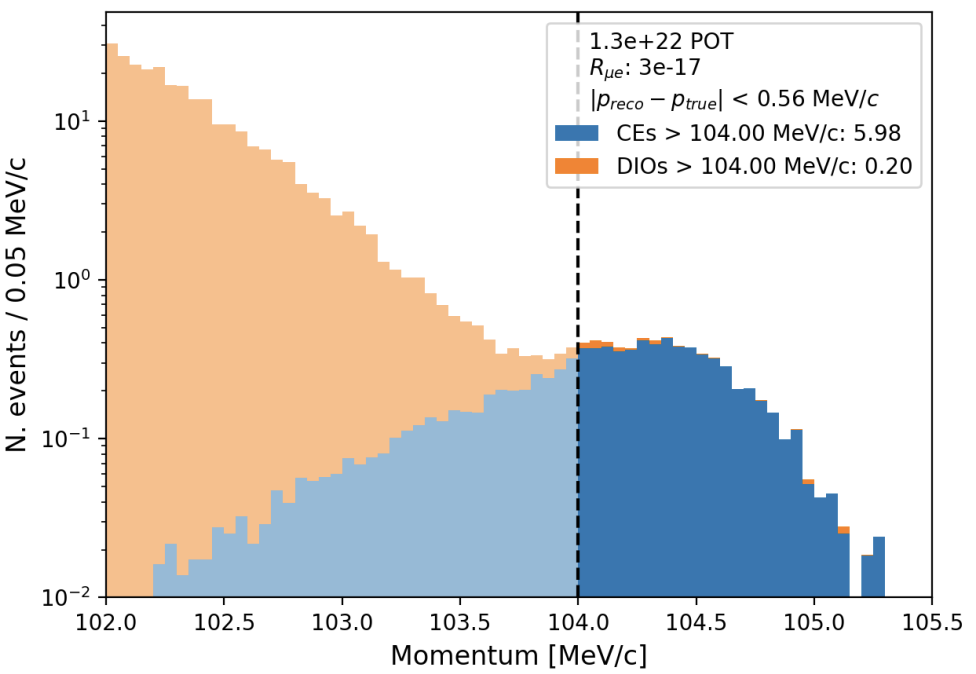}
                           \caption{}
     \end{subfigure}
     \hfill
     \begin{subfigure}[b]{0.45\textwidth}
         \centering
         \includegraphics[width=\textwidth]{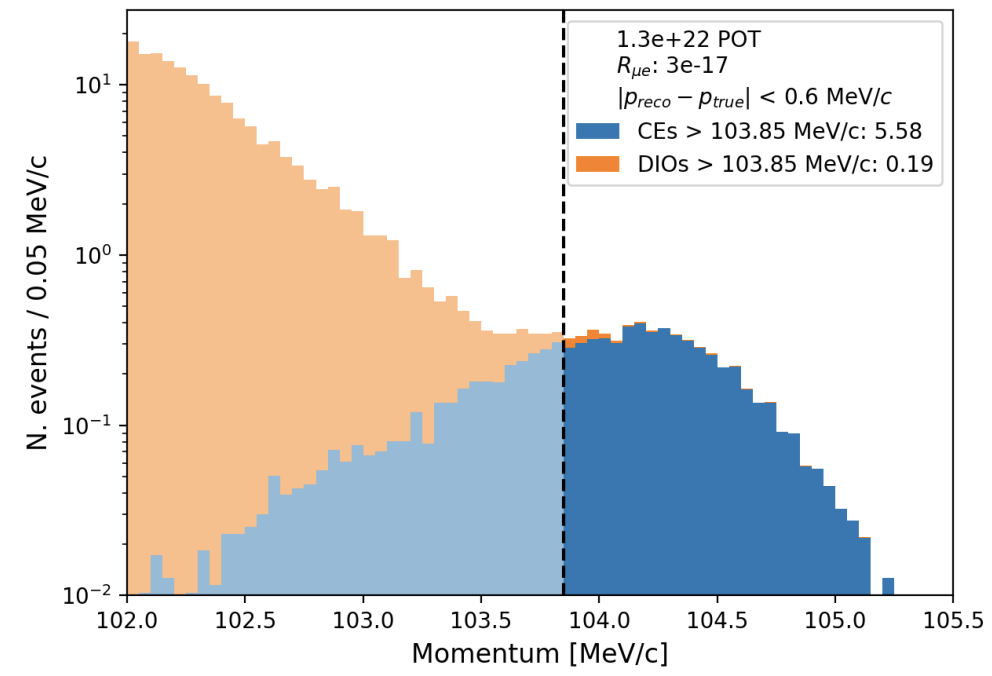}
                           \caption{}
     \end{subfigure}
     \hfill   
\end{center}
\caption{ (a) CE signal vs DIO tail spectrum for 15 $\mu$m Mu2e straws. (b) CE signal vs DIO tail spectrum for 8 $\mu$m prototype Mu2e-II straws.}
\label{fig:strawSensitivity}
\end{figure}

The Mu2e-II tracking environment is even more challenging due to an order of magnitude change in occupancy and radiation dose. With an expected improvement of 100 times POT, more muons are stopped compared to the Mu2e, which increases the DIO background, making it the dominant background of the experiment. Improved momentum resolution is critical in separating much of the DIO background tail from the signal conversion electrons. Removing material from the active tracking area allows for limited improvement, as shown in Figure \ref{fig:strawSensitivity}, but encounters mechanical limits and increases construction difficulty. Due to this limitation, changes in the geometry, drift gas and electronics are needed to fulfill Mu2e-II requirements while improving the momentum resolution.  

\begin{figure}[H]
\begin{center}
         \centering
         \includegraphics[width=0.5\textwidth]{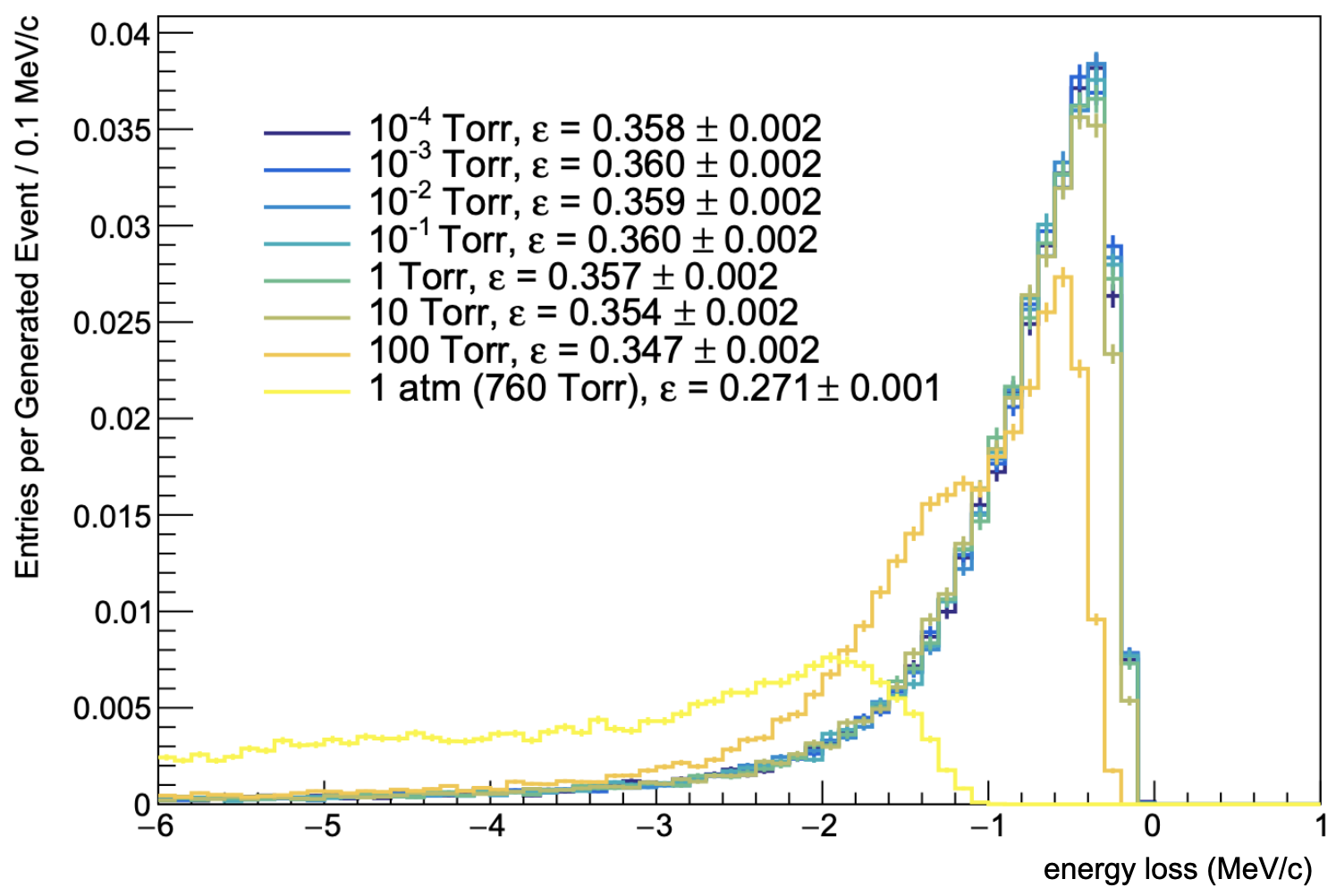}
\end{center}
\caption{A comparison of true MC conversion electron energy loss for different detector solenoid vacuum configurations.}
\label{fig:vacuum}
\end{figure}

Current R\&D is taking place on the production of 8 $\mu$m thin straws that are built by two layers of 3 $\mu$m Mylar wound in a spiral along the straw axis with 2 $\mu$m adhesive keeping the seams together. These ultra thin straws pass mechanical requirements for the Mu2e tracker and increase the sensitivity 10\% by reducing the mass from the active tracking area of the detector. The prototype straws do not have a metallization layer at this point and a reduction in the thickness of the inner and outer metallization layers are in discussion. The thickness of the outer layer of metallization has a profound effect on the leak rate of the straws. Other production methods such as ultrasound welding used in COMET straws~\cite{VOLKOV2021165242} or microforming of extremely thin metal tubes~\cite{MENG2022353} should be investigated as alternatives to the Mu2e style straws. In addition, the latest simulation studies argue that the Mu2e experiment could run at lower bore vacuum as far as up to $10^{-1}$ Torr as shown in Figure \ref{fig:vacuum}. Running at lower vacuum may lead to additional technical issues but could be considered in order to expand the Mu2e-II tracker design space.

From experience with the mechanical construction of the Mu2e tracker panels and track reconstruction simulations, changes could be made to geometry and materials to improve on the original design. One idea is to add a third layer of straws to the panel, which improves left/right ambiguity and pattern recognition. This would make the sealing of the middle layer of straws quite difficult. Sealing channels should be incorporated into such a design and the latest 3d printing technologies and new materials should be investigated to provide a strong inner ring compared to the plastic inner ring of the Mu2e tracker panel. Straw size could be changed by optimizing construction complexity to charge load on the wire and momentum resolution. An optimized straw size could allow for a simpler tracker panel production. Ideas for improving the construction also exists but they depend on the changes to the tracker panel geometry and layout.  

Another key element in the tracker design is the type of drift gas used in the detector. A different gas ends up changing critical parameters like signal threshold, gain, drift velocity, spatial resolution, diffusion and more. ArCO$_2$CF$_4$~\cite{MURTAS2020106421} and HeCH$_{10}$~\cite{WU2021165756} are possible candidates that are being used in other drift chambers. Signal simulation should be conducted to make sure there are no red flags associated with other gases with respect to Mu2e-II requirements. Information on the aging effect of these gases or potential new mixtures are scarce in the literature. Aging studies should start during the Mu2e era to come to an agreement on the gas of choice before the production starts for Mu2e-II. 

\begin{figure}[H]
\begin{center}
         \centering
         \includegraphics[width=0.5\textwidth]{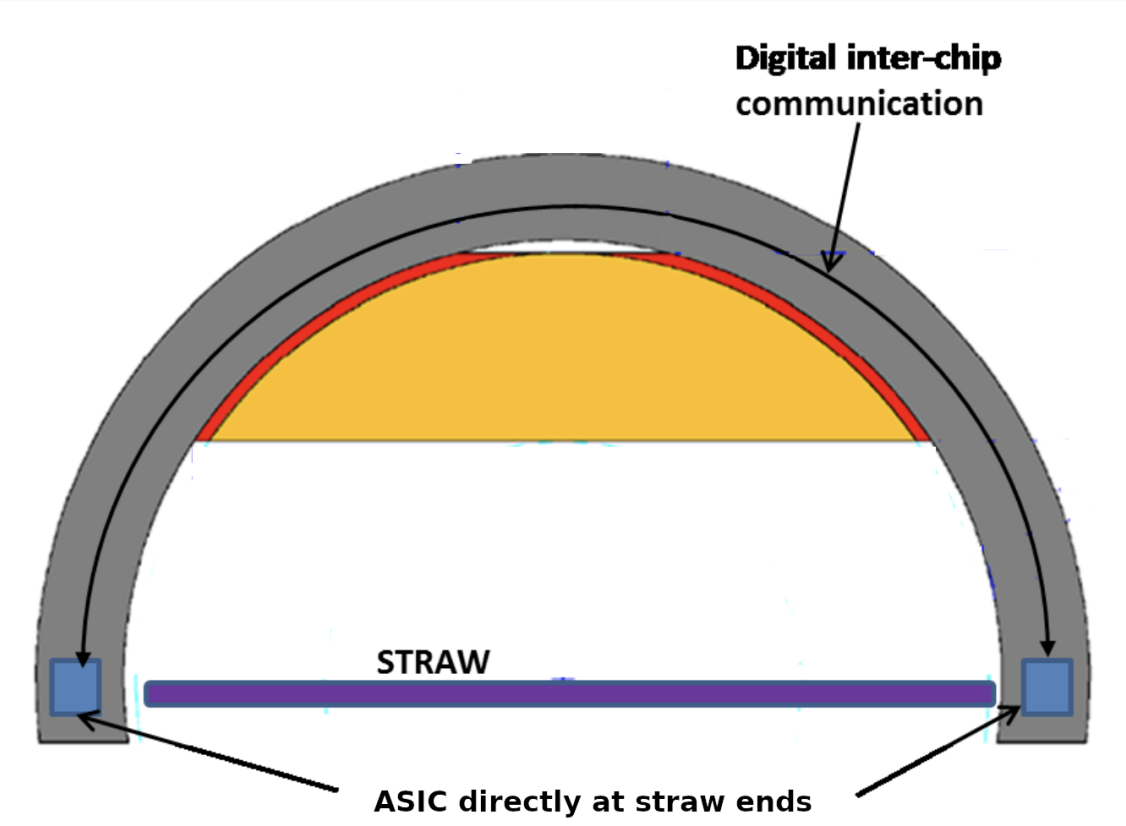}
\end{center}
\caption{A cartoon for the usage of ASIC chips for the readout of the Mu2e-ii straws. Switching to ASIC chips open up new possibilities for mechanical design, power delivery and cooling.}
\label{fig:asicCartoon}
\end{figure}

Finally, the tracker electronics should be reconsidered for Mu2e-II. An initial estimate of 750 hadrons/cm$^2$/yr (hadrons $>$ 30 MeV) is 12 times larger than the Mu2e calculations and necessitates an even stricter requirement on the radiation hardness of the electronics. The current safety factor on electronics is 12 times, therefore the new electronics need to be verified for higher dose. In the meantime, the biggest challenge is the occupancy. The hit rate on an average Mu2e straw is 100 kHz and the predicted Mu2e-II hit rate is 1.6 MHz. Not only the Mu2e-II tracker electronics need to process at this high rate, triggering on a single hit cluster is also desired to improve momentum resolution. Using ASIC chips for signal processing seems to be the future direction to handle the increased rates and a simple cartoon of a tracker detector using these chips can be seen in Figure \ref{fig:asicCartoon}. ASIC chips also require less power and occupy less space. This gives more flexibility to the mechanical design of the tracker panel and how power is delivered. Another modern approach to signal processing is to pass hit classification and filtering to FPGA using AI/ML~\cite{Chiarello_2017}. This is a developing and very popular R\&D area and research should be conducted to find the best use scenario for the Mu2e-II tracker detector. 

\subsubsection{Alternative tracker designs}\label{sec:trackeralternate}

\begin{figure}[H]
\begin{center}
         \centering
         \includegraphics[width=0.5\textwidth]{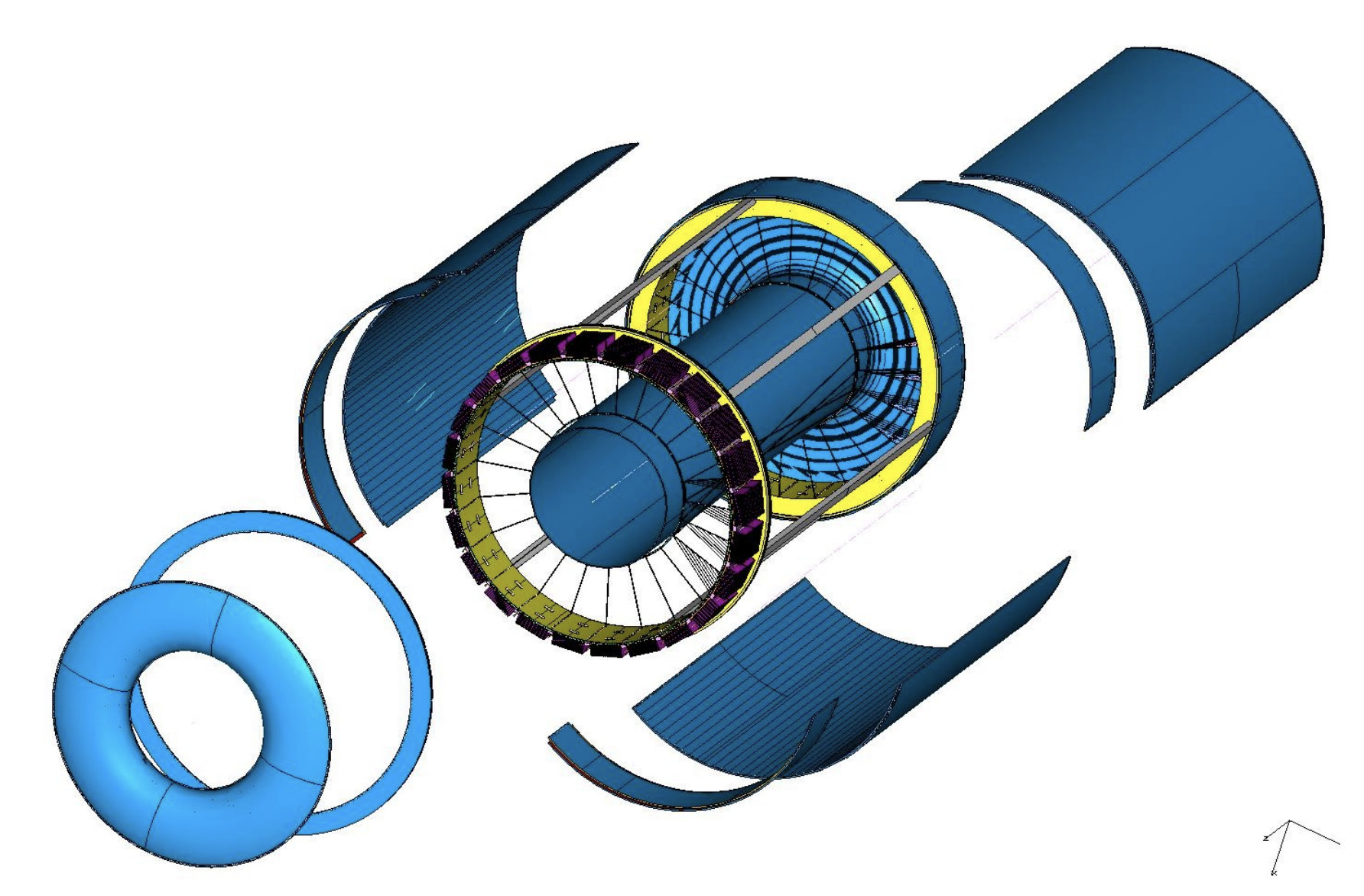}
\end{center}
\caption{I-Tracker design for an alternative to the Mu2e like tracker. In this design, drift cells made out of sense wires and field wires are strung between spokes aligned in the beam direction within a gas vessel that makes the seal against the vacuum.}
\label{fig:iTracker}
\end{figure}

The ``I-tracker'' design has been proposed during the early days of the Mu2e project~\cite{i-tracker}. In this design, square drift cells made out of a sense wire centered within field wires that are strung along the beam direction as shown in Figure~\ref{fig:iTracker}. Wires are precisely positioned into a metal frame that locates them down to 20 $\mu$m. The metal frame is installed within an ultra light gas vessel that provides the gas seal. This approach alleviates a lot of construction and mechanical issues with the straw detector, primarily the wire positioning and gas leaks. To control the multiple scattering within the gas environment, He is proposed. He as a drift gas is slower compared to the ArCO$_2$ used in the Mu2e tracker, however it potentially offers a different way to deal with the increased hit rates of the Mu2e-II environment. The latest studies aim to reduce the drift cell size to a 3 mm $\times$ 3 mm square. The effect of this change to the momentum resolution should be studied with respect to the Mu2e-II rates.   

The ultra light gas vessel idea is also applicable to a Mu2e like tracking detector. Similar to the I-tracker, the aim is to reduce the leak requirement on the tracker straws and to ease the construction of a tracker panel with straws. The workshop did not focus on this idea~\cite{altTracker}, however with the recent developments on the vacuum requirements of the Mu2e tracker and the interest in the I-tracker for the Mu2e-II, a fusion of the I-tracker and the straw tracker remains an interesting prospect.

\subsection{Calorimeter} 
\label{sec:mu2ecalorimeter}

The	Mu2e calorimeter~\cite{Calo:Atanov2018} consists of 1348 pure CsI crystals, comprising two disks, read out with custom SiPMs developed with Hamamatsu company. 
The calorimeter has a robust rate performance at Mu2e rates but may be	challenged by Mu2e-II instantaneous rates that	are two to	three times higher. The x10 integrated radiation dose on the calorimeter readout electronics motivates the study of appropriate rad-hard readout electronics	at a level informed by the HL-LHC detector upgrades. An alternative calorimeter design has been developed based on BaF2 crystals readout with  solar-blind UV sensitive SiPMs that efficiently collect the very fast UV component ($\sim$ 220 nm) of the scintillation light while suppressing the slow component near 310 nm. This alternative design would be considerably more robust against Mu2e-II rates but requires the development and commercialization of the required solid-state photo sensors, which is ongoing. The Mu2e-II calorimeter should have the same energy ($<$10\%) and time ($<$500 ps) resolutions as in Mu2e, aiming to provide a standalone trigger, a track seeding, and PID as before. However, the Mu2e-II environment presents two challenges	to the calorimeter system:
\begin{enumerate}
    \item The pileup with respect to CE seems to scale linearly with beam intensity, so to keep the same level we have in Mu2e (15\%) with 150 ns we need to rescale the new signal length. The signal length for Mu2e-II should be 75 ns.
    \item Under the assumption that the total integrated dose (TID) from the beam flash in the calorimeter from 800 MeV protons scales as the number of stopped muons with respect to the Mu2e 8 GeV beam, a factor 10 increment is expected. The $\times$10 increase in the integrated dose (neutron fluence) corresponds to 10 kGy ($1\times10^{13}$ n/cm$^2$/sec) for both crystals and sensors motivates consideration of more radiation tolerant crystals and	sensors such as Barium fluoride (BaF2) crystal and Solar blind SiPMs.
\end{enumerate}

\begin{figure}[H]
    \centering
    \includegraphics[width=1\linewidth]{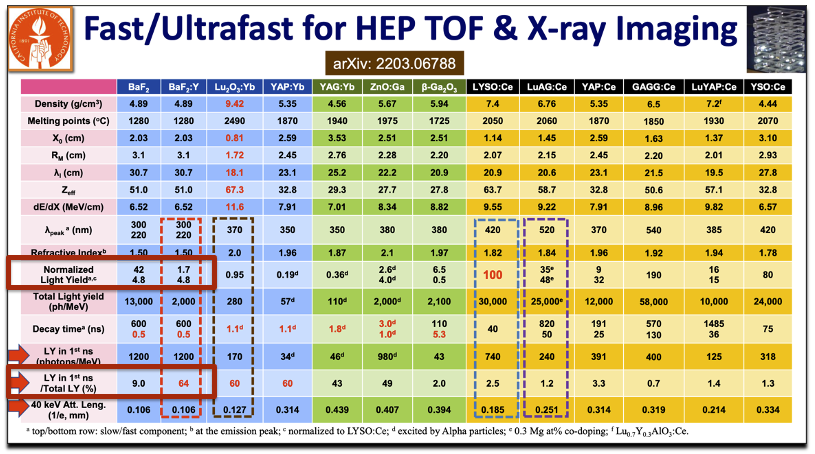}
    \caption{Main parameters of crystals used in HEP; comparison with the BaF2 and BaF2 Yttrium doped crystal.}
    \label{fig:calo_table}
\end{figure}

Supported by the DOE HEP ADR program, the Caltech crystal lab has been developing yttrium doped ultrafast barium fluoride crystals~\cite{Zhu2017} to face the challenge of high event rate and severe radiation environment. In 2017, they found that yttrium doping in BaF$_2$ is effective in the suppression of its slow scintillation component with 600 ns decay time while maintaining its ultrafast sub-ns scintillation component unchanged. In a collaboration with SICCAS and the Beijing Glass Research Institute, yttrium-doped BaF$_2$ crystals of up to 19 cm length were successfully grown, and showed a factor of ten suppression in the slow component (see Figure~\ref{fig:calo_table}). They also found that 25 cm long BaF$_2$ crystals are radiation hard up to 120 Mrad. This R$\&$D will continue in collaboration with crystal vendors. Support is requested to develop high-quality yttrium doped BaF$_2$ crystals of large size for the Mu2e-II calorimeter.\\ 

\begin{figure}[H]
    \centering
    \includegraphics[width=0.55\linewidth]{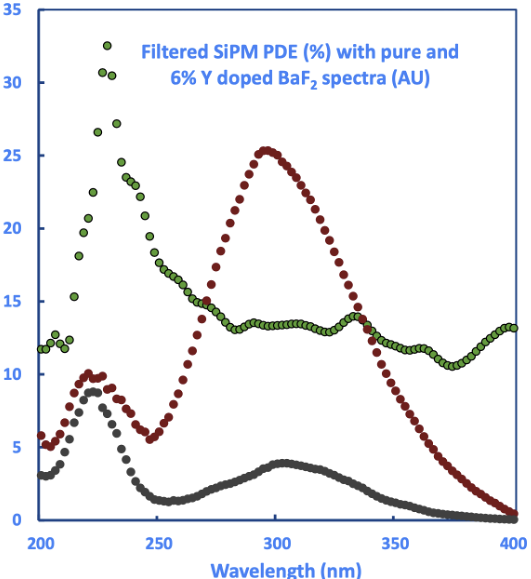}
    \caption{Green: 3 layers-SiPM PDE; Red: pure BaF2 emission; Grey: 6\% Y doped BaF2.}
    \label{fig:calo_sipm_uv}
\end{figure}

The development of a photosensor capable of efficiently reading out the fast component of barium fluoride is an urgent component of R$\&$D for Mu2e-II. Working with FBK and JPL, the Caltech group is developing a large area SiPM (nominally the same size as that used in Mu2e) that incorporates an integrated ALD filter having high efficiency at the 220 nm fast component and substantial extinction of the 300 nm slow component~\cite{Hitlin2020a}, as shown in Figure~\ref{fig:calo_sipm_uv}. A second phase of this development will further incorporate a delta-doped layer that will improve the rise and decay times of the SiPM response, as was demonstrated with the Delta-doped RMD APDs. The initial batch of wafers has been furnished by FBK to JPL. These have had the ALD filter applied and are now being returned to FBK for the next step in processing. Several rounds of development are anticipated, motivating a request for R$\&$D funds from Mu2e-II.\\

\begin{figure}[H]
\centering
\includegraphics[width=0.9\linewidth]{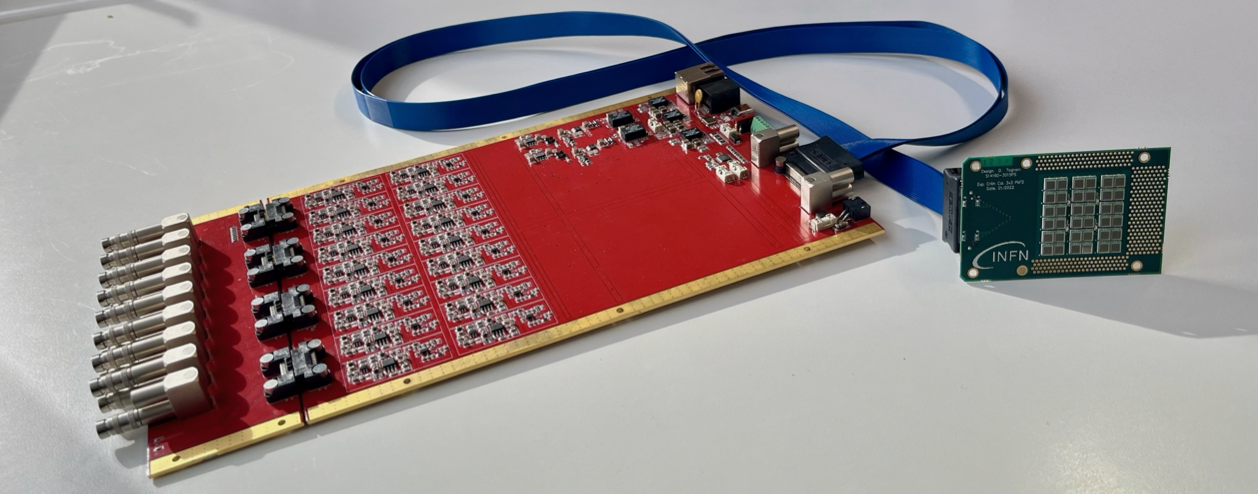}       
 \caption{Crilin calorimeter electronics: controller board (left) and SiPM board (right) matrix.}
    \label{elec1} 
\end{figure}

An extremely fast electronics has been developed by the INFN-LNF group with the collaboration of D. Tagnani (INFN-Roma3) in 2022 for the Crilin calorimeter~\cite{2022crilin}. Crilin’s FE electronics are composed of two subsystems: the SiPM board (Figure~\ref{elec1}, right) and the Mezzanine Board (Figure~\ref{elec1}, left). The SiPM board houses a layer of 36 photo-sensors so that each crystal in the matrix is equipped with two separate and independent readout channels, the latter being composed of a series connection of two Hamamatsu S14160-3010PS SMD sensors. The 10 $\mu$m SiPM pixel size, along with the series connection of two photo-sensors, were selected for a high-speed response, short pulse width and to better cope with the expected total non-ionizing dose (TNID) without showing an unmanageable increase in bias current during operation.

All bias voltages and SiPM signals for each readout channel are transported between the SiPM board and the Mezzanine Board by means of individual 50 $\Omega$ micro-coaxial transmission lines. Decoupling capacitors for each channel, along with a PT1000 temperature sensor, are also installed onboard. This readout scheme can be adapted to the Mu2e-II calorimeter together with an alternative proposal of using 8 cm  LYSO crystals. The proposal has many advantages:
\begin{itemize}
    \item 8 cm length LYSO is enough to achieve O(5\%) energy resolution;
    \item the equivalent noise energy is not a problem and there is good longitudinal response uniformity;
    \item no expected degradation in performance after 10$^{13}$ neutrons/cm$^2$; 
    \item SiPMs already exist, no R$\&$D needed,
    \item the high LYSO light yield permits the use of the SiPMs at low over-voltage and thus enhances the radiation resistance and reduces the power dissipation;
    \item so long as a front-end amplifier is not needed, there are no problems with the irradiation level of electronics.
\end{itemize}

Although this backup solution seems to be practical, simulation studies are needed to verify its rate capability in the Mu2e-II environment. 

\subsection{Cosmic Ray Veto} 
\label{sec:mu2ecrv}

The Mu2e experiment expects one signal-like event per day induced by cosmic rays. These are cosmic muons interacting somewhere in the detector material, producing an electron that by coincidence mimics a signal. The cosmic ray veto (CRV) detector suppresses this background. It consists of four layers of scintillating counters~\cite{Mu2e:2017lae} with a cross-section of $5 \times 2$~cm$^2$ that are read out through wave-length shifting fibers~\cite{Dukes:2018scs} connected to $2 \times 2$~mm$^2$ silicon photomultipliers (SiPM)~\cite{Blazey:2019vdr}. The CRV covers the full detector region of the experiment, all sides, as well as the top. Localized CRV hits coincident in multiple layers (three out of four in most regions) trigger an offline $\sim 125$~ns (to be optimized) veto in the signal window. The Mu2e CRV needs to suppress the cosmic ray background by a factor of a few $1000$ with efficiencies up  to 99.99\% in the most sensitive areas while keeping the dead time low in order to reach single-event sensitivity on the order of  $2.5\times 10^{-17}$ ($6\times 10^{-17}$ 90\% C.L.) with around three years of running.  An upgrade to the CRV system is required for Mu2e-II as described below~\cite{Mu2e-II:2022blh}.  

The duty cycle of Mu2e-II is expected to be a factor of three higher, resulting in the need for a three-fold higher suppression of the cosmic ray background. At the same time, the higher beam intensity of Mu2e-II will lead to higher noise (non-cosmic ray hits) rates of more than a factor of three. This poses a challenge to keep the dead time low. 

If the Mu2e CRV were used for Mu2e-II, dead times on the order of 50\% would be expected. The dead time arises from noise hits that occur in coincidence and trigger veto windows. There are two main sources for such fake hits: a) secondaries from the primary production beam, and b) secondaries from stopped muons. It has been shown in simulations that the secondaries from the primaries production beam can be suppressed efficiently by improved barite- and boron-loaded concrete shielding. To address the higher rate from secondaries from stopped muons, the channel rate needs to be reduced, which reduces the false confidence rate, reducing the deadtime. To reduce the channel rate, a finer segmented detector concept is proposed (see below). 

\begin{figure}[H]
    \centering
    \includegraphics[width=0.48\linewidth]{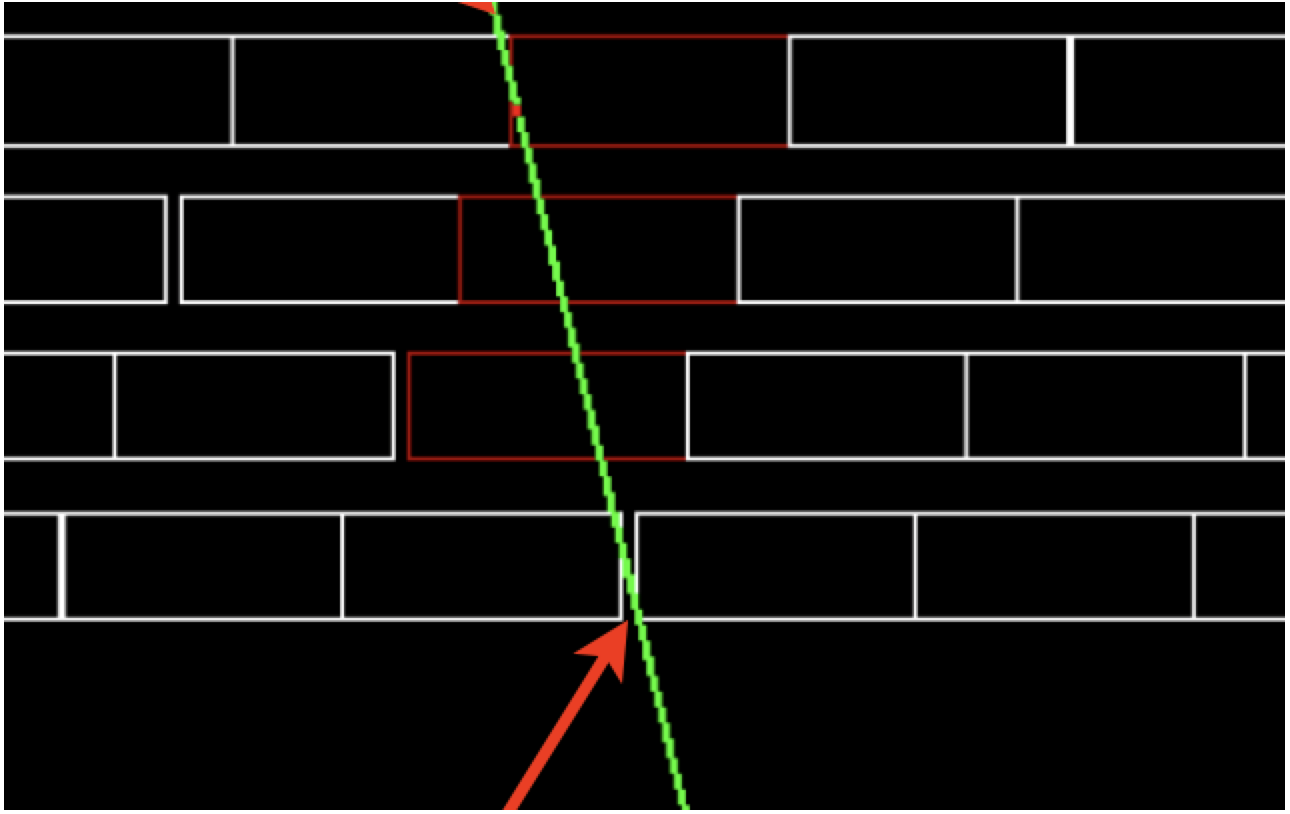}
    \includegraphics[width=0.48\linewidth]{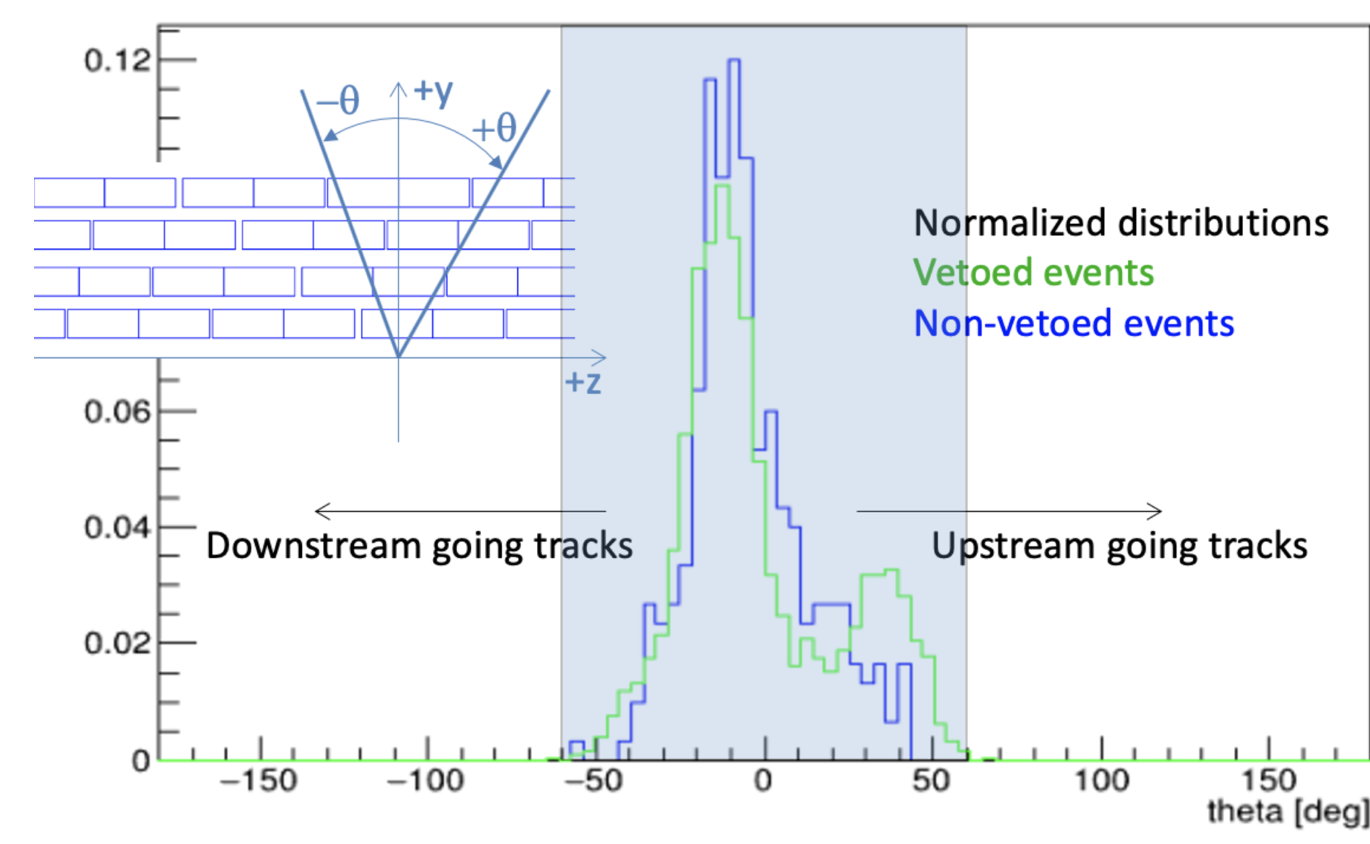}
    \caption{Left) example of a cosmic muon in a CRV gap. Right) azimuthal-angle distribution of cosmic rays resulting in event-like events.}
    \label{fig:crv:gap}
\end{figure}

There are different sources of inefficiency in the CRV. Although the Mu2e CRV is designed to minimize gaps by offsetting the different layers, there remain some corner cases where the cosmic muons manage to sneak in through gaps. Figure~\ref{fig:crv:gap} (left) shows such an example. The cosmic ray background from such gaps in Mu2e-II with the Mu2e CRV design is estimated to be $0.22\pm0.15$. It turns out that most of the cosmic muons producing a signal-like electron have an azimuth-angles (angle with respect to ``up'') smaller than 60$^\circ$ as shown on the right side of figure~\ref{fig:crv:gap}. 

\begin{figure}[H]
    \centering
    \includegraphics[width=0.8\textwidth]{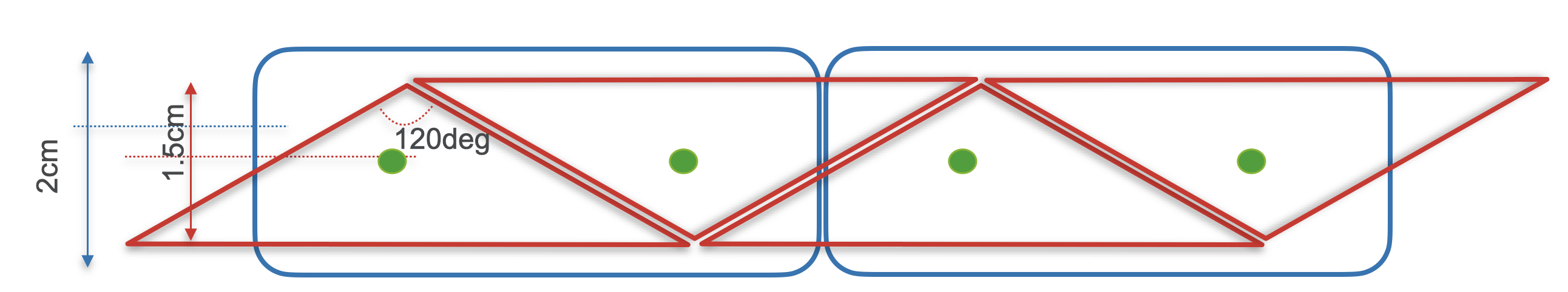}
    \caption{Sketch of the proposed triangular shaped counter design in red overlayed on the shape of the Mu2e CRV design in blue. }
    \label{fig:crv:triangle}
\end{figure}

This motivates the proposal for an improved CRV design consisting of triangle-shaped counters with an angle of 120$^\circ$ stacked together. Figure~\ref{fig:crv:triangle} shows a sketch of the proposed design. This design not only reduces the rate per channel due to its higher granularity, but it also reduces the inefficiencies due to gaps. For cosmic rays with an azimuth-angle smaller than 60$^\circ$ it's impossible to travel along a counter gap. In addition, cosmic rays at these shallow angles deposit more energy in other layers due to the longer path length through the module. With this improved design, the cosmic ray background in Mu2e-II from gaps is estimated to be below $0.1$ events. 

A second source of inefficiency is uninstrumented areas of the detector. The opening for the transport solenoid is the most prominent example. The cosmic ray background from this opening for Mu2e-II is estimated to be $0.08\pm0.02$. So far this was considered irreducible. However, it turns out that additional shielding would allow to reduce this contribution by about a factor of 2.

A third source of inefficiency is an insufficient light yield. An extensive program is ongoing to characterize and understand the aging of the Mu2e CRV detector. Due to these aging effects, at least parts of the CRV will need to be replaced for Mu2e-II. There are multiple options to increase the overall light yield for a Mu2e-II detector:
\begin{itemize}

    \item Moving from 1.4~mm to 1.8~mm wavelength shifting fibers increases the light yield by 24\%. This was already done for some of the most critical top modules for Mu2e~\cite{Solt:2023uxp}.
    \item More modern SiPMs (for example S14160) have significantly higher photon detection efficiency. Not only is the overall efficiency improved, but they are also more sensitive to the wavelength of the wavelength-shifting fibers. 
    \item It was shown that potting the fibers channels with silicon resin improved the light yield by 40\%~\cite{Artikov:2017fmr}. R\&D on how to seal the channels to avoid damaging the readout due to leaking resin is ongoing.
\end{itemize}

An additional background source is neutral cosmic particles that produce a signal-like electron by coincidence. This contribution is not negligible for Mu2e-II. Adding 6 feet of concrete shielding above the target reduces this background to $0.02\pm0.002$. 

The readout bandwidth of the Mu2e CRV is limited by the 10~MB/s link between front end boards (FEB) and readout controllers (ROC). In Mu2e, the off-spill window will be used to transmit the on-spill data which is expected to be suppressed by the online-trigger by two orders of magnitude. In Mu2e-II no off-spill time will be available, either the bandwidth needs to be significantly improved or the trigger will need to achieve a significantly higher suppression. An additional challenge in Mu2e arises in the FEB event builder to provide the bandwidth to account for the beam instabilities that require an overhead of up to a factor of 2. To achieve this, the most active channels are sparsified. While the beam is expected to be much more uniform in Mu2e-II the event builder might already be saturated in nominal conditions. Detector side changes will be needed to use the same readout scheme. The proposed higher granularity helps also in this aspect.

In summary, at least parts of the Mu2e CRV can not be used for Mu2e-II due to aging effects and radiation damage in the SiPMs. We propose an improved triangular-shaped CRV design with finer granularity reducing the single-channel rate, which reduces the dead time. At the same time, the triangular shape improves the detector's efficiency. The light yield of the counters is also enhanced with this geometry. For the CRV at Mu2e-II, enhanced high-Z shielding is the most crucial part. It reduces the readout noise, which is crucial to control the dead time. In addition, improved shielding around the transport solenoid opening reduces muons sneaking in through that gap and additional shielding on top of the target suppresses cosmic background from neutrals.  With this improved detector design and shielding, we predict that the CRV background can be limited to the level summarized in Table~\ref{tab:crv:background}.  

\begin{table}[H]
    \centering
    \begin{tabular}{l|r}
         source & background \\
         \hline
         inefficiencies from gaps in counters & $<$ 0.1 \\
         inefficiencies from uninstrumented gaps & $0.08\pm0.02$ \\
         neutrals & $0.02\pm0.002$ \\
         \hline
         \bf total & $0.20\pm0.08$ \\
    \end{tabular}
    \caption{Expected cosmic ray background budget for Mu2e-II with the proposed changes. Optimizing the the momentum cuts reduces the total to $0.17$. From~\cite{Mu2e-II:2022blh}.}
    \label{tab:crv:background}
\end{table}

Moving forward, we plan to seek funding to build triangular prototypes that should be installed in the running Mu2e experiment. An R\&D program is needed to improve the counter profile, explore the possibility of coating for enhanced reflectivity, and improve or fill the fiber channels. In addition, the aging behavior of such new counters needs to be studied. Mu2e-II prototypes installed in Mu2e will benefit the running experiment by providing an additional handle on determining the CRV efficiency. In parallel, dedicated simulation efforts with triangular-shaped counters are needed. We propose to procure some enhanced shielding to start building up experience. The shielding design needs to be optimized.  Obviously, running Mu2e will reduce uncertainties related to the CRV design needed to meet requirements for Mu2e-II. However, detector R\&D must be started now, in order to be ready to meet the challenges of Mu2e-II.

\subsection{Trigger and DAQ}

Obtaining higher sensitivity in Mu2e-II imposes more performant requirements on the trigger and data acquisition (DAQ) compared with Mu2e. We make the following assumptions:
\begin{enumerate}
    \item Compared with Mu2e, Mu2e-II will have twice the number of detector channels and five times the number of pulses on target, leading to a ten times higher data rate.
    \item The Mu2e-II event size may be three times the expected Mu2e event size of 200 kB, because of the greater channel count and more background hits.
    \item We assume that we can support a factor of two in tape capacity over Mu2e, leading to 14 PB/yr.
    \item The required trigger reduction factor in Mu2e-II is $~$3000:1.
\end{enumerate}

Mu2e-II does not have the large 1400 ms gaps between batches of spills that Mu2e does. That is, Mu2e-II will have a steady event stream, so there is no ``catch-up'' time. Buffering can be used to handle local fluctuations in event rate, but not gaps in beam from the accelerator. Two scenarios may be considered for cost-effectiveness: ({\it i\/}) Large CRV buffers and a software trigger; ({\it ii\/}) Small CRV buffers and a hardware trigger.

An important decision is which detector subsystems are triggered, possible approaches include:
\begin{itemize}
    \item Same as Mu2e -- Stream all tracker and calorimeter data; software trigger for CRV based on the tracker and calorimeter;
    \item Stream calorimeter; hardware trigger for tracker and CRV based on the calorimeter;
    \item High-level software trigger for storage decision.
\end{itemize}

The radiation levels at the detector will be higher than for Mu2e. Mu2e-II will likely not want to design its own rad-hard links. Instead, we will probably want to make use of technology developed for CMS/ATLAS.

The generic data readout topology is anticipated to be a multi-stage TDAQ system, consisting of the front-ends, data concentrators, event builders, and storage decisions. The data concentrator aggregates small front-end fragments into larger chunks for efficient event building. Data is switched from the concentrator layer to the event builder layer so that full events arrive at the event builder layer and are buffered. At the event builder preprocessing or filtering could occur.
In the storage decision available decision nodes make high level storage decisions.

Four TDAQ LOIs were contributed in the Snowmass 2021 process:
\begin{itemize}
    \item A 2-level TDAQ system based on FPGA pre-processing and trigger primitives;
    \item A 2-level TDAQ system based on FPGA pre-filtering; Features of these first two approaches include: Mu2e already using FPGAs in the ROCs and DTCs; FPGA can offer flexibility for algorithm development; these solutions are more tightly coupled to the sub-detector readout systems.
    \item TDAQ based on GPU co-processing. Features include data transfer is challenging and importing C-style algorithms is not simple
    \item Triggerless TDAQ based on a software trigger (scaling up the Mu2e system). Features include cooling in the DAQ room and data transfer and processing become very challenging.
\end{itemize}

FPGA algorithm development could make use of C-style coding High Level Synthesis (HLS), offering advantages over
manual VHDL or Verilog development. CMS is investing in the HLS approach to FPGA algorithm development. There is also a hls4ml collaboration developing machine learning tools using HLS.
In principle helix pattern-recognition can be coded on FPGAs, and one could use very powerful FPGAs if located outside the detector solenoid.

R\&D will benefit greatly from Mu2e experience. The current Mu2e trigger algorithms can be used
on commercial FPGA boards to perform feasibility studies. A successful demonstration would consist of delivering a demonstrator that could be operated parasitically in the Mu2e TDAQ towards the end of Run 2.
\subsection{Physics and sensitivity}

The aim of Mu2e-II is to achieve a single event sensitivity of $\mathcal{O}(10^{-18})$. Designing an experiment capable of this level of sensitivity involves a multi-faceted approach. On-going R$\&$D focuses on three main aspects: 

\begin{itemize}
  
    \item stopping target design, designing a target that is optimal for conversion while minimizing energy losses of possible signal electrons;
      \item production target design, optimizing pion production while designing a technically feasible target; 
    \item detector design, specifically designing a tracking system capable of the required momentum resolution.
\end{itemize}

This discussion is divided into two sections. Section~\ref{sens_pheno} details phenomenological considerations needed to inform our choice of stopping target, and how we present our final result. Section~\ref{sens_tech} details technical considerations that inform the design of the experiment while ensuring we maintain the required signal-to-background ratio and momentum resolution. 

\subsubsection{Phenomenological Considerations}
\label{sens_pheno}

\noindent \textbf{Improved understanding of nuclear dependence of conversion rate}
\vspace{0.5cm}

 Measurements of the atomic number dependence of the rate of muon-to-electron conversion are detailed in Refs.~\cite{Kitano2002, Cirigliano:2009bz, Heeck:2022wer}. In the event of a charged lepton flavor violation (CLFV)  signal at Mu2e and/or COMET, we can elucidate the Lorentz structure of New Physics coupling at Mu2e-II by obtaining additional conversion measurements in other nuclei. Mu2e and COMET both intend to use an aluminum (Al-27) target, Mu2e-II must choose a material that is complementary to Al-27, but which is also technically feasible in the Mu2e-II experimental set-up.

Reference~\cite{Kitano2002} is the most widely cited treatment of the atomic number dependence of $\mu N \rightarrow eN$,  which has been extended by Refs.~\cite{Cirigliano:2009bz,Heeck:2022wer}. 
 During the workshop a new approach to the estimation of the atomic number dependence of the conversion rate was presented. Barrett moments~\cite{Barrett1970} are utilized to include muonic $X$-ray measurements of nuclear charge distributions in addition to electron scattering data. The latter alone was used in previous studies~\cite{Kitano2002, Cirigliano:2009bz, Heeck:2022wer}. By including muonic $X$-ray data we can account for deformation in nuclei, which is particularly apparent for higher-$Z$ materials. Nuclear deformations are parametrized by a three-parameter Fermi distribution, which takes into account the effect of permanent quadrupole moments. In addition, the new approach uses a deformed relativistic Hartree-Bogoliubov method to account for neutron distributions, this differs from Ref.~\cite{Kitano2002, Cirigliano:2009bz, Heeck:2022wer} which simply scaled the proton distributions by a factor of $N/Z$.

In general, the presented treatment differs from Ref.~\cite{Kitano2002,Cirigliano:2009bz,Heeck:2022wer} in several ways:
\begin{itemize}
    \item \textbf{More data through inclusion of muonic $X$-ray data which helps account for deformations:} the use of muonic $X$-ray data substantially enlarges the sample size in the regime above $Z$=60, where many nuclei have substantial quadrupole deformations. Elastic electron scattering and muonic $X$-ray data are combined using Barrett moments, and devise a procedure to incorporate the effect of permanent $Y_{20}$ deformations on the effective nuclear skin thickness.
    
    \item \textbf{Accounting more accurately for neutron distributions:} a model based on a deformed relativistic Hartree-Bogoliubov calculation is used to estimate the neutron distributions. This model allows the inclusion of a wider variety of nuclei and explores isotopic effects on the conversion rate.
    
    \item
    \textbf{Accounting for isotope abundance, and feasibility of single isotopes:} recognizing that separated isotopes are hard to obtain in sufficient quantities to make practical stopping targets of $\sim 100$ g. Instead, elements that are comprised of a single stable isotope or in which the dominant stable isotope is greater than $\sim 90\%$ abundant are explored. Vanadium is proposed as the best option for Mu2e-II.
    
\end{itemize}

 A detailed publication describing the result of the work is imminent, and we leave the explicit details to that document.

\vspace{0.5cm}

\noindent \textbf{Normalization:  Presenting conversion results  }

\vspace{0.5cm}
The conventional approach to the normalization of $\mu \to e$ conversion experiments, quoting the conversion rate (experimental limit or theory prediction) normalized to
the measured rate of $\mu$ capture on a given nucleus, has been in place for more than seventy years. As the current round of experiments approach a sensitivity that may yield a signal,
this convention should be re-examined, particularly because future experiments will likely focus on the $Z,A$ dependence of the conversion process. A talk at the Workshop proposed a revised convention for presenting both theoretical and experimental results on $\mu \to e$ conversion going forward that addresses the shortcomings of the historical approach.
\par
Normalization to the muon capture rate is not precisely analogous to the idea of a branching fraction (the number of decays into a particular mode, divided by all decays), which would be to divide the conversion rate in the field of a particular nucleus by all possible fates of the muon ($\mu\rightarrow e$ conversion, DIO or nuclear capture). However, normalization to muon capture was effectively codified by Weinberg and Feinberg in 1959~\cite{Weinberg1959}; essentially all results or predictions on muon to electron conversion have henceforth been presented in the form: 

\vspace{-20pt}
\begin{center}
$${R_{\mu e}}\,{\rm{or\, }}{B_{\mu e}}{\rm{(Z)\ or \, CR}}({\mu ^{_ - }}{\rm{N}} \to {e^ - }{\rm{N) }}
=\frac{{\Gamma ({\mu ^{_ - }}{\rm{ + }}\,{\rm{N}} \to {e^ - }{\rm{ + }}\,{\rm{N)}}}}{{\Gamma ({\mu ^{_ - }}{\rm{ + }}\,{\rm{N}} \to {\rm{all\ captures)}}}}\, .$$
\end{center}
\par
Compilations of the history of experimental limits on CLFV processes typically place the 90\% confidence level limits for decays and conversion on the same plot, ignoring the fact that they are normalized differently. The decays are reported as true branching fractions, while the conversion rate limits are on the fraction of muon captures resulting in production of a monoenergetic electron, which does not account for all fates of a muon in a muonic atom. Indeed, the lifetime of such a muon is determined in varying proportions by the conversion rate, a BSM process, by the nuclear capture rate, an incoherent Standard Model process, and by the lifetime of the decay-in-orbit muon, which is modified from the free decay rate by the atomic binding energy, the so-called Huff factor (0.993 for aluminum, 0.981 for titanium and 0.850 for gold)~\cite{suzuki1987}.
\par
There is a simpler alternative to normalization that hews more closely to what both experiments and theoretical calculations actually do, and has benefits in understanding the $Z,A$ dependence of the conversion process. Conversion experiments are normalized to the number of muon stops in the nuclear target within the sensitive time window. This requires knowledge of the total muon lifetime in a particular muonic atom, as well as of a set of relevant experimental efficiencies.
The number of conversion signal events (or to this point, absence of events) is divided by the number of muons stopped in the target (as measured by, {\it e.g.}, counting $2P-1S$ muonic x-rays or other transitions, such as delayed $\gamma$s from muon capture. These experiments do not, in general, measure the muon capture rate).
The number of candidate muons for conversion is determined by muon decay as well as by nuclear muon capture, that is, by the total lifetime in the muonic atom. Thus, for example, the experimental measurement of the total muon lifetime, $864.0 \pm 1.0$ ns in aluminum~\cite{suzuki1987}, with its associated uncertainties, unavoidably enters the calculation of the experimental efficiency and therefore the calculation of the $\mu \to e$ conversion rate. Since the overlap of the muon atomic wave function with the nuclear proton and neutron distribution influences the effective lifetime, the BSM physics and the Standard Model nuclear physics are inextricably mixed. Thus, the measured rate (or limit on the rate) {\it ab initio} depends in part on the muon capture lifetime. The muon nuclear capture rate {\it grosso modo} follows Wheeler's~\cite{Wheeler1949} $Z_{\rm eff}^4$ law, but in detail shows the effect of nuclear shell structure on nuclear size. 
The Weinberg-Feinberg convention reports results as a ``capture fraction", by analogy to a branching fraction, by dividing the measured rate once again by the muon capture rate, thereby exaggerating the effect of nuclear shell model structure. 
\par
From a theory perspective, a model calculation of the rate of conversion effectively yields an absolute rate (more specifically a rate characterized by $G_F^2$ and mass-scale coupling factors). The convention has then been to divide this BSM rate by the experimentally measured SM muon capture rate. Thus a BSM conversion rate calculation that is effectively on the same footing as a calculation of a BSM decay rate is presented as a hybrid ratio of the calculated rate of the coherent conversion process divided by the experimental measurement of a partially incoherent SM muon capture process, analogous to a branching fraction. This hasn't mattered in any practical sense to this point. It is, however, comparing a calculable coherent process to a difficult-to-calculate (and therefore usually measured) process. When comparisons of decay and conversion sensitivity are made there are issues, as there are in comparison of CLFV rates for different nuclei. 
\begin{figure*}[t!]
    \centering
    \includegraphics[width=1.0\textwidth]{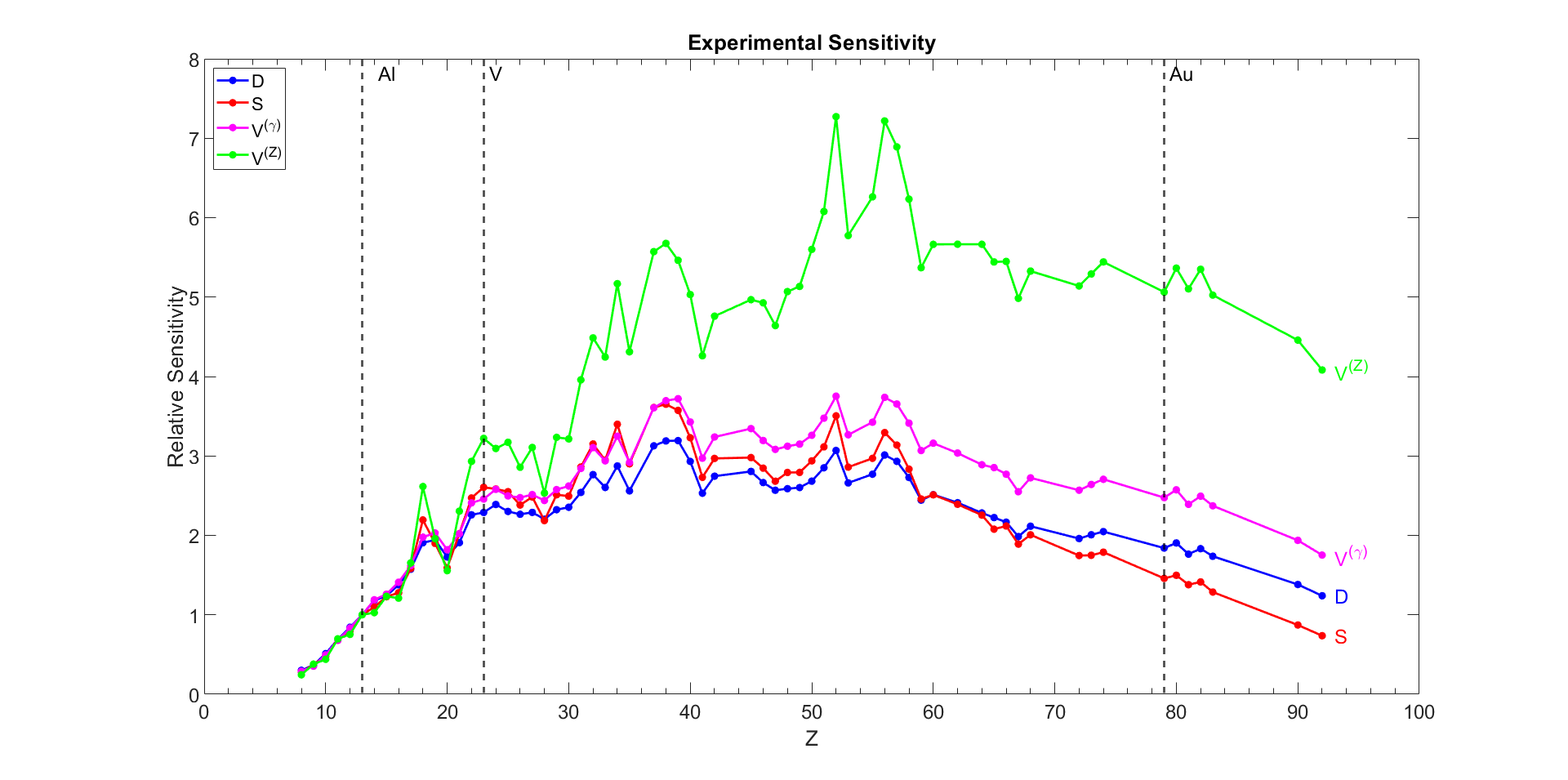}
    \caption{Sensitivity of CLFV experiments as a function of atomic number, relative to sensitivity in aluminum. The normalization is to total muon lifetime, rather than to muon capture. }
    \label{fig:sensitivity_MZ_nat}
\end{figure*}
\par
The method of normalization matters in two different ways. Plots of the  chronological improvement of limits on CLFV processes typically plot  the limits for $\mu \to e \gamma$, $\mu \to 3e$ and $\mu \to e$ conversion on the same scale, even though the decay results are true branching fractions, while the conversion rates are fractions of the $\mu$ capture rate. It would be preferable to normalize the conversion rate to  the fate of all stopped muons, which is what is actually measured. The ratio of muon decay in orbit to $\mu^-$ capture is a strong function of $Z$, which yields a somewhat different dependence on $Z$ for the two normalization approaches. The proposal is therefore to present conversion results as they are actually measured and compare these results with what is actually calculated. Thus the normalization would be to all fates of the muon in the atom:
\vspace{-25pt}
\begin{center}
$${\rm CR}({\mu ^{_ - }}{\rm{N}} \to {e^ - }{\rm{N) }} 
=\frac{{\Gamma ({\mu ^{_ - }}{\rm{ + }}\,{\rm{N}} \to {e^ - }{\rm{ + }}\,{\rm{N)}}}}{{\Gamma_{\rm total} ({\mu ^{_ - }}{\rm{ + }}\,{\rm{N}} \to {\rm{all)}}}}\, .$$
\end{center}

We have done a new comprehensive study of the ($Z,A$) dependence of $\mu \to e$ conversion~\cite{Borrel2023} that extends previous work in several areas. In addition to employing the normalization to the total muon lifetime instead of the $\mu$ capture lifetime, it includes nuclear size and shape data from muonic X-rays, accounts for nuclear quadrupole deformations and treats proton and neutron distributions separately. Figure~\ref{fig:sensitivity_MZ_nat}
shows the results, relative to the conversion rate in aluminum. While variations due to nuclear shell structure inevitably remain, they are reduced from previous treatments that normalize to $\mu$ capture.

\vspace{40pt}
\noindent \textbf{Exotic physics signatures }
\vspace{0.5cm}

Arguably, high-intensity muon facilities have underappreciated potential to search for light weakly coupled new physics. As an illustration of these ideas, we discuss a proposed search for two-body decays to light new particles with both pions and muons at Mu2e. Modern muon facilities offer unprecedented statistical samples of both muons and pions in a controlled, thin-target limit, with high-resolution detectors~\cite{Lee:2018wcx, Bernstein:2019fyh, MEGII:2021fah, Hesketh:2022wgw}. Their flagship physics goals are often highly specialized due to the kinematic ``smoking gun'' signatures of CLFV occurring in small pockets of phase space that the Standard Model struggles to populate. Despite their highly specialized design,  muon facilities have unique capabilities to search for physics beyond the Standard Model (BSM).

Having a broad portfolio of physics in addition to a specialized flagship measurement is beneficial for modern HEP experiments. As a relevant example, the neutrino community has increasingly embraced searches for dark sector physics~\cite{MiniBooNEDM:2018cxm}. Similarly, experiments designed for precise measurements of neutrino oscillation parameters such as DUNE, JUNO, and Super Kamiokande also search for supernova neutrinos and proton decay~\cite{DUNE:2020zfm,Super-Kamiokande:2020wjk,DUNE:2020ypp,Super-Kamiokande:2021jaq,JUNO:2022qgr}. A broad physics use case (beyond CLFV) is beneficial because it both maximizes the physics impact of the experiments themselves and increases connections across different subfields of HEP.

As a concrete example of the opportunities discussed above, we now present a proposal for Mu2e to search for two-body decays $\mu^+\rightarrow e^+X$ ~\cite{Plestid_forthcoming}, \cite{Plestid_Ack}. The ability to perform a successful search in this channel relies crucially on the sample being $\mu^+$ rather than $\mu^-$. The statistical sample is so high at Mu2e $\sim 10^{18}$ muons stopped, that even calibration data (which would be $\mu^+$ rather than $\mu^-$) with one-millionth the statistics, e.g.\ $10^{12}$ or $10^{13}$ stopped muons, may offer world leading constraints. This illustrates an important point, that even in ``sub-optimal'' configurations that sacrifice orders of magnitude in statistical power, modern muon facilities may still offer unparalleled reach for certain models of new physics. We project that Mu2e can overcome existing constraints from the TWIST-II experiment at TRIUMF~\cite{TWIST:2014ymv} and provide world leading limits on the branching ratio of $\mu^+ \rightarrow e^+ X$~\cite{Plestid_forthcoming}.

\begin{figure}
\includegraphics[width=0.65\linewidth]{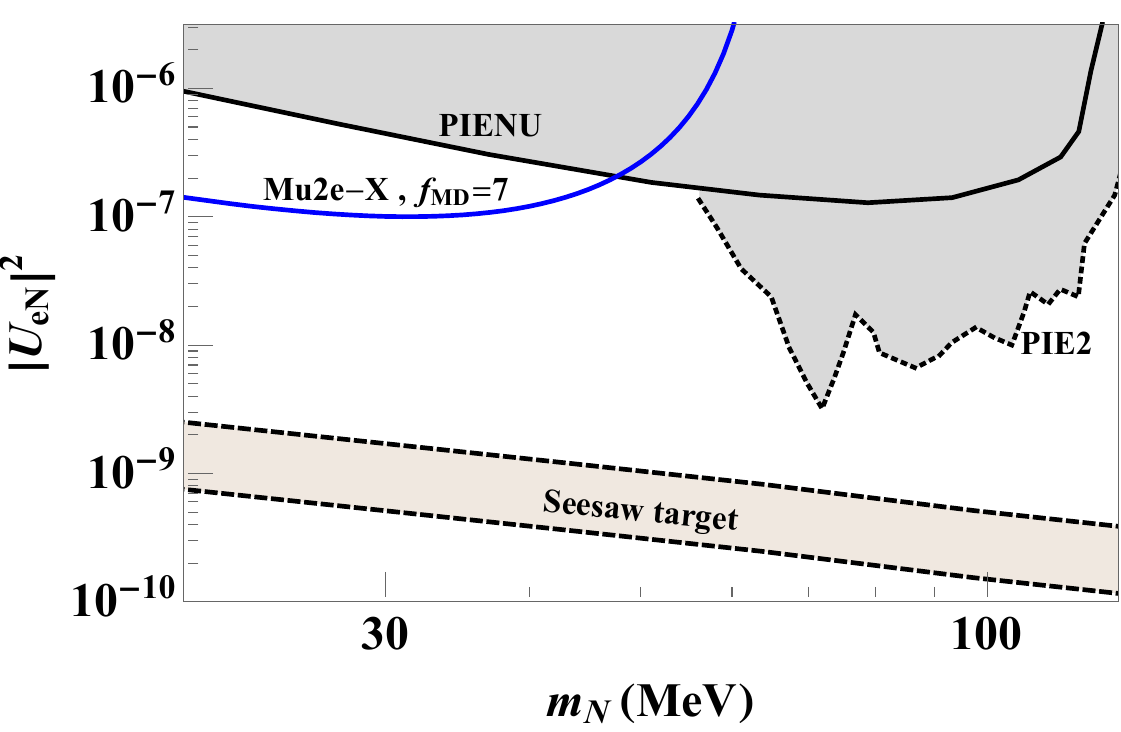}
\caption{Projected sensitivity to right-handed neutrinos mixing with electron flavor. Mu2e calibration data with a momentum degrader that improves sensitivity by a factor of 7 (reduction of muon decay in flight background by a factor of 5 and an increase of stopped pions by a factor of 3). The see-saw target corresponds to parameter space that would produce acceptable neutrino masses in a type-I see-saw framework. To appear in Ref.~\cite{Plestid_forthcoming}} 
\label{fig:Mu2eHNL}
\end{figure}

Charged pions also offer exciting new physics opportunities via $\pi^+\rightarrow e^+ X$. Pions are typically seen as a hindrance; for instance Mu2e designed their beam structure around a timing cut to extinguish backgrounds from radiative pion capture~\cite{Bernstein:2019fyh}. Nevertheless, a modified data acquisition strategy that focuses on early periods of the beam structure can enable Mu2e, or other muon facilities, to search for BSM physics in pion decays. The experiment's thin target and high-resolution detectors differentiate them from other high-statistic pion sample. For instance pion decay at rest ($\pi$DAR) facilities~\cite{LSND:1996jxj,COHERENT:2015mry,JSNS2:2021hyk} often have huge statistical samples, but use a thick target and copious shielding making a search for a mono-energetic electron impossible. By way of contrast, this type of signal is {\it precisely} what the Mu2e tracker is designed to look for. A modified $B$-field of $76\%$ removes all Michel electrons, and enables good energy reconstruction for $60-70~{\rm MeV}$ electrons. Using a bump-hunt strategy we project that Mu2e could place world-leading limits on right-handed neutrinos mixing with electrons (Fig.~\ref{fig:Mu2eHNL}), competitive even with the highly specialized PIONEER proposal~\cite{PIONEER:2022yag}. The signal is a monoenergetic positron from $\pi^+\rightarrow e^+ N$ with $m_N\gtrsim 20~{\rm MeV}$ to avoid overlap with the peak from $\pi^+\rightarrow e^+\nu_e$ (which may also be used as a source of calibration). 

In summary, muon facilities offer many unexplored sidebands that can be exploited for new physics. The statistical samples and modern detector quality allow even a single day's worth of calibration data to be impactful. More effort should be invested into studying novel signatures of BSM physics that may be unearthed with muon facility data.

\subsubsection{Technical Considerations}
\label{sens_tech}
\noindent \textbf{Designing a Pion Production Target }
\\

The sensitivity of the Mu2e II experiment critically depends on the number of muons stopped in the target over the course of the experiment. The muon beam is generated from low momentum pions produced in the production solenoid, where the rate of stopped muons directly depends on the rate of low momentum pions produced in primary proton interactions in the production target. This production rate depends both on the production target design as well as the pion production in the target material.

The rate of stopped muons per 800 MeV primary proton on the production target simulated using the Mu2e II Offline, which uses GEANT4 for particle transport, is shown in Table \ref{tab:sens_stopping_rates} for both the tungsten and carbon conveyor target designs, as well as using the Mu2e Hayman tungsten target for comparison. The carbon conveyor production target is estimated to have a 25\% higher muon stopping rate than the tungsten conveyor target, though, as discussed in Section \ref{sec:target_particle_yields}, MARS and FLUKA do not predict the higher particle production rates using the carbon conveyor target that GEANT4 predicts.
The Mu2e-era Hayman tungsten production target, which could not survive under the Mu2e II operating conditions, but also was not optimized for the trajectory of 800 MeV protons in the production solenoid's magnetic field, has a higher muon stopping rate of about $10^{-4}$ per proton on target. This suggests that targets optimized for Mu2e II may be able to achieve stopping rates on the order of $10^{-4}$ per proton on target.

\begin{table}
  \centering
  \begin{tabular}[t]{l|c}
    Target            & R(muon stops / POT) \\
    \hline
    Tungsten conveyor & $( 7.2 \pm 0.1)\times 10^{-5}$ \\
    \hline
    Carbon conveyor   & $( 9.0 \pm 0.1)\times 10^{-5}$ \\
    \hline
    Mu2e Hayman       & $(10.3 \pm 0.3)\times 10^{-5}$ \\
  \end{tabular}
  \caption{Stopped muon rates in the stopping target per primary proton on the production target for different production target designs using 800 MeV primary protons. The ``Mu2e Hayman'' target refers to the Mu2e-era tungsten production target design.}
  \label{tab:sens_stopping_rates}
\end{table}

\begin{figure}
    \centering
    \includegraphics[width=0.8\linewidth, trim = 20 0 30 0, clip]{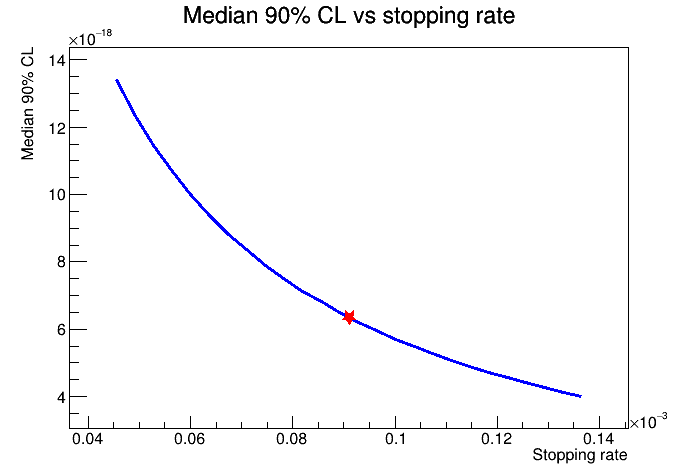}
    \caption{The Mu2e II expected 90\% CL in the absence of a signal as a function of the stopped muon rate. Selection efficiencies are assumed fixed, where only the beam-related backgrounds and the signal rates are varied.}
    \label{fig:sens_sensitivity_vs_stopping_rate}
\end{figure}

A change in the muon stopping rate leads to a direct change in the single event sensitivity for $\mu^- \rightarrow e^-$ and the rate of beam-related backgrounds, while leaving the rate of cosmic ray-related backgrounds unaffected, assuming little change in the reconstruction efficiency and no change in selection. Figure \ref{fig:sens_sensitivity_vs_stopping_rate} shows the expected sensitivity of the Mu2e II experiment for varying muon stopping rates. This does not account for changes in the signal search selection to account for the change in the beam-related background rates, which would improve the sensitivity, or any change in the detector pileup due to changes in the particle production rates, which may improve or diminish the sensitivity.

\vspace{0.5cm}
\noindent \textbf{Designing a detector system }
\vspace{0.5cm}

Several aspects pertaining to the design of the tracker, calorimeter, and cosmic ray veto system are detailed in their respective sections. We do not re-iterate them here but it is, of course, crucial that we have excellent momentum resolution, and minimize multiple scattering and energy losses and cosmic backgrounds in order to achieve our sensitivity goals. There are already ongoing R$\&$D efforts to understand possible tracker configurations. So far these have focused on the use of thinner straws (8$\mu$ m compared to 15 $\mu$ m in Mu2e). More exotic tracker designs could also help, and there are opportunities to contribute. For the calorimeter, more radiation resistant material is necessary, barium fluoride will replace the cesium iodide used in Mu2e. There is an ongoing R$\&$D effort for the best sensor technology to use with barium fluoride. There is also a need to replace the cosmic ray veto system. This is crucial to ensuring the elimination of cosmic ray background.

 The physical design of the pion production target has already been discussed. There is also a need to re-design the muon-stopping target, which could be further optimized for improved sensitivity. The design is a compromise, a heavier target stops more muons, but also means more energy straggling and multiple scattering which can decrease momentum resolution. 
\newpage

\section{Advanced Muon Facility}
The Advanced Muon Facility (AMF) is a proposal for a next-generation muon facility at Fermilab that would exploit the full potential of the PIP-II accelerator to deliver the world's most intense $\mu^+$ and $\mu^-$ beams. This facility would enable broad muon science with unprecedented sensitivity, including a suite of CLFV experiments that could improve the sensitivity of planned experiments by orders of magnitude, and study in detail the type of operators contributing to NP in case of an observation (e.g. high-Z target in conversion experiments). 

The AMF complex would use a fixed-field alternating gradient synchrotron (FFA) to create a cold, intense muon beam with low momentum dispersion. Short intense proton pulses are delivered to a production target surrounded by a capture solenoid, followed by a transport system to inject the muons produced by pion decays into the FFA ring. The phase rotation trades time spread for momentum spread, producing a cold, monochromatic muon beam. During that time (${\cal O}(1) \ \rm \mu s$), the pion contamination is reduced to negligible levels, and the FFA injection / extraction system effectively cuts off other sources of delayed and out-of-time backgrounds. The phase rotation requires very short proton pulse, and a compressor ring is required to rebunch the PIP-II beam.

The following sections summarize the discussions held on the compressor ring, the FFA synchrotron, the conversion and decay  experiments, and other opportunities with high intensity muon beams. Prioritized R\&D tasks for each topic and synergies with other efforts are also highlighted.
\subsection{Proton Compressor ring}

\subsubsection{PAR Proposal}

\begin{figure}[htb]
\centering
  \includegraphics[width=0.55\textwidth]{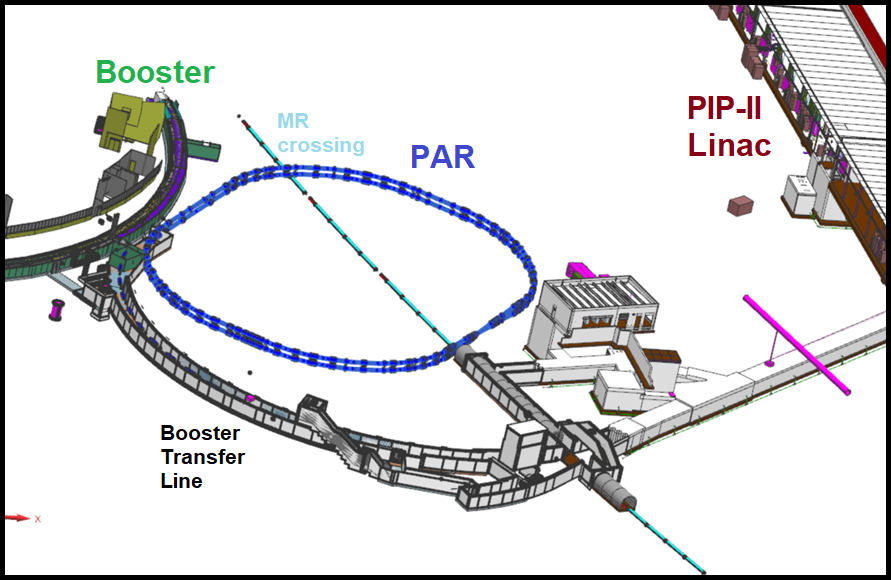}
  \caption{PAR ring (blue) located at the end of the Booster Transfer Line (grey) leading from the PIP-II Linac (red) to the Booster (green). Adapted from B. Pellico and S. Dixon in~\cite{ACE_Workshop}.}
  \label{fig:PARsite}
\end{figure}

A primary objective of the PIP-II linac upgrade is to improve Fermilab's Booster performance and thereby increase beam power to DUNE/LBNF. It will be CW-capable, leaving a large portion of beam power available for new GeV-scale experimental programs. The PIP-II era accumulator ring (PAR) has been proposed to enable the PIP-II linac to better perform both roles; to improve Booster performance and to provide a platform for new  HEP experiments. In particular, a beam dump program exploring dark sector (DS) and neutrino physics called PIP2-BD~\cite{toupsDM} could operate with PAR proton pulses. For the Booster role, the circumference of PAR needs to match that of the Booster, but the location at the end of the Booster Transfer Line constrains the available area. The PAR design uses a novel "folded figure-8" architecture, allowing it to fit within a smaller footprint, see Fig.~\ref{fig:PARsite}.  A key advantage of PAR for the Booster program is to use a longer injection section with an extraction line for unstripped H- particles. 

Due to the folded figure-8 design, PAR can accommodate two extraction sections and two RF systems. Consequently, PAR can extract to the Booster at a 20 Hz rate and extract to PIP2-BD at a 100~Hz rate or above. The two RF systems would allow PAR to operate with 44~MHz for bucket-to-bucket transfer to Booster mode or at less than 10~MHz for pulse compression to the PIP2-BD program. A third operating mode is to capture the beam at 44~MHz but extract single-bunch pulses at a high rep rate ($\sim 800 \rm\, Hz$). Table~\ref{table:PAR_Modes} below gives a summary of the projected beam modes available from PAR.

A self-consistent PAR lattice design has already been developed with several key features. There is a 10 m uninterrupted injection straight section that allows for the ability to safely extract unstripped H$^{-}$ particles while also being suitable for higher energy beams (at least 1 GeV). Also within this injection straight is a 28~m dipole-to-dipole length with $\sim\pi /2$ phase advance, leaving room for downstream collimation. In the long crossover section, there is a 12" parallel shift between the top and bottom rings which means there is no shared beampipe required in the crossover region. In the other long straight section, there is a cluster of quadrupole magnets that together are referred to as a phase trombone. This phase trombone allows the tune to be changed $\pm75$ degrees without impacting the beta functions at other locations around the ring. Suitable RF cavity, magnet, and kickers designs have also been identified. Tracking simulations show that the lattice is stable and features adequate betatron tunespace.

\begin{table} [htb]
\setlength{\tabcolsep}{18pt}
\renewcommand{\arraystretch}{1.5}
\begin{center}
\begin{tabular}{ | c | c | c | c | }
 \hline
 \multicolumn{4}{|c|}{PAR Power Modes} \\
 \hline
  & Nominal - To Booster & PIP2-BD (h=4) & AMF (h=1) \\
 \hline
 Intensity/pulse (ppp) & 8-16e12    & 1-2.5e12 & 0.1-0.2e12  \\
 Energy (GeV) & 0.8    & 0.8 &  0.8 \\
 Rep. Rate/pulse (Hz) & 100    & 200-400 & 400-800 \\
 \bf Power (kW) & \bf 100-210    & \bf 30-130 & \bf 5-21 \\
 Pulse length (ns) & 2000    & 385-170 & 22-18 \\
 \bf Exp. duty factor & \bf 2e-4   & \bf 7.7-6.8e-5 & \bf 0.9-1.4e-5 \\
 Beam capture Rate (MHz) & 44    & $<$10 & 44 \\
 \hline
\end{tabular}
\caption{Three projected beam modes available from PAR accelerator.}\label{table:PAR_Modes}
\end{center}
\end{table}

\subsubsection{Towards a Compact PIP-II Accumulator Ring for AMF}

A detailed design of the PAR has been completed and is in the process of being documented (a summary is available above). However, the same level of detail is not presently available for a more compact and higher power accumulator ring better suited for the AMF proton compressor. In~\cite{toupsDM}, a set of modest self-consistent parameters for Compact PIP-II Accumulator Ring (CPAR) is articulated, whereby CPAR would deliver 90~kW of 1.2~GeV protons in less than 20~ns pulses.

In Ref.~\cite{CLFV:Snowmass}, Prebys presents a framework for optimizing proton compressor performance for AMF program and lays out the accelerator requirements for achieving 1~MW beam power from a proton compressor delivering 12.2~ns pulses. To achieve the maximum 12.2~ns pulse intensity within a space-charge tuneshift limit, the proton compressor ring should (firstly) be as compact as possible and (secondly) should have as large as possible transverse acceptance. Following this framework, a ring more compact than ORNL SNS ring but with similar transverse acceptance could provide 400~kW of beam power in 12.2~ns pulses.

The Snowmass paper~\cite{CLFV:Snowmass} also laid out the proton compressor extraction scheme. The RF frequency of the ring should correspond to 12-25~ns RF buckets, with every RF bucket or every other RF bucket filled. The proton pulses are extracted bunch-by-bunch with 10-40~ns kicker rise/fall times. Since the injection scheme will require filling each bunch simultaneously,~\cite{toupsDM} proposes to extract multiple times for each injection. Indeed a (third) critical design strategy should be to have the highest possible extraction kicker repetition rate to alleviate space-charge and injection requirements.

Two other design strategies for the proton compressor ring should be considered. The beam energy can be increased above the value discussed above (fourth strategy) to overcome the space-charge limit to pulse intensity (even accounting for the fact that the ring could be less compact or have a smaller acceptance). The cost optimization of increasing the beam energy relative to other design strategies (such as transverse acceptance) is still to be determined. Lastly (fifth strategy), the machine acceptance should be much larger horizontally than vertically, to allow the strongest dipole field strength while accepting the greatest number of particles. The proton compressor ring design can use alternate gradient focusing to accommodate an open mid-plane in the accelerator bending arcs.

A more detailed design and optimization of the AMF proton compressor is currently underway. Our most recent design projection shows that a 1~MW proton compressor with 10-20~ns pulses may be achievable in a 1.2~GeV ring with a 95\% normalized emittance of 100 $\pi$ mm mrad in a 150~m circumference (with 800~Hz extraction and 100~Hz injection rates).

Design work includes the development of the proton compressor lattice (layout of magnets), dipole geometry, location of critical devices, assessment of injection and extraction strategies. While preliminary calculation of the H- injection parameters suggest that the ring is compatible with H- foil injection, this program would clearly benefit from the development of H$^{-}$ laser stripping injection technology~\cite{Cousinau:PRAB}.

\subsubsection{PIP2-BD Experimental Program}
Theoretical work has highlighted that sub-GeV dark sector models are able to explain cosmological dark matter abundance and a broad class of these models can be tested with accelerator-based fixed-target experiments. Additionally the observation of coherent elastic neutrino-nucleus scattering (CEvNS) by the COHERENT experiment provides a novel and powerful experimental tool for beyond the Standard Model neutrino physics. The PIP2-BD program looks to access this physics with GeV-scale, high-power, low-duty beams.

The proton compressor for the proposed AMF facility would make an excellent proton source for a beam dump physics program called PIP2-BD~\cite{toupsDM}. Aside from the proton source and the beam dump, the PIP2-BD experiment requires only a 100t LAr detector.

\begin{figure}[htb]
\centering
  \includegraphics[width=0.55\textwidth]{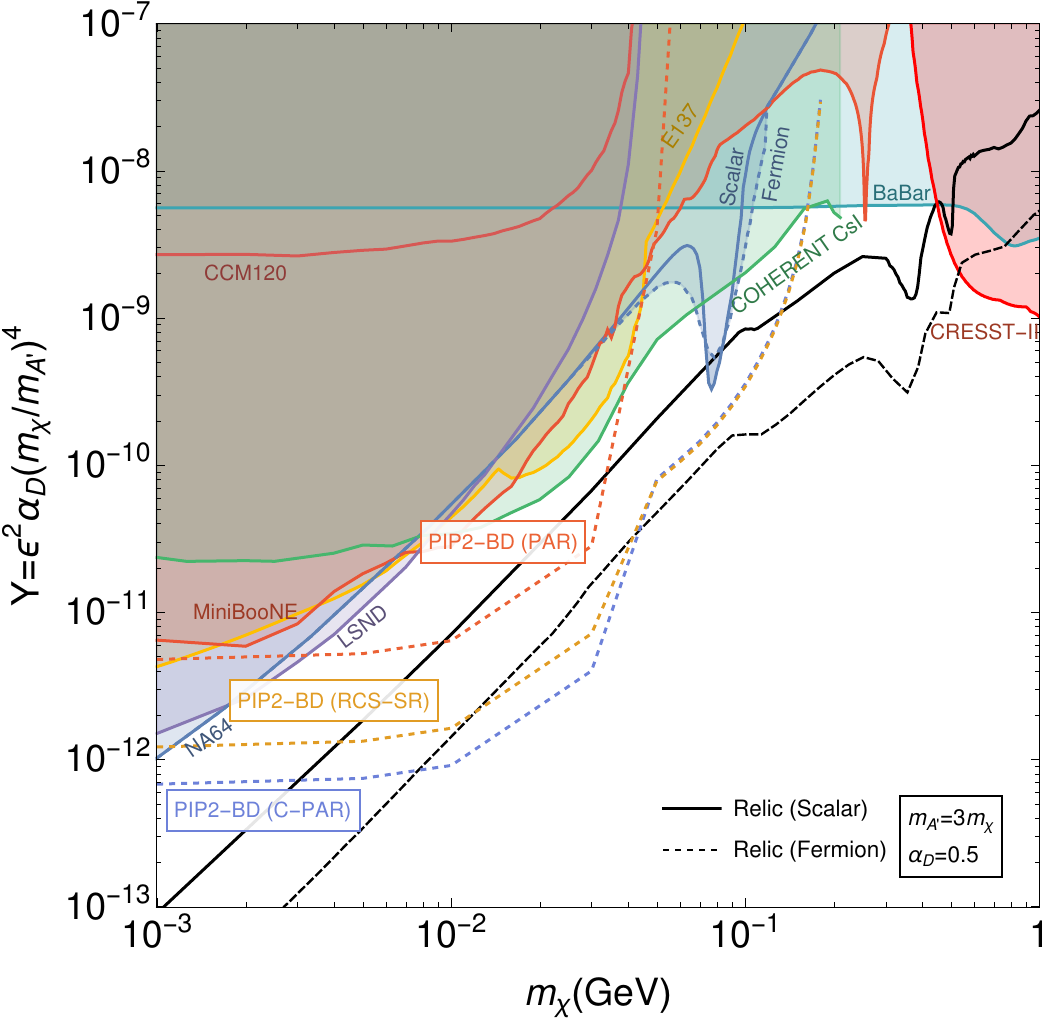}
  \caption{The 90\% C.L. sensitivity to the vector portal DM model for the three accumulator ring scenarios. Both RCS-SR and C-PAR are able to search a large parameter space reaching the expected thermal relic density for both scalar and fermion DM. Figure and caption source~\cite{toupsDM}}
  \label{fig:darkmatter}
\end{figure}

Three scenarios were developed for the PIP2-BD Snowmass paper, all more modest than that proposed for AMF, given in the table below~\ref{table:PIP2BD}. As one example of the physics reach, the exclusion plots for vector portal dark matter (DM) models is shown for the three scenarios in Fig.~\ref{fig:darkmatter}.

The PIP2-BD program provides a natural staging scenario for the AMF program. First, a 0.8~GeV compact proton accumulator ring delivers beam to the PIP2-BD program at 100-300 kW. Next, the proton accumulator ring is upgraded (in energy, extraction rate, and/or pulse-length) and begins beam delivery to the AMF program at 300-1000~kW. In this case, the design of the initial proton accumulator ring needs to account for the beam requirements of both the initial beam program and subsequent upgrade.

\begin{table} [h]
\begin{center}
\begin{tabular}{|p{1.5cm}|p{1.5cm}|p{1.5cm}|p{1.5cm}|p{1.5cm}|}
\hline
Facility & Beam energy (GeV) & Repetition rate (Hz) & Pulse length (s) & Beam power (MW)\\
\hline
PAR & 0.8 & 100 & $2 \times 10^{-6}$ & 0.1 \\
C-PAR & 1.2 & 100 & $2\times 10^{-8}$ & 0.09\\
RCS-SR & 2 & 120 & $2 \times 10^{-6}$ & 1.3 \\
\hline
\end{tabular} \caption{The parameters of three possible accumulator ring scenarios considered to bunch the beam current from the PIP-II linac. See text for more details.}
\label{table:PIP2BD}
\end{center}
\end{table}

\subsection{Muon FFA}
\subsubsection{Motivation for an FFA ring}
\begin{figure}
    \centering
    \includegraphics[width=\textwidth]{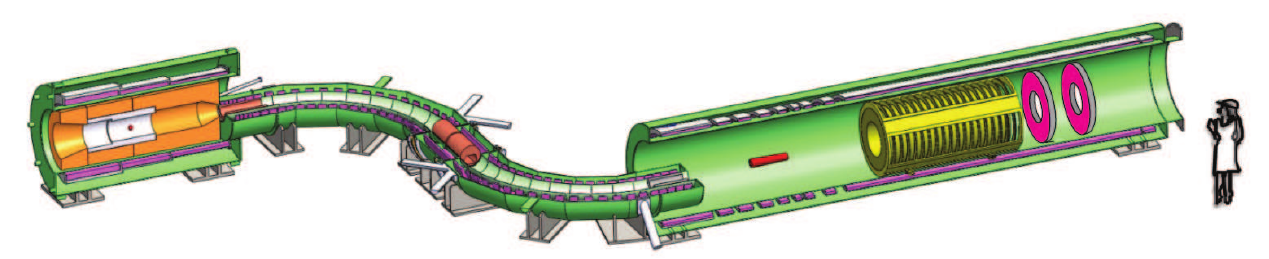}
    \caption{\label{fig:mu2eFig} The three-part solenoid of the Lobashev scheme, as implemented in Mu2e.}
\end{figure}

Muon-to-electron conversion experiments such as Mu2e or COMET use the Lobashev scheme~\cite{Abadzhev:1992nx}. In that scheme, a proton beam strikes a target inside a solenoid. Produced $\pi$'s decay into $\mu$'s and a graded magnetic field directs the $\mu$'s through a ``transport solenoid" and into a third ``detector solenoid."  Both Mu2e and COMET use this scheme; Fig.~\ref{fig:mu2eFig} shows Mu2e for definiteness.

This scheme has two fundamental limitations.  First, some of the pions born in the first Production Solenoid survive to reach the final Detector Solenoid.  When those pions interact with the conversion material (normally referred to as a stopping target, Al for either Mu2e or COMET), they can undergo radiative pion capture, $\pi^-N \rightarrow \gamma N^{\prime}$.  That photon can either convert in the stopping target or internally convert, yielding an electron in the signal region. Since pions decay with a 26 ns lifetime, a delayed live gate exponentially reduces the pion contamination ($\mathcal{O}(10^{-11})$ in Mu2e.) Second, the initial proton beam produces a ``flash" of electrons from $pN \rightarrow \pi^0 \rightarrow \gamma \rightarrow e^+e^-$ that would overwhelm any detector. Again, a delayed live gate is the natural solution.

\begin{figure}
    \centering
    \includegraphics[width=0.8\textwidth]{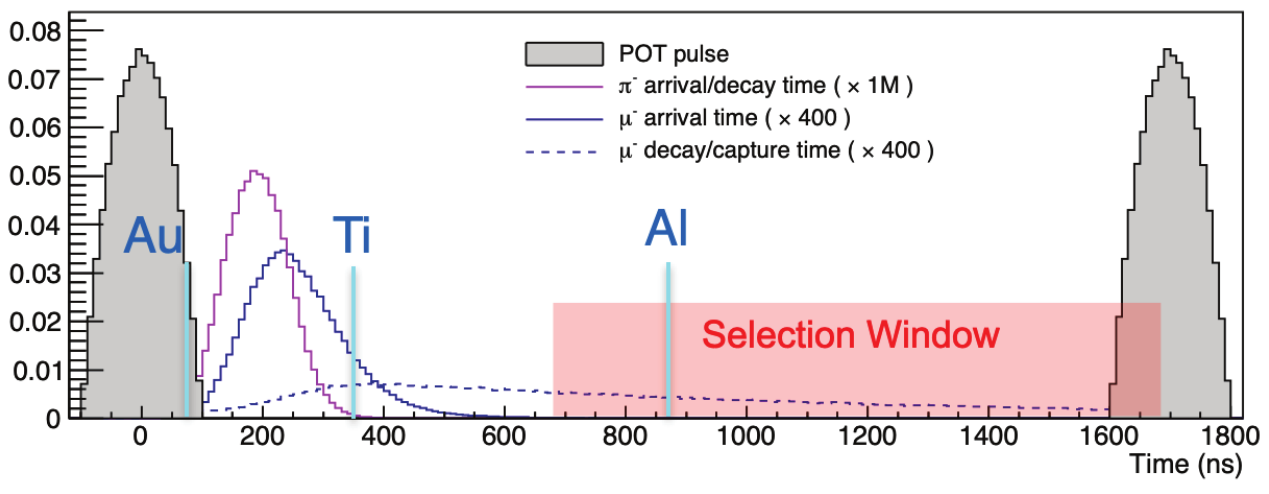}
    \caption{A timeline for Lobashev-style experiments, shown for Mu2e.  A $\approx 250$ ns wide proton pulse strikes a target, producing charged pions and a ``beam flash" of electrons. The muonic lifetimes of Au, Ti, and Al are shown.}
    \label{fig:timeline}
\end{figure}

Unfortunately, this delay limits the reach of the experiments, yielding a second limitation. Fig.~\ref{fig:timeline} shows the timeline for Mu2e with the lifetimes of muonic Au, Ti, and Al.  We see $\tau_{\rm Au}$ is well within the beam pulse and the flash period and is therefore not a practical target in this scheme.  Ti is possibly workable but still presents difficulties. Going to high-$Z$ materials such as Au are needed to either probe the nature of any CLFV interaction or set the most stringent limits, so these two limitations are a limit on the technique and prevent us from exploring what might be the most interesting physics. 

A possible way to go beyond this limitation is to use a fixed-field alternating gradient synchrotron (FFA)~\cite{Symon:1956pr} as a muon storage ring.  Pions would decay in the ring leaving a nearly pure muon beam and only the relatively small amount of ``beam flash" near the central momentum of the FFA could be circulated.  The FFA can also use RF to perform ``phase rotation," trading the pulsed time structure and broad momentum spread from the compressor ring for more continuous beam with a small momentum spread of a few percent.  Such a cold, pure muon beam is ideal for exploring high-$Z$ materials with a short muonic lifetime.  A cold beam at low energy will lose energy quickly through $dE/dx$ and come to rest in a thin target. The CSDA range in Al for a Mu2e-type beam at 40 MeV/c with $T=7.3$ is about $\times 3$ greater than a 29 MeV/c stopped beam at $T=4$, and the Lobashev-style Mu2e beam extends to nearly 100 MeV/c with a $\times 25$ greater CSDA range.  We see that as the beam energy decreases, it becomes easier to find a conversion or decay point than for a Mu2e-type beam at about 40 MeV/c.

FFAs, although invented in the 1950's, have attracted considerable attention relatively recently due to their unique properties~\cite{Collot:2008zz}. In particular, they can provide lattices with very large acceptances in both the transverse and longitudinal planes, which is essential for muon beam applications. The scientific interest in FFAs has increased significantly since the construction of the first proton machine with RF acceleration in 2000~\cite{Aiba:2000}. Since then several FFA machines have been successfully constructed and operated, and several possible applications were developed. The review of FFA developments was presented and discussed during the workshop.

\subsubsection{FFA ring}

\begin{figure}[ht]
 \begin{center}
\includegraphics[width=0.6\textwidth]{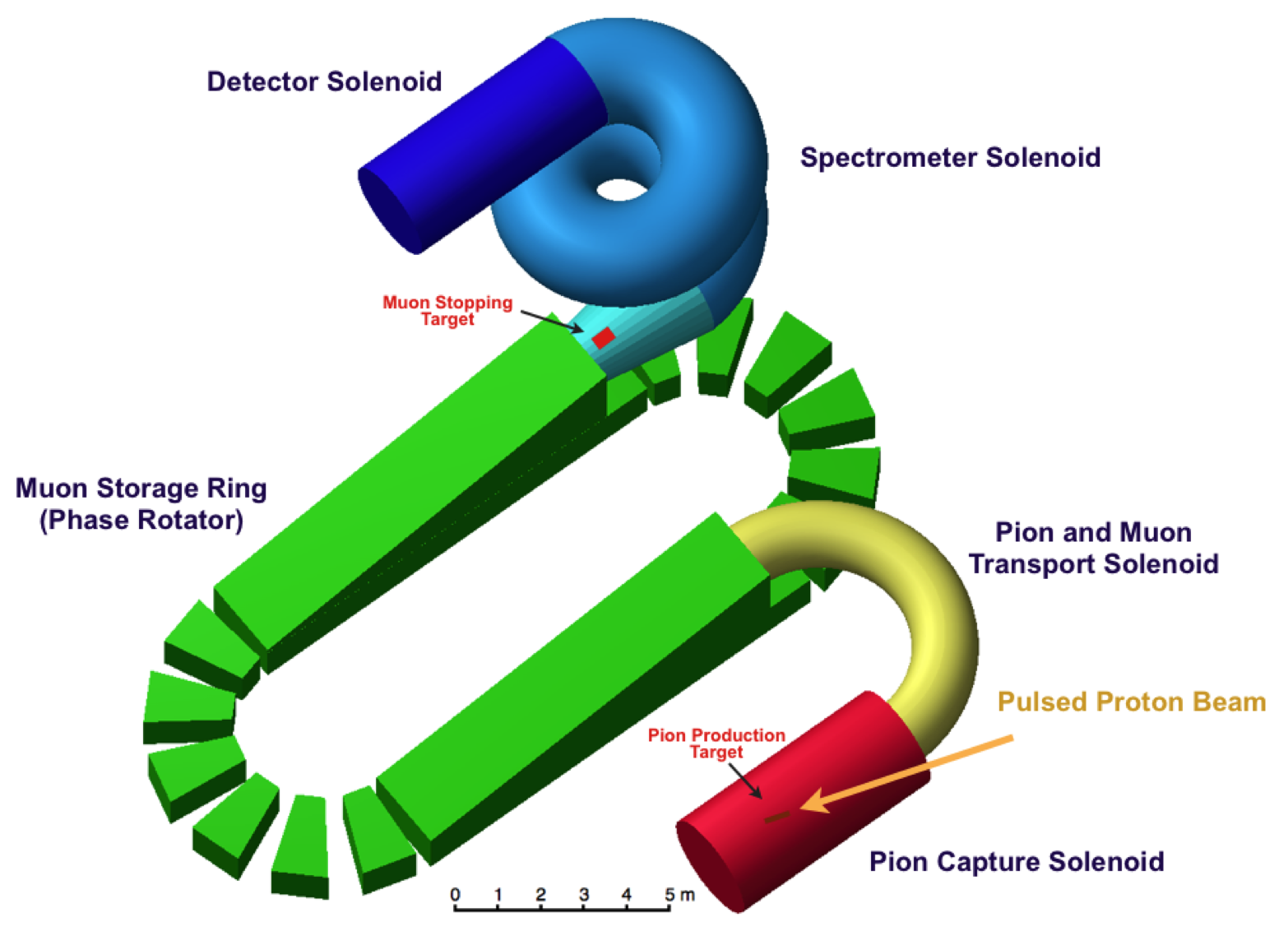}
\end{center}
    \caption{The conceptual layout of the PRISM system.}
    \label{fig:prism}
\end{figure}

A FFA ring was proposed for the next generation muon-to-electron conversion experiments~\cite{KUNO2005376} to provide the large required acceptance for the muon beam, which is produced as tertiary beam with large emittances and momentum spread. This experiment requires a high intensity compressed proton bunch to be sent to the pion production target, which is immersed in a high field capture solenoid. The pions produced from proton interactions decay into muons, which are transported and injected into a small FFA ring. In this ring, the beam undergoes the longitudinal phase space rotation using RF system to transform the initial short muon bunch with a large momentum spread of $\sim\pm20\%$ into a long bunch with the momentum spread reduced by about an order-of-magnitude. The use of the phase rotation motivated the name for this experimental system --- the Phase Rotated Intense Source of Muons (PRISM). The ring RF system can operate at harmonic number one, requiring low frequency RF cavities based on Magnetic Alloy technology, which allows to mix different RF frequencies creating the saw-tooth shape for the voltage phase dependence used to maximize the efficiency of the phase rotation~\cite{Ohmori:2008zza}. The narrow momentum spread beam is then extracted and sent to the muon stopping target. The conceptual layout of the experiment is shown schematically in Fig.~\ref{fig:prism} and main parameters of the facility are given in Table~\ref{tab:PRISM:parameters}.

\begin{table}
    \centering
\begin{tabular}{lcc}
\hline
\hline
Proton beam power & $\sim$1~MW\\
Proton beam energy & $\sim$GeV\\
Proton bunch duration& $\sim$10~ns\\
Target type & solid\\
Pion capture solenoidal magnetic field& 10-20~T\\
Reference muon momentum & 45~MeV/c (or lower)\\
Momentum acceptance& $\pm$20\%\\
Minimal transverse physical acceptance (H/V)& 3.8/0.5~$\pi$.cm.rad\\
RF voltage & 3-5.5~MV\\
RF frequency & 3-6~MHz\\
Repetition rate & 100-1000~MHz\\
\hline
\hline
\end{tabular} 
 \caption{Main parameters of the PRISM facility.}
 \label{tab:PRISM:parameters}
\end{table}

Several lattice solutions have been proposed for PRISM~\cite{Alekou:2013eta}, with an initial baseline based on a scaling DFD triplet~\cite{Sato:2008zze}, which was successfully constructed and verified experimentally. The studied solutions included both scaling and non-scaling FFA lattices, consisting of regular cells or with a racetrack geometry.

\begin{figure}
  \begin{center}
    \includegraphics[width=0.8\textwidth]{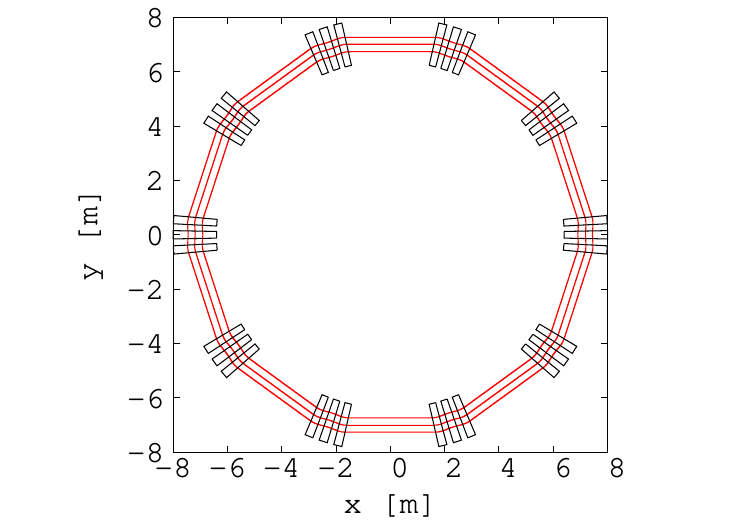}
  \end{center}
  \caption{
    Schematic drawing of the muon storage ring. The lattice is based on the scaling Fixed Field Alternating gradient (FFA) triplets with FDF symmetry. 
  }
  \label{Fig:FFA:ring}
\end{figure}

\begin{figure}
  \centering
    \includegraphics[width=0.49\textwidth]{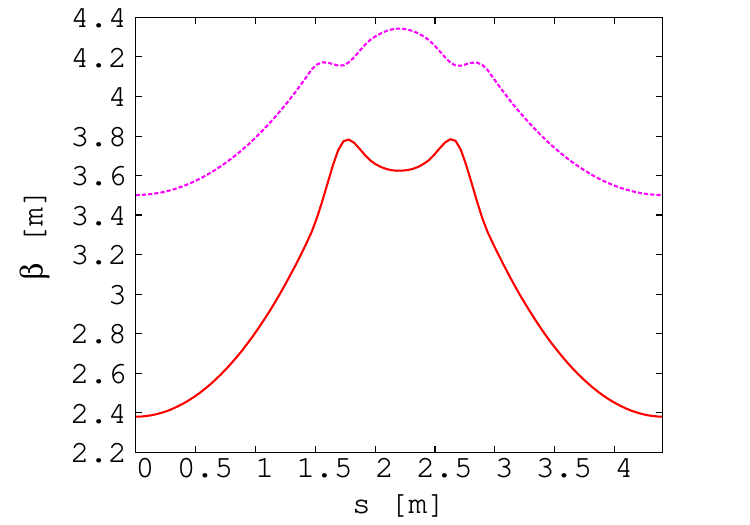}
       \includegraphics[width=0.49\textwidth]{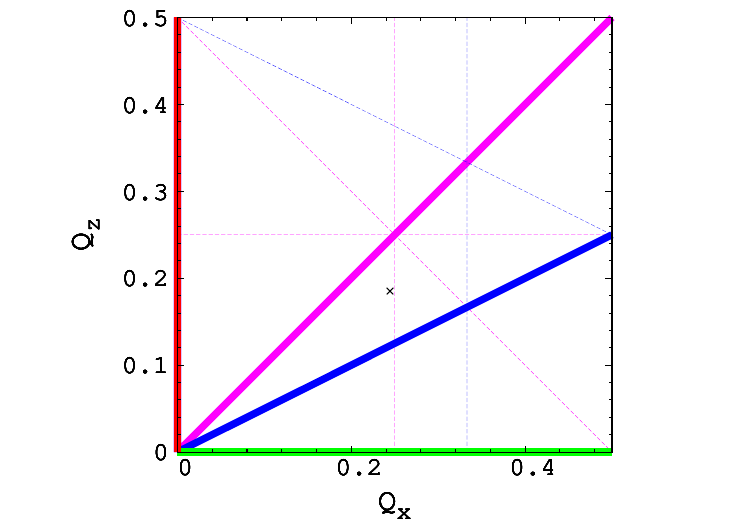}
\caption{Left: horizontal $\beta$ function (red curve) and the vertical one (purple curve) in a symmetric cell of the baseline FFA ring for PRISM. Right: working point (tune per cell) of the baseline FFA ring solution. The second and the third order resonance lines are represented by the blue and purple lines, respectively. The bold lines denote the systematic resonances.}
\label{Fig:FFA:optics}
\end{figure}

\begin{figure}
  \centering
    \includegraphics[width=0.49\textwidth]{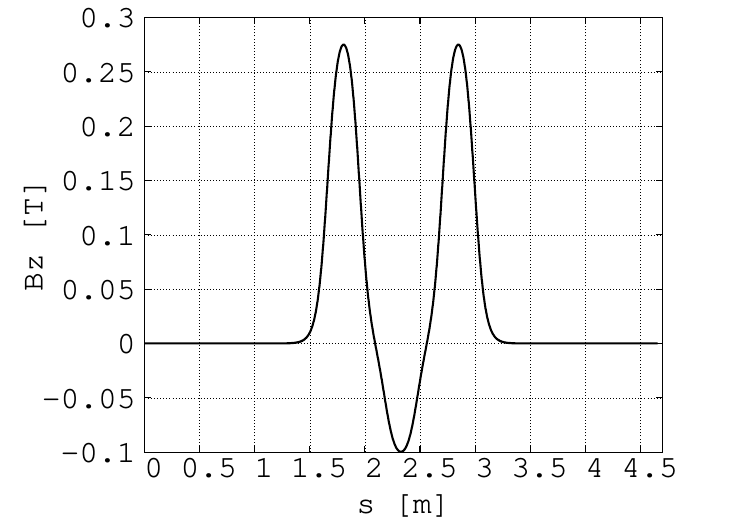}
       \includegraphics[width=0.49\textwidth]{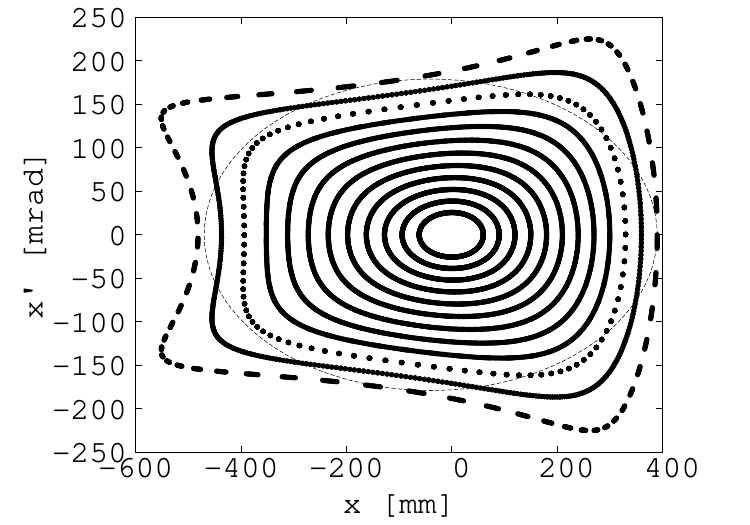}
\caption{Left: vertical magnetic field on the reference radius on the median plane of the baseline FFA ring. The field corresponds to the reference momentum of 68\,MeV/c for historical reasons and can easily be scaled down to a lower momentum. Right: horizontal dynamical acceptance studies in the FFA ring at the reference momentum. Particles are tracked over 100 turns with different amplitudes in the plane of study including a small off-set from the closed orbit in the other plane. The black ellipse represents the acceptance of $77\,\pi$\,mm\,rad.}
\label{Fig:FFA:tracking}
\end{figure}

The current baseline, proposed to take advantage of recent advances in FFA accelerators, was discussed in details during the workshop. It is based on a regular FDF scaling lattice with ten identical cells, as shown in Fig.~\ref{Fig:FFA:ring}. The FDF symmetry can provide the required large dynamical acceptance, while simultaneously increasing the drift length for injection/extraction needs. The symmetrical optics in the ring with identical cells avoids large variations of the $\beta$ functions, which can drive dangerous resonances. The use of the scaling FFA principle makes possible, thanks to its intrinsic zero-chromatic properties, to keep the optics quasi-identical for off-momentum particles. In particular, the tune working point is  independent of momentum and situated away from the resonance lines, which can severely diminish the dynamical acceptance. The $\beta$ functions in one of the symmetric lattice cells and the working point (tune per cell) of the baseline FFA ring are shown in Fig.~\ref{Fig:FFA:optics}. The magnetic field on the median plane of the FFA ring along the reference radius, described using the Enge model of the fringe fields, is shown in Fig.~\ref{Fig:FFA:tracking}. The performance of the FFA ring was verified in tracking studies. In order to incorporate tracking through the combined-function magnets, taking into account the fringe fields and large amplitude effects, a code used for the full FFA machine developed previously (FixField code) was used~\cite{ Lagrange:2018triplet}. It is a step-wise tracking code based on Runge-Kutta integration, using Enge-type fringe fields. The results of the multiturn tracking shows that the horizontal dynamical acceptance of the machine is very large and exceeds an impressive figure of $77\,\pi$\,mm\, rad, as shown in Fig.~\ref{Fig:FFA:tracking}~(right-hand plot). The vertical dynamical acceptance is still being optimized, with the goal to achieve at least $5\,\pi$\,mm\, rad. The vertical dynamical acceptance is typically smaller in horizontal FFAs, and the vertical physical acceptance is nevertheless limited by the injection needs. The main parameters of the baseline FFA ring solution for PRISM can be found in Table~\ref{tab:FFA:parameters}.

\begin{table}
    \centering
\begin{tabular}{lcc}
\hline
\hline
Reference radius & 7~m\\
Length of one straight section & 3.15~m\\
Initial momentum spread & $\sim\pm$20\%\\
Final momentum spread & $\sim\pm$2\%\\
Reference muon momentum & 45~MeV/c (or lower)\\
Reference tunes per cell ($q_h$, $q_v$) & ($0.245$, $0.185$)\\
Number of cells in the ring & 10\\
Field index k & 4.3\\
Harmonic number & 1\\
\hline
\hline
\end{tabular} 
 \caption{Selected parameters of the FFA storage ring.}
 \label{tab:FFA:parameters}
\end{table}

\subsubsection{Beam transport and injection into an FFA ring}

A significant challenge of the PRISM system resides in the design of the efficient beam transport from the decay solenoid, where the muon beam is formed, and its subsequent injection into the FFA ring. 
The beam in the solenoid is very strongly focused, has symmetric coupled optics in both transverse planes, and a large natural chromaticity. Dispersion is either zero or very small even in the presence of a bending field in the curved solenoids. This beam needs to be transported with negligible losses into the FFA, which has zero chromaticity, decoupled asymmetric optics with intermediate focusing strength, and has relatively large dispersion function.

\begin{figure}
  \begin{center}
\    \includegraphics[width=0.95\textwidth]{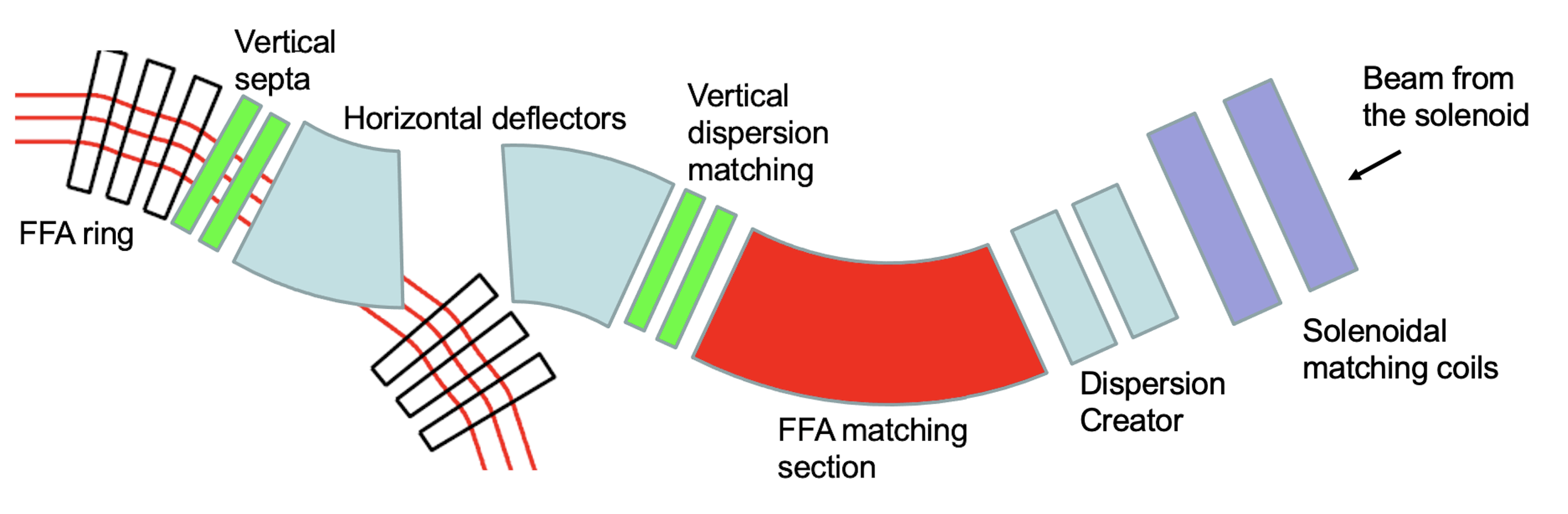}  \end{center}
  \caption{
    Conceptual layout of the muon beam transport from the decay solenoid and injection to the FFA ring. The beam is moving from right to left. Injection kickers located in the ring are not shown.
  }
  \label{Fig:FFA:injection}
\end{figure}

A conceptual design for the beam transport and injection system has been proposed (Fig.~\ref{Fig:FFA:injection}). The beam first needs to exit the solenoid while the beam dynamics is under control. The proximity of the solenoid may also saturate the downstream Alternating Gradient (AG) iron dominated magnets, which should be avoided, and the reduction of the solenoid field needs to be controlled. A system of two solenoidal coils is proposed to perform the beam matching from the quasi-uniform decay solenoid into the downstream accelerator, controlling the beam size and divergence, while a more complex system may be required. Located downstream from this matching section, the dispersion creator consists of two rectangular dipoles with equal, but opposite bending angles, which generates the initial dispersion required for the FFA lattice. This is followed by the scaling FFA matching section, which aims to control the matching of the $\beta$ functions and the final dispersion into the values needed for the FFA ring. Finally, a system of bending magnets and septa is used to introduce the beam into the FFA ring with a vertical offset with respect to the circulating beam. The beam is bent horizontally and the dispersion flips sign due to the strength of the bending magnets needed. The final horizontal magnet is a Lambertson-type septum~\cite{Paraliev:2018ski}. The horizontal septum needs to be followed by a vertical septum, which brings the beam closer to the closed orbit in the ring. The vertical magnets upstream and downstream of the horizontal septum provides the matching of the vertical dispersion function to zero in the ring. The beam is then passed through one cell of the FFA, where the offset in position is transformed into a vertical divergence offset, which is cancelled by the kicker magnets finishing the injection process by placing the beam on the circulating closed orbit. Further studies are needed to demonstrate the full feasibility of the described beam transport and injection system, and significant R\&D is required to address its full optimisation and to design the hardware. The extraction system and the beam transport from the FFA ring into the experiment can be realized as a reverse copy of the injection system, but may be significantly simpler as the momentum spread of the beam is a factor of ten smaller. The main challenge is the rise time of the extraction kicker(s), which needs to be shorter than the injection ones as the bunch at extraction is significantly longer. However, the use of conventional magnets like quadrupoles in the transport line downstream of the extraction point may be feasible due to the reduced momentum spread.

The FFA can operate with both signs of muons, one at a time, if the machine fields can be reversed. However, it might be possible to apply the concept of the singlet FFA lattice~\cite{39289}, in which beams of both signs can circulate in the same direction simultaneously. Further studies are needed to investigate this idea.

\subsection{Conversion experiments}

The physics case for muon-to-electron conversion experiment at AMF was strongly highlighted during the workshop, with potential sensitivity gains up to two orders of magnitude beyond what Mu2e-II aims to realize. Key design elements draw inspiration from the PRISM/PRIME concept~\cite{KUNO2005376}, which introduced a fundamental shift in the experimental technique searching for muon conversion. The introduction of a phase-rotated, slow muon beam in an FFA ring alleviates several limitations of the scheme adopted by current collaborations~\cite{bartoszek2015mu2e,comet-tdr}, and opens the possibility to measure the conversion rates using high-Z target material. Such measurement would be critical to study the source of New Physics~\cite{Kitano2002}, should a signal be observed. The following sections review the challenges of the current experimental approach, the broad requirements for a conversion experiment at AMF, and the main R\&D efforts to develop a future detector concept.

\subsubsection{Current approach and a new FFA design} 
The COMET and Mu2e experiments are based on similar concepts to produce muons and detect conversion electrons, albeit with some differences. In both cases, muons are produced by a proton beam hitting a primary target in a solenoid, and transported to a stopping target to be captured by Al nuclei. COMET includes a C-shape magnetic spectrometer between the stopping target and the detector, filtering out neutral particles, low-energy electrons and positively charged particles. This results in a lower occupancy and radiation dose in the detector, but precludes the possibility of measuring both positively and negatively charged tracks at the same time. By contrast, the Mu2e(-II) detector is (will be) placed just downstream of the stopping target, and features an annular design to be insensitive to low-energy particles. While this geometry presents more challenges in terms of track reconstruction and radiation dose, it enables the simultaneous measurements of $\mu^- \rightarrow e^+$ and muon-to-electron conversion.

The sensitivity of these experiments is limited by several factors, including the background due to the beam flash and pion decays (also preventing measurements with high-Z targets), the background induced from out-of-time protons (characterized by the beam extinction factor), the tracker momentum resolution to distinguish muon DIOs from the signal, the cosmic-induced background, the radiation dose in the detector, and the trigger latency.

To overcome some of these limitations, a new design based of a fixed field alternating gradient synchrotron (FFA) has been proposed~\cite{KUNO2005376}. The pion/muon beam emerging from the primary production target is directed into a FFA ring and undergoes phase rotation, trading time resolution for energy resolution. A momentum spread of $\sim 2\%$ is reached after six turns. During that time (${\cal O}(1) \ \rm \mu s$), the pion contamination is reduced to negligible levels. Using a primary proton beam energy below the antiproton production threshold would also eliminate the corresponding background, and the FFA injection / extraction system effectively cuts off other sources of delayed and out-of-time backgrounds. 

As a result an ultra pure, cold, monochromatic muon beam is available at the exit of the FFA. Such a beam would enable measurement with short muonic atom lifetimes, i.e. high-Z target material. In addition, the average beam energy can be significantly lower than that available in COMET/Mu2e, potentially as low as $20-30 \MeV$. Almost all muons could be stopped in a  thinner stopping target, reducing energy loss fluctuations and improving the conversion electron momentum resolution. 
   
Limitations arising from the DIO background, cosmic induced background and secondary particles produced from muon captures still need to be addressed. Adding a magnetic spectrometer after the stopping target would filter neutral and low-energy particles, significantly reducing the detector occupancy and radiation dose. The PRIME proposal includes a 540 degrees spectrometer (the so-called ``Guggenheim scheme"), but other designs could be considered. While this approach isn't charge symmetric, it should be noted that there is no need to determine the RPC background in-situ with positrons.

\subsubsection{Momentum resolution requirements}
\begin{figure}[ht!]
\centering
\includegraphics[width=0.6\textwidth]{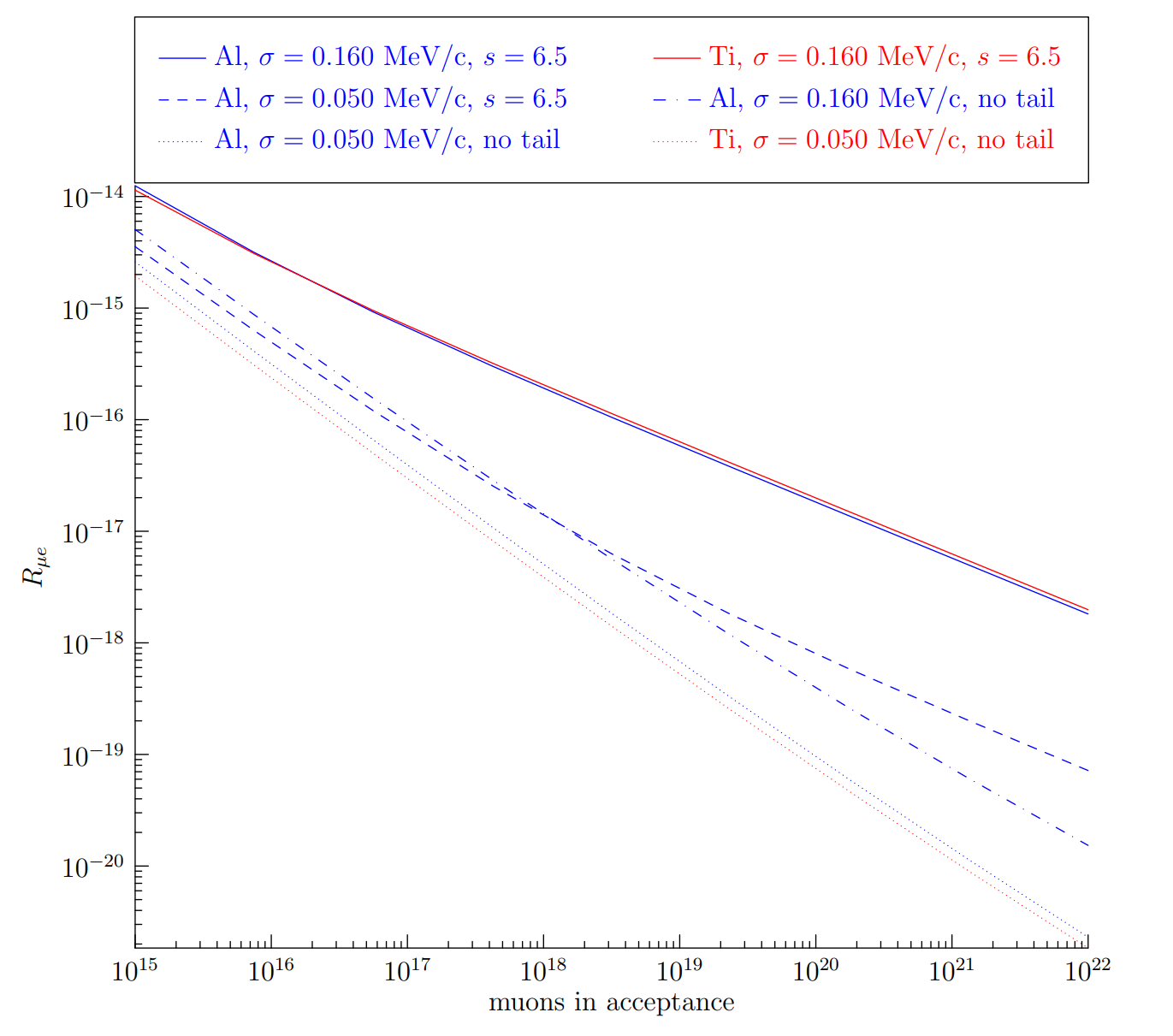}
\caption{The relationship between experimental sensitivity, signal resolution, and muons in acceptance for an Al-27 target (blue) and Ti-48 target (red).}
\label{fig:amf_ce_signal_res}
\end{figure}

The DIO background scales with the number of stopped muons and can only be distinguished from signal electrons by their momenta. A method to approximately quantify the relationship between signal resolution and statistical power of a conversion experiment has been recently developed, assuming all other sources of background are negligible~\cite{sensitivity-scaling}. A resolution function is convolved with both the signal and DIO spectra. This function comprises two components: a Landau distribution with width $\sigma$ describing the core and low side tail - this broadly accounts for the effects related to charged particles traversing the detector material - and a power law with power $s$ to model the high side tail arising from mis-reconstructed tracks. The results for a few scenarios on Al-27 and Ti-48 are shown in Figure~\ref{fig:amf_ce_signal_res}. For moderate gain in muon statistics ($\mathcal{O}(10^{17})$), improving the core resolution from 0.160~\MeV/c to 0.05~\MeV/c has a larger impact than removing the tail. However, reducing the tail has the largest effect once reaching higher muon statistics ($\mathcal{O}(10^{21}-10^{22})$). In order to achieve sensitivity below the $10^{-19}$ level, improvements on both the core and the tail are necessary.

\subsubsection{Tracking}
The quantitative analysis described above demonstrates the necessity of improving momentum resolution to surpass the expected sensitivity of Mu2e-II. 
In both Mu2e(-II) and COMET, tracking detectors in a constant magnetic field are employed to sample the electron trajectory for this purpose. In Mu2e(-II), a low-mass ionizing straw tube tracker is designed with a hole through the center to maintain acceptable occupancy levels. Conversely, the COMET detection section features a C-shaped filtering spectrometer that allows for a straw tube tracker without a central hole. Considering the occupancy issue is exacerbated by increasing muon beam intensity, the PRIME design assumes the inclusion of a spectrometer between the stopping target and tracking detector for the next-generation experiment. The general consensus reached during the workshop discussion is that AMF should plan for a filtering spectrometer as well.

Assuming the presence of a magnetic spectrometer, the tracking environment of the AMF conversion experiment would differ significantly from Mu2e(-II). Several improvements in the tracking environment at AMF highlight the benefits of the FFA magnet and spectrometer solenoid in the detection region. These advantages include eliminating the need for a hole in the center of the detector, virtually eliminating beam flash background particles and radiation, facilitating straightforward shielding against muon capture products (such as neutrons, heavy charged particles, and gammas), filtering to allow only highly momentum electrons to reach the detector, and creating a minimal impact on the tracker when changing the stopping target material due to the large physical separation between the stopping target and tracker. However, a few downsides exist, such as the potential need for additional shielding materials, which could make accessing the tracking detector for repairs more challenging, and the loss of the ability to measure positively charged particles.

To work towards leveraging the improved tracking environment and ultimately enhancing the momentum resolution, several tracking technologies were discussed. Continuing the legacy of Mu2e(-II) and COMET by utilizing a straw tube proportional tracker is attractive for several reasons. This technology allows for a highly segmented detection volume, offers good intrinsic momentum resolution, and benefits from the extensive expertise developed by the collaborations involved in current conversion experiments. However, manufacturing a straw tube tracker is challenging, especially if efforts are made to minimize tracker mass by using thinner straws. Additionally, the large number of straw tubes used in the detector provide many potential sources of gas leaks. An alternative option is employing a multi-wire proportional chamber (MWPC) for tracking, which inherently has less mass due to its single large gas volume. Furthermore, MWPCs are easier to manufacture compared to straw trackers. One major drawback of MWPCs is their inferior segmentation compared to straw trackers. Consideration may be given to exploring Gas Electron Multiplier (GEM) trackers that construct a high voltage potential across holes in a polymer sheet. GEM trackers offer advantages such as ease of manufacture, flexibility in geometric design, and a single large gas volume. However, drawbacks of GEM trackers include limited experience in their use and a potentially higher intrinsic mass. Exploring newer technologies represents another viable avenue, as a novel technique may better suit the needs of AMF, although this route very likely entails the most substantial R\&D efforts.

It is important to note that the development of a tracking detector for the conversion experiment holds potential synergies with positive muon experiments conducted at AMF. R\&D efforts in tracking technology could also benefit from collaborations with the Instrumentation Frontier, particularly the Micro-Pattern Gaseous Detector sub-group and the Solid State Detectors and Tracking sub-group.

\subsubsection{Simulation Scheme}
Targeted simulations are a crucial tool to guide the design process forward. While most of the Mu2e-II simulation efforts so far have been reasonably straightforward extensions of the mature Mu2e effort, AMF will explore a much broader landscape of possibilities and it will be difficult to directly re-use the Mu2e simulation code. However, there are some benefits to starting fresh and using experiences from Mu2e(-II) to organize the simulations in a sensible and scalable way from the beginning.

The simulation scheme should be derived from the current and anticipated future needs. We must consider the wide range of study that will need to be pursued, and the many different software packages that will be employed in the various stages of the simulation. Given these requirements, it was advocated to develop compartmentalized simulations. A sketch of the simulation stages and potential software tools is given in Table~\ref{tab:amf_ce_sim}.

\begin{table}[ht]
\begin{tabular}{p{4cm}|p{5cm}|p{3cm}}
Simulation Stage                    & Subtasks                                                            & Potential Software    \\
\hline
Proton beam\newline $\rightarrow$ Production Target & Beam transport                                                      & G4Beamline  \\ & &  \\ 
Pion production                             & Radiation dose, power/heat, \newline pion transparency, extinction           & Geant4/Mars \\ & &  \\
FFA (muon transport\newline and preparation)        & Accelerator dynamics, insertion, extraction, flash                   & G4Beamline  \\ & &  \\
Muon stopping                               & Stopping target, muon intensity, daughter production                & Geant4      \\ & & \\
Muon daughter transport                     & DIO collimation, signal selection, solenoid design, EM field design & G4Beamline  \\ & &  \\
Muon daughter detection                     & Detector solenoid, detector \newline design, pileup                          & TrackToy
\end{tabular}
\caption{A sketch of the different stages of simulation that a conversion experiment at the AMF may need. Details of the stages and potential software to use is included.}
\label{tab:amf_ce_sim}
\end{table}

One crucial element of compartmentalization is the connection of the different stages of the simulation. 
This is something that was not done for Mu2e, and we should make an effort to integrate this aspect form the start. 
A simple data format should be sufficient for the early studies, for example the HDF5 file format, as the produced samples should have a small footprint. Ultimately, a well thought out scheme should be devised.

\subsubsection{Cosmic Rays}

Cosmic ray muons can decay or interact with material near the stopping target to mimic signal electrons. In fact, the cosmic ray background is expected to be at the level of 1 conversion-like electron / day, the largest source in the Mu2e experiment~\cite{Mu2e:2022ggl}. A dedicated cosmic ray veto (CRV) detector is used to maintain this background at an acceptable level, but its performance must be further improved to extend the sensitivity of conversion experiments. There are several reasons to believe this can be achieved. 

First, the AMF detector region could be farther away from the primary target region. This region is close to the CRV in Mu2e(-II), resulting in large particle fluxes (mostly neutrons and gammas) producing false coincidences in the CRV and causing dead time. Keeping the detector region well separated from the production region will mitigate this issue. Second, high rates in the CRV near the stopping target could be significantly reduced if the concrete shielding blocks are enriched with high-Z materials~\autoref{sec:mu2ecrv}. Work is planned for Mu2e-II to fabricate and test the performance of such shielding blocks. Third, increased overburden would reduce the hadronic component of cosmic rays, a contribution becoming significant for Mu2e-II. Finally, the muon beam frequency at AMF would be much lower than foreseen for Mu2e-II, significantly reducing the exposure to cosmic ray background.

\subsubsection{R\&D Projects}
The discussions during the parallel session yielded a number of R\&D projects which are enumerated below and categorized by relative priority.

{\bf High Priority}
\begin{itemize}
    \item What type of beam can the FFA provide? Do we need to plan for an induction linac to lower the central momentum? This would have a large impact on designs.
    \item Map out the acceptance of different proposed experiment configurations (e.g. C-shaped vs. S-shaped spectrometer solenoid in the detection region)
    \item Utilize the sensitivity tool to study a few higher-Z target materials (e.g. Au).
    \item Which tracking technologies are people interested in pursuing? Determine the required R\&D needed for these options.
    \item Develop a scheme for cohesive, compartmentalized simulations and work out scheme for connecting different stages of the simulation.
    \item Determine consequences of running without $e^{+}$ (e.g. $\mu^{-}\rightarrow e^{+}$ signal channel, calibrations).
\end{itemize}

{\bf Medium Priority}
\begin{itemize}
\item Update the PRISM/PRIME diagram to include the other experiments of AMF that would utilize the FFA.
\item Start exploring CRV requirements and design. Keep cosmic-induced background in mind when exploring detector designs.
\item Explore shielding and design improvements.  
\item Understand what background particles come out of the FFA.
\end{itemize}

{\bf Low Priority}
\begin{itemize}
    \item Determine which tools should be used for the different steps in the simulation/analysis chain.
    \item Explore exotic solutions to e$^{+}$ measurements (e.g. assay the stopping target, multiple tracking paths).
    \item Assess feasibility of track back-extrapolation in different designs.
\end{itemize}

\subsection{Decay experiments}

One of the most appealing perspectives of AMF is the possibility of gathering in the same laboratory a vast community working on different experiments searching for New Physics in muon interactions, in order to take advantage of possible synergies. In this respect, besides the design of a next generation of muon to electron conversion experiments, it is critical to scrutinize, already in this conceptual design phase, both the machine and detector requirements posed by muon decay experiments. A similar effort was promoted in 2021 by the Particle Physics Laboratory of the Paul Scherrer Institut (PSI, Switzerland), in view of the upgrade of the muon beam lines foreseen in 2027-2028 (the High Intensity Muon Beam project, HIMB) and summarized in a written document~\cite{Aiba:2021bxe}.

\subsubsection{Beam requirements}

Searches for rare muon decays require intense beams of positive muons (to avoid the energy spectrum deformation produced by the capture of negative muons in the nuclear field) with a continuous time structure (to minimize the accidental time coincidence of multiple muon decays). Moreover, the kinematical resolutions are limited by the interaction of the decay products in the muon stopping target, the thickness of which has to be minimized.

The standard solution is the use of surface muons ($28.5 \MeV$ muons produced by pion decays at rest on the surface of the proton target), slowed down in a thin degrader and stopped in a thin target, with thicknesses optimized in order to have the Bragg peak of the muons fully contained in the latter. In this strategy, starting from a relatively low momentum is critical in order to have a small straggling of the total range and hence the possibility of using a thin stopping target (few hundred microns of plastic materials). As an alternative, in particular if the maximum available beam rate is too high to be tolerated by the experiments, one can consider to stop only a fraction of the muons in a thinner target, letting the rest of the beam to go through it. As a drawback, one has to prevent muons decaying right after the target since the products would contribute to the backgrounds without enhancing the signal. It implies that vacuum (and hence vacuum-compatible detectors) has to be used around the target, while both ongoing experiments (MEG II and Mu3e) are performed in a helium volume at atmospheric pressure.

\subsubsection{Experimental signatures and general detector requirements}

The two golden channels for LFV in muon decays are $\mu^+ \to e^+ \gamma$ and $\mu^+ \to e^+ e^+ e^-$, but other, more exotic decays are also of interest in some specific New Physics model~\cite{Renga:2019mpg}. In the search for the $\mu^+ \to e^+ \gamma$ decay, using muons at rest, a positron and a photon are searched for, emitted back-to-back, each with an energy equal to half the mass of the muon, $\sim 52.8 \MeV$. When very high muon beam intensities are used (up to $5 \times 10^{7}\,\mu$/s in MEG II), the dominant source of background is the accidental coincidence of positrons and photons from two different muon decays. Hence, the time difference between the two particles is an important discriminating variable for signal against background, along with the particle energies and their relative angle. These kinematic variables also allow to suppress the sub-leading background coming from the radiative muon decay $\mu^+ \to e^+ \nu_e \overline \nu_\mu \gamma$. 

The $\mu^+ \to e^+ e^+ e^-$ search is performed requiring two positrons and one electron from a common vertex, with an invariant mass equal to the muon mass. The three-track vertex strongly suppresses the accidental coincidences, so that the $\mu^+ \to e^+ e^+ e^- \nu_e \overline \nu_\mu$ background also plays an important role.

Both experiments, as well as the more exotic searches, require the reconstruction of electrons and/or positrons with a relatively low momentum, around and below $ 50 \MeV$. In this range, magnetic spectrometers with tracking detectors overwhelm the calorimeters for the measurement of the energy, providing at the same time the necessary resolution on angles and vertex. The reconstruction is strongly affected by the interaction with the materials of and surrounding the target, and with the materials of the detectors. For this reason, extremely light trackers have to be used. Large volume gaseous detectors (drift chambers and time projection chambers) give the best compromise of single hit resolution and material budget, but poor granularity and significant aging rates will make their use problematic with the higher rates of the future facilities. State-of-the-art, 50~$\mu$m-thick monolithic silicon pixel sensors (MAPS) can provide the necessary rate capabilities with an acceptably low material budget, and the next generation of experiments could take advantage of even thinner devices. Both silicon and large gaseous detectors need to be complemented with faster detectors to achieve the time resolution that is necessary for the rejection of accidental backgrounds.

Photons can be reconstructed using calorimeters, with an appropriate choice of the scintillating material. The LXe calorimetry developed for the MEG experiment provides excellent performances at low energies, but cost and handling of large xenon detectors make problematic the construction of detectors with large acceptance. The use of innovative crystals have been investigated in the light of future precision physics experiments~\cite{Papa:2023voa}, although large scale production and cost can be an issue also for these kinds of materials. For some specific applications like $\mu^+ \to e^+ \gamma$, where the accidental background limits in any case the acceptable rate of detected photons, an alternative solution is the conversion of photon in thin layers of materials and the reconstruction of the outcoming $e^+e^-$ pair in a magnetic spectrometer, because the higher energy resolution that can be achieved with this technique and a higher rate of muons can be exploited to compensate for the very low efficiency of the conversion process. 

\subsubsection{\texorpdfstring{$\mu^+ \to e^+ e^+ e^-$}{mu+ to e+e-e+}}

At present, the best limit on the branching ration (BR) of $\mu^+ \to e^+ e^+ e^-$ comes from the SINDRUM experiment, $BR(\mu^+ \to e^+ e^+ e^-) < 1.0 \times 10^{-12}$, and dates back to the late 1980s~\cite{SINDRUM:1987nra}. A new experiment called Mu3e is under construction at PSI~\cite{Mu3e:2020gyw}. The experiment has been already designed to be ultimately operated at HIMB, with more than $10^9\,\mu$/s, aiming at a final BR sensitivity around $10^{-16}$.

\begin{figure}[ht!]
\centering
\vspace{-9.5cm}
\includegraphics[width=\textwidth]{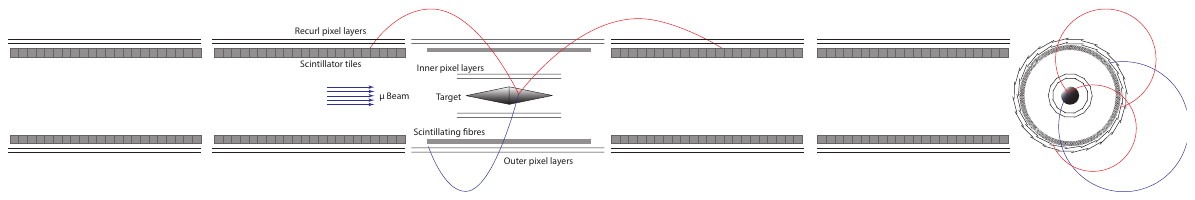}
\vspace{-10.5cm}
\caption{Sketch of the Mu3e experiment at PSI.}
\label{fig:mu3e}
\end{figure}

As a consequence, the detector is already designed to cope with an environment very similar to the one expected at AMF. The detector is composed of cylindrical tracking stations made of HV-MAPS, complemented by layers of plastic scintillating tiles and fibers readout by silicon photomultipliers (SiPM), within a $\sim 1$~T magnetic field. A sketch of the detector is depicted in Figure~\ref{fig:mu3e}. 

The HV-MAPS have $80 \times 80 \rm \, \mu m^2$ pixels and $50$~$\mu$m thickness, resulting in a material budget corresponding to $X/X_0 = 0.011\%$ per layer. In the high occupancy environment produced by muon stopping rates exceeding $10^{9}$~$\mu$/s, the relatively poor time resolution of the HV-MAPS, $O(50~\rm ps)$, would compromise the possibility of a correct association of hits to different tracks. So, the scintillation detectors are critical to isolate hits produced by the same particle.

The two innermost tracking layers will provide an accurate vertexing with $< 200$~$\mu$m resolution, while the geometry of the outer layers is optimized to get $< 500 \keV$ momentum resolution. The scintillation detectors will also measure the time of the particles with a resolution better than 100~ps, allowing for an effective rejection of accidental backgrounds.

\subsubsection{\texorpdfstring{$\mu^+ \to e^+ \gamma$}{mu+ to e+ gamma}}
\label{sec:meg}

The best limit is much more recent, $BR(\mu^+ \to e^+ \gamma) < 4.2 \times 10^{-13}$, published by the MEG collaboration in 2016~\cite{MEG:2016leq}, and the first result of the upgraded experiment MEG II~\cite{MEGII2018} is expected within 2023. Then, the future of this quest is much more unsettled. It was investigated in Ref.~\cite{next_meg} and more recent advances were considered in the HIMB science case document~\cite{Aiba:2021bxe}.

By contrast to $\mu^+ \to e^+ e^+ e^-$, the photon reconstruction cannot provide by any means the sub-millimeter resolution that is necessary for an effective two-particle vertexing, so that the $\mu^+ \to e^+ \gamma$ search is much more limited by the accidental background. Since it originates from accidental time coincidences, it scales with the square of the muon beam rates, and it poses a severe limitation on the maximum rate that can be proficiently exploited. Indeed, for a given beam rate, if the other discriminating variables ($e^+ \gamma$ energies, relative angle and relative time) do not provide enough separation to keep the background yield at $O(1)$ over the full lifetime of the experiment, the sensitivity, scaling with $S/\sqrt{B}$, becomes independent of the beam rate, the increase of which would only overcrowd the detectors without any real statistical advantage. In the MEG II experiment, the optimal beam rate is found to be around $5 \times 10^{7}\,\mu$/s, below the maximum rate already achievable at PSI.

Additionally, the detectors of the MEG II experiment are not designed to cope with the very high occupancy produced by a muon beam rate increased by a factor 10 to 100 at future facilities: both in the positron spectrometer (instrumented with a high-granularity drift chamber) and in the LXe calorimeter, pileup events are already impacting the reconstruction of the events at $5 \times 10^{7}\,\mu$/s; moreover, at $10^{9}\,\mu$/s, aging effects in the drift chamber would become unmanageable if specific solutions are not found to strongly suppress them.

New options need to be investigated, aiming at an improvement of the resolutions (to suppress the accidental background) and an increase of the rate tolerance of the detectors. A few technical solutions were already investigated in the HIMB science case document. 

On the positron side, a silicon detector \emph{\`a la} Mu3e would provide the necessary high rate capabilities, although the momentum resolution could be slightly worse, due to a less favorable compromise between material budget and number of hits. In this respect, a next generation of HV-MAPS, thinned down to 25~$\mu$m, and an optimization of the detector geometry and magnetic field could allow to fill the gap against gaseous detectors. To now, the silicon detector solution looks as the best option for the next-generation experiments, although some ideas still circulate in the community, to improve the rate capabilities of gaseous detectors with new geometries (transverse drift chambers, transverse drift tubes \emph{\`a la} Mu2e, radial time projection chambers), new wire materials and new, hydrocarbon-free gas mixtures.

On the photon side, the conversion technique seems to be the most promising one to fully exploit higher beam rates. The basic idea is that, with a beam rate up to 100 times higher, but a total conversion efficiency not much larger than a few per cent, the total photon rate (and hence the accidental yields) would not significantly increase, while the better resolutions achievable in the photon reconstruction would allow to more effectively reject the backgrounds. Moreover, a larger acceptance could be achieved with lower cost, and the directional information extracted from the $e+e-$ pair would allow to build a $e\gamma$ vertex, which would provide a further handle to reject the accidental coincidences.

\begin{figure}[ht!]
\centering
\includegraphics[width=0.7\textwidth]{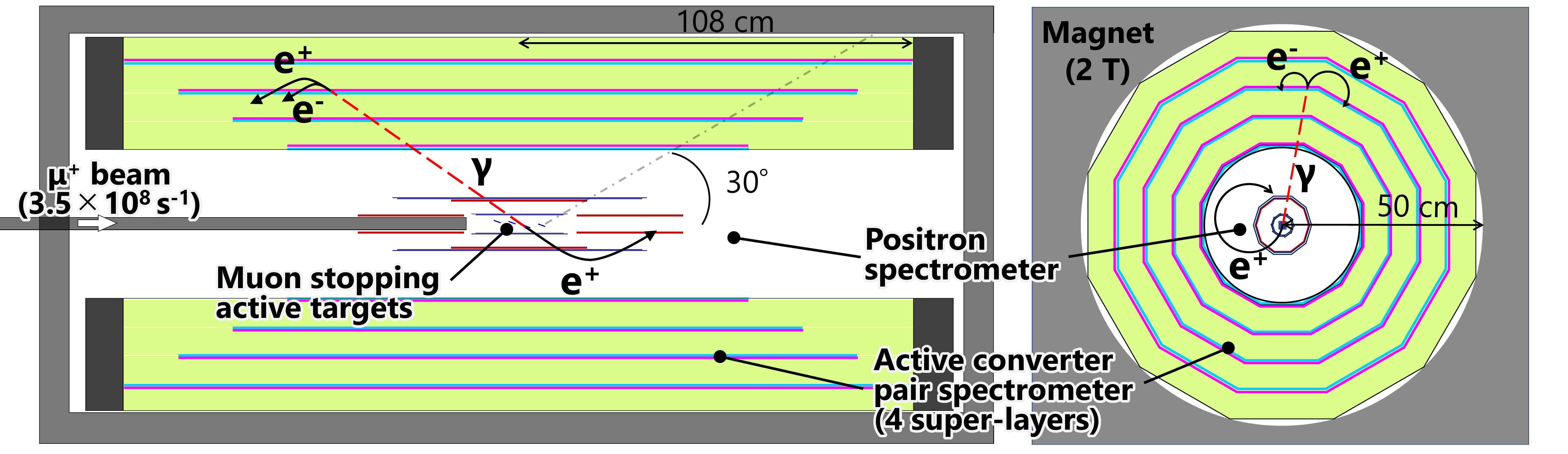}
\vspace{1cm}\\
\includegraphics[width=0.5\textwidth]{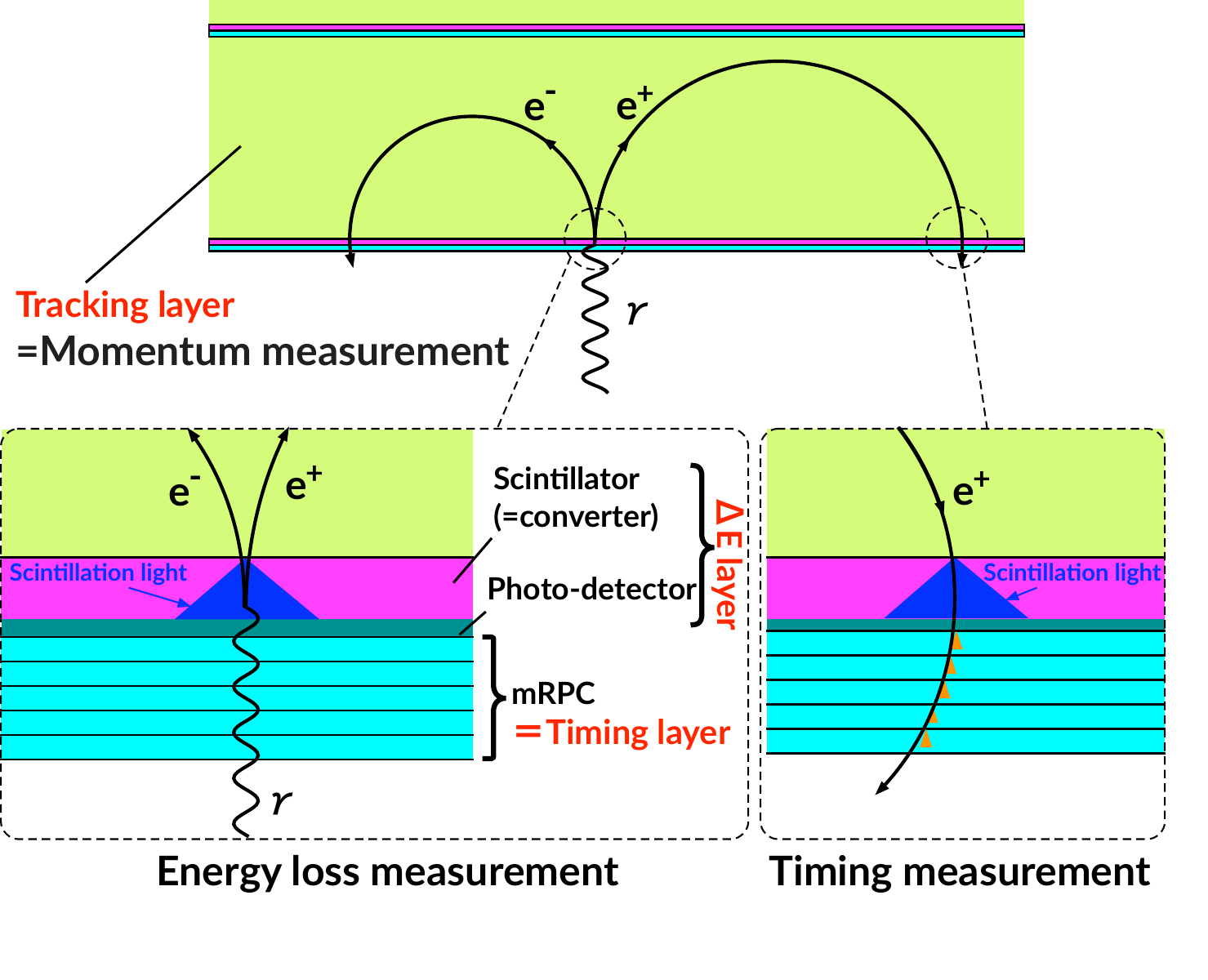}
\caption{The possible layout of a $\mu^+ \to e^+ \gamma$ experiment based on a silicon positron tracker and a photon conversion detector (top), with a detail of a possible photon detector concept (bottom).}
\label{fig:gamma_conversion}
\end{figure}

In the last couple of years some significant advance was made in the understanding of the performance of the conversion technique and in the conceptualization of a photon detector for a future experiment. Figure~\ref{fig:gamma_conversion} shows a possible arrangement for a photon detector based on the conversion technique. 

Instead of a thin and dense passive converter, a scintillating crystal instrumented with SiPM is considered. It would allow to measure the energy deposit in the converter itself, that is the main limiting factor to the energy resolution when a passive converter is used. 

Tracking the $e^+e^-$ pair with excellent resolution and good efficiency will require at least two or three layers of position-sensitive detectors in close proximity to the conversion layer (within about 1.5~cm for a 1~T magnetic field). Although silicon detectors would fit this design, the necessity of stacking multiple conversion layers to increase the efficiency would make this option impractical from the technical and economical point of view, considering that tens of square meters of detectors would be needed to guarantee a good acceptance. However, it can be shown that the geometry of drift chambers with stereo wires do not fit this design, while time projection chambers would be too long to provide the necessary resolutions when operated with low-mass gas mixtures. The option that is currently under consideration is a time projection chamber with radial drift. Simulations are on going to assess the performance of a detector with such a geometry.

Finally, the conversion technique should be adopted without detriment to the good photon time resolution ($\sim 70$~ps) characterizing the MEG and MEG II experiments. It could be achieved with the addition of gaseous detectors with fast timing, like multi-gap resistive plate chambers (mRPC), just in front of the active converter. These detectors would be light enough to not deteriorate the performance of the active converter, while providing the necessary time resolution on the $e^+$ and $e^-$, after they recurled through the spectrometer.

Simulations and test beam measurements on scintillating crystals for the active converter are on-going. Preliminary results have been shown, indicating that four layers of 3~mm thick LYSO crystals could give a conversion efficiency of 10\% and an energy resolution of about $140 \keV$, to be combined with the resolution of the $e^+e^-$ tracker, the simulation of which is on going.

Other innovations were also briefly discussed at the workshop, namely an active target to get a direct information on the positron vertex and multiple thinner targets in vacuum.

\begin{figure}[ht!]
\centering
\includegraphics[width=0.5\textwidth]{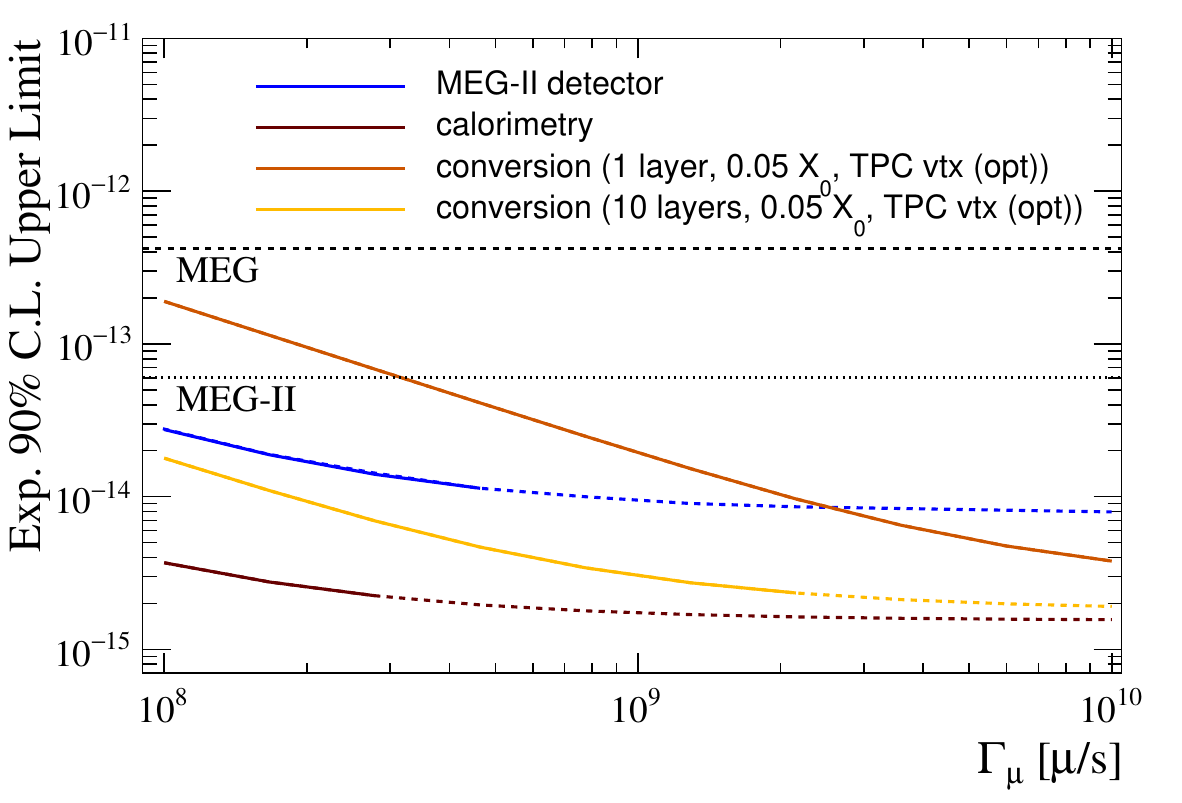}
\caption{Sensitivity of next-generation $\mu^+ \to e^+ \gamma$ experiments under different assumptions for the detector design. See Ref.~\cite{next_meg} for a detailed description of the underlying assumptions.}
\label{fig:next_meg_sens}
\end{figure}

All these studies give a preliminary indication that the performance envisaged in Ref.~\cite{next_meg} could be realistic, so that a BR sensitivity approaching $10^{-15}$ could be reached with beam intensities exceeding $10^{9}~\mu$/s, as shown in Figure~\ref{fig:next_meg_sens}.

\subsubsection{Exotic decays}

More exotic channels can be considered, where the muon decays to new, non standard particles, that can in turn decay into standard particles or remain invisibile (either because they are stable or their product do not interact with the detector). A review of several channels and searches can be found in Ref.~\cite{Renga:2019mpg}.

As a case study, the decays of muons with lepton-flavor violating axion-like particles (ALPs) in the final state were discussed at the workshop. The two channels of interest are $\mu^+ \to e^+ a$ and $\mu^+ \to e^+ a \gamma$, where the ALP $a$ escapes the detection.

When the mass of the ALP is zero or very low compared the energy resolutions of the detector, the search for $\mu^+ \to e^+ a$, which is based on the reconstruction of a monochromatic positron, is affected by large systematic uncertainties because a peak at the kinematical end point of the muon decay spectrum can be easily faked by a miscalibration of the momentum scale (e.g. in the calibration of the magnetic field). The most recent limit, from the TWIST collaboration~\cite{TWIST:2014ymv}, is the result of a careful exploration of possible miscalibrations at the $\keV$ level, and strong improvements are difficult to envisage.

For higher masses, similar searches in experiments tailored for $\mu^+ \to e^+ \gamma$ are typically limited by the momentum acceptance of the positron spectrometer, which is optimized for the highest momenta, but there is an intermediate range, where the TWIST search is statistically limited, that could be explored in more detail already in MEG II. It should be noticed that, although the trigger of MEG II requires the coincidence of a positron and a photon, the large majority of the acquired events is of accidental nature, and hence the observed positron can still be used to reconstruct signatures without photons.

\begin{figure}[ht!]
\centering
\includegraphics[width=0.5\textwidth]{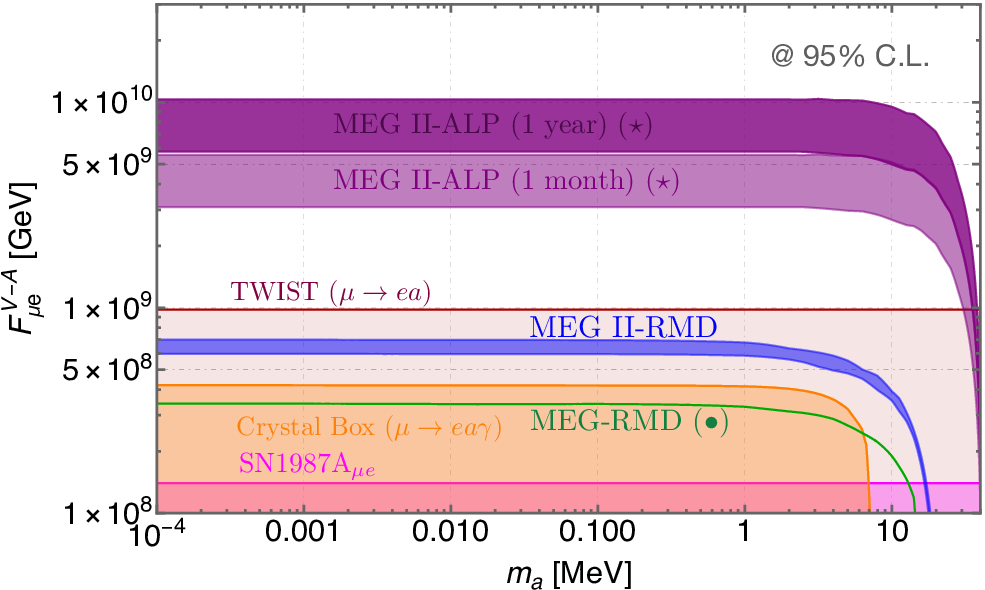}
\caption{Sensitivity of MEG II, with a dedicated trigger strategy and a reduced muon beam intensity, to left-handed (V-A) ALP couplings, compared to the current limits. See Ref.~\cite{Jho:2022snj} for details.}
\label{fig:axions}
\end{figure}

Anyway, having in mind the possibility of reconstructing the photon with high resolutions, the $\mu^+ \to e^+ a \gamma$ channel can be even more interesting in MEG II and future $\mu^+ \to e^+ \gamma$ experiments, considering the background discrimination power provided by the reconstruction of a missing mass. This possibility was examined in detail in Ref.~\cite{Jho:2022snj}. In this case, the available statistics is limited by the trigger requirements for the $\mu^+ \to e^+ \gamma$ search, which restrict the available phase space to a small region at high energies and large $e\gamma$ angle. The outcome of the studies done so far is that it would be advantageous to allocate for this specific search a relatively small data taking period (a few percent of the total beam time), during which the trigger requirements are released, and the beam rate reduced to enhance the prompt over accidental yield ratio. This new data taking strategy would strongly enhance the sensitivity of MEG II to $\mu^+ \to e^+ a \gamma$, and with 1 month of data taking, assuming for instance V-A currents, the current limits for the ALP couplings would be overcame by a factor up to 5, as shown in Fig.~\ref{fig:axions}. Detailed simulations and trigger rate estimates are ongoing within the collaboration to confirm these results.

Being the sensitivity of these searches also limited by the background from accidental coincidences, the same considerations done for $\mu^+ \to e^+ \gamma$ at future high-intensity facilities also apply in this case.

\subsection{Other experiments}
\subsubsection{Session Summary}

In this session of the workshop, we discussed the current status and future plans of various muonium experiments and proposals. Muonium (commonly referred to as M or Mu, by particle or nuclear physicists, respectively) is a $\mu^{+}e^{-}$ bound state that is produced when $\mu^{+}$ stop in matter. It is a purely leptonic, hydrogen-like atom where the nucleus consists of a $\mu^{+}$. This makes it an excellent venue for precision tests of QED since, e.g., nuclear size effects do not need to be considered. Muonium experiments typically employ ``surface'' muon beams: beams of antimuons produced when positive pions decay at rest inside the production target; this two-body decay yields a forward-going, quasi-monoenergetic ($\approx\,$4\,MeV, or $\approx$\,28\,MeV/$c$) $\mu^+$ beam that is 100\% backward-polarized and can be stopped to produce muonium in thin conversion targets.

Three categories of fundamental-physics measurements were discussed at the Workshop: the search for M--$\overline{\rm M}$ mixing, precision M spectroscopy measurements, and the measurement of the gravitational acceleration of M in the Earth's gravitational field. Published results from the first two measurement categories date from over 20 years ago~\cite{MACS:1999,LambShift:1984,1s-2s:2000,hyperfine:1999}, so the field is ripe for renewed efforts. New spectroscopy measurements are in progress: the MuSEUM~\cite{MuSEUM:2021,MuSEUM-Nishimura:2021} hyperfine-splitting experiment at J-PARC and the Mu-MASS~\cite{MuMass-Crivelli:2018,MuMass-Ohayon:2021} 1s--2s experiment at PSI. The MACE M--$\overline{\rm M}$ mixing experiment is proposed at the Chinese Spallation Neutron Source, and the LEMING M gravitational experiment is in an R\&D stage at PSI~\cite{LEMING}. The PIP-II linac under construction at Fermilab should enable the world's highest-intensity muonium beam. All three categories of experiments thus appear worthy of pursuit at Fermilab once PIP-II is operational, assuming a suitable high-power PIP-II target facility is built and the currently planned accelerator is upgraded for continuous operation. 

\subsubsection{Muonium Theory}


Rare muonium decays of interest include ${\rm M}\to e^+e^-$, ${\rm M}\to\gamma\gamma$, and ${\rm M}\to\nu_e\overline{\nu}_e$, all three of which would be dominated by new physics if they occur. In effective field theory, compared to e.g.\ $\mu\to3e$, they probe different combinations of Wilson coefficients, so are worth searching for whether or not $\mu\to3e$ is observed.

\begin{figure}[tb]
    \centering
    \includegraphics[height=1.75in,trim=0 5 0 5,clip]{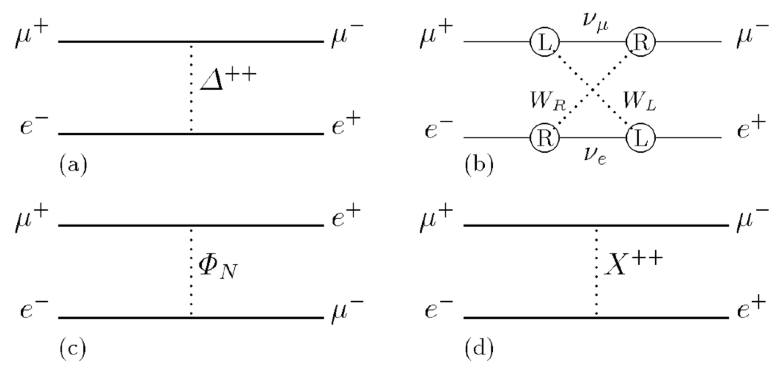}
    \caption{Examples of new-physics diagrams contributing to M--$\overline{\rm M}$ mixing: exchange of (a)~a doubly charged Higgs boson $\Delta^{++}$, (b)~heavy Majorana neutrinos, (c)~a neutral scalar $\Phi_N$, or (d)~a bileptonic flavor diagonal gauge boson $X^{++}$ (from~\cite{MACS:1999}).}
    \label{fig:M-Mbar-diagrams}
\end{figure}

Conversion of muonium to antimuonium\,---\,the simultaneous conversion of an antimuon to a positron and an electron to a muon\,---\,would be an example of double charged-lepton flavor violation. While not forbidden (neutrino oscillation being well established), the rate of such conversion via virtual neutrino oscillation is so low as to be essentially unobservable. Representative new-physics diagrams contributing to muonium--antimuonium oscillations are shown in Fig.~\ref{fig:M-Mbar-diagrams}. Since some new-physics mechanisms favor M--$\overline{\rm M}$ mixing over muon-to-electron conversion, the former is not necessarily suppressed with respect to the latter. As with rare decays, since Mu2e and M--$\overline{\rm M}$ mixing are differently sensitive to new physics, both should be sought as sensitively as possible. 

\subsubsection{Muonium--Antimuonium Conversion Experiment (MACE)}

The MACE experiment proposes to search for muonium--antimuonium oscillation at a high-intensity muon beamline at, e.g., the High Intensity Heavy-ion Accelerator Facility (HIAF) or the Chinese Spallation Neutron Source (CSNS). 

\begin{figure}[tb]
    \centering
    \includegraphics[height=2.25in,trim=0 60 0 60,clip]{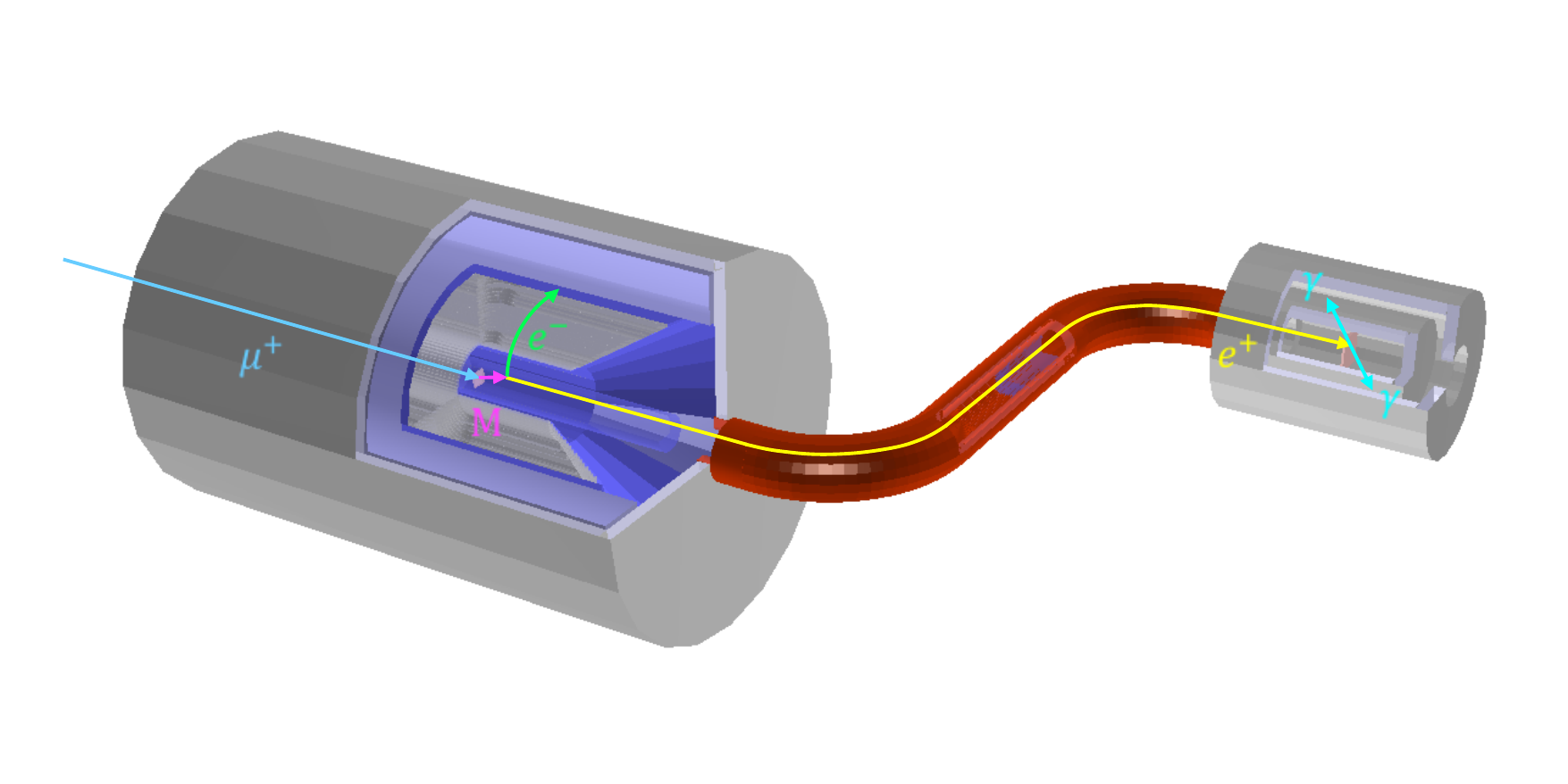}
    \caption{3D cutaway view of  proposed MACE apparatus. Antimuons are incident from the left on an aerogel target in which many of them stop and convert to muonium. If a muonium atom converts to antimuonium, its decay electron is reconstructed by the surrounding cylindrical drift chamber and its slow positron is accelerated and sign- and velocity-selected in the 
    S-shaped solenoid channel, conveying it to a microchannel plate in which it annihilates, sending gammas to be detected in the surrounding calorimeter.}
    \label{fig:MACE-3D}
\end{figure}

The existing limit ($P_{\rm M\overline{M}} < 8.3 \times 10^{-11}$ at 90\% C.L. in an 0.1\,T magnetic field) was set by the MACS experiment at PSI in 1999~\cite{MACS:1999}. The goal of MACE is to improve on MACS sensitivity by more than two orders of magnitude, by increasing the rate of muonium formation and improving apparatus resolutions. The design (Fig.~\ref{fig:MACE-3D}) employs a $\mu^+\to {\rm M}$ conversion target made of silica aerogel placed within a cylindrical drift chamber surrounded by a solenoid magnet. The aerogel is perforated (``laser ablated'') with an array of small blind holes to increase its surface-to-volume ratio, enhancing the probability that M atoms produced within escape to the surrounding vacuum~\cite{Beer:2014}. ${\rm M\to\overline{M}}$ conversion leads to an atom consisting of a muon bound to a positron, producing a fast electron and slow positron once the muon decays\,---\,opposite in electric charge to the decay products of muonium. The fast electron is reconstructed and its sign and momentum measured in a cylindrical drift chamber. The slow atomic positron left behind when the muon decays is accelerated electrostatically and sign- and velocity-selected in a curved solenoid channel, which transports it to a microchannel plate (MCP), in which it annihilates, sending a pair of $511 \keV$ gamma rays into the surrounding calorimeter.

Optimization of the muonium yield and detector performance is under study. The muonium vacuum yield for various target designs is simulated using the random-walk method. The simulation has been  validated and is believed to be reasonably accurate. This simulation provides a basis for  optimization of the perforated silica aerogel target and indicates that a surface muon beam with a relatively small momentum spread (2.5\% or less) can  increase the yield of muons to 1.5\% or higher, to be compared with the 0.5\% yield in MACS. Together with the higher expected muon intensity, this yields an improvement of about two orders of magnitude compared to MACS. On the other hand, the background rejection capability of MACE depends on the spatial resolutions of the detector\,---\,in particular, those  of the spectrometer and MCP. The wire structure of the drift chamber has been modeled, and an initial reconstruction algorithm has been implemented. By examining various readout layer layouts, the geometry of the drift chamber can be further optimized. Current results demonstrate vertex resolution in the range 5--10~mm. By improving the reconstruction algorithm, the resolution of the drift chamber is expected to be further improved, with a goal below 5~mm.

The design has been simulated in Geant4 to characterize its performance~\cite{MACE-Snowmass:2022}. The limiting backgrounds in MACS were accidental coincidences and the rare decay $\mu^+\to e^+e^+e^-\nu_e\overline{\nu}_\mu$ (which can fake the signal mode if one of the positrons is missed). MACS had one event passing its elliptical time--vertex cut (cut radii of 4.5~ns and 12~mm, respectively). For a background-free sample with two orders of magnitude more M events,  the time, vertex distance of closest approach, and energy resolutions should be substantially tightened with respect to those in MACS. Preliminary MACE simulations indicate resolutions of 6.5~mm in distance of closest approach (comparable to) and 0.65~ns (better than that in MACS) in electron--positron time difference. Further resolution improvement is desired and is the subject of ongoing studies. 

\subsubsection{Muonium Spectroscopy}

Muonium spectroscopy experiments aim to measure one of three transitions: the ground state hyperfine splitting, the 1s--2s interval, or the Lamb shift. The hyperfine and 1s--2s splittings together with $g-2$ allow an independent, muon-only determination of the fine-structure constant $\alpha$. 

The hyperfine splitting is currently being measured by the MuSEUM experiment~\cite{MuSEUM:2021,MuSEUM-Nishimura:2021} at J-PARC with an uncertainty goal of 5\,Hz (1.2\,ppb). This would exceed the precision of the theoretical prediction and so can be viewed as a measurement of the ratio $m_\mu/m_e$, which dominates the systematic uncertainty of the prediction. In MuSEUM, muonium is created by stopping highly polarized muons in krypton gas. Positrons will be preferentially emitted in the spin direction (upstream). A perpendicular microwave magnetic field is applied in order to flip the spins. The 
hyperfine splitting is determined 
by measuring the ratio between upstream and downstream positron emission as a function of microwave frequency. 

The 1s--2s splitting (and Lamb shift) 
are currently being measured by Mu-MASS~\cite{MuMass-Crivelli:2018,MuMass-Ohayon:2021} at PSI. The uncertainty goal for the 1s--2s splitting is 10\,kHz (4\,ppt), which will enable a 1 ppb measurement of $m_\mu$. In Mu-MASS, muonium is formed in silica, then excited to the 2s state and ionized with a pair of lasers. The muon is then transported away from the creation point and its emitted Lyman-alpha photon is detected in an MCP detector.

\subsubsection{Low-Energy Muons at Fermilab}
We briefly describe the MeV Test Area (MTA) beam line at FNAL, which could be used for low-energy muon and muonium R\&D (and perhaps for competitive physics experiments as well), and expectations for the PIP-II superconducting linac, whose construction is in progress.

\begin{figure}
    \centering
    \includegraphics[height=3.5in]{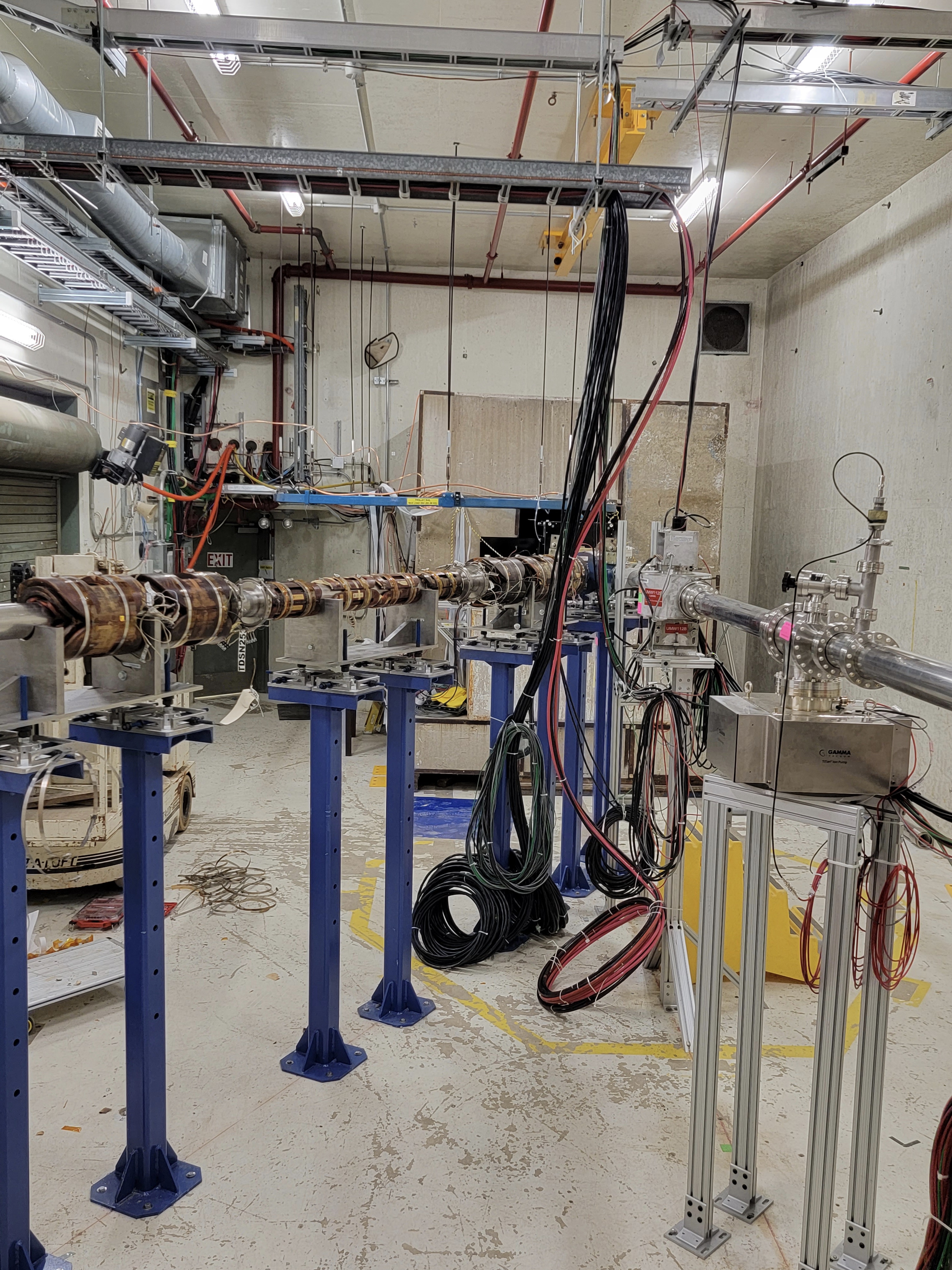}
    \caption{Photo of Fermilab MeV Test Area showing 400\,MeV proton beamline at right and pion--muon quadrupole channel at left. The protons enter traveling into the page at the right edge of the photo, and muons (produced both within the target and via pion decay in flight) exit out of the page at the left edge. A target holder is situated at the intersection of the two beam lines, followed by a small solenoid to capture charged particles produced in the target.}
    \label{fig:MTA-beam}
\end{figure}

The MTA is situated near the end of Fermilab's 400\,MeV H$^-$ Linac and  shielded well enough to accept a large fraction of the protons accelerated by the Linac (although a shielding assessment demonstrating this has yet to be finalized). A new MTA beam line was recently designed and installed to serve an approved muon-catalyzed-fusion R\&D program; it naturally produces low-energy pions and muons of both signs. The layout is shown in Fig.~\ref{fig:MTA-beam}. Geant4 simulation studies using a 3-cm-long rectangular W target predict  surface-muon production at a rate of order $\sim$\,10$^{-9} \mu^+/$proton-on-target (POT) into the acceptance of the beam line, which is centered at a 146.4$^\circ$ production angle. Measurements made in the 1970s~\cite{Cochran:1972} suggest that, compared to the thin graphite targets used at surface-muon beam lines at spallation-neutron facilities such as PSI, RAL, and TRIUMF, a Ta or W target should produce more charged pions and surface muons per POT by factors of 3 ($\pi^+$ and $\mu^+$) and 8 ($\pi^-$ and $\mu^-$). (These factors are not borne out in Geant4 studies, but Geant4 is not expected to be reliable for such cases; measurements are therefore needed, and 
planned.) Studies to optimize the target configuration and beam-line acceptance are in progress and may yield higher muon rates~\cite{Bungau:2013}. However, for  purposes of surface-muon (and muon-catalyzed-fusion) R\&D, the rates so far simulated are more than sufficient.

The PIP-II linac will of course accelerate to 800\,MeV, twice the energy of Fermilab's existing Linac. While neither 800 nor 400\,MeV is optimal for surface-muon production, the predicted rate decrease per POT with respect to the peak at 590\,MeV is only about 15\%~\cite{Bungau:2013}. Furthermore, especially  if it is upgraded for CW operation, most of the protons that the PIP-II linac could accelerate will not be destined for the Main Injector and could enable a low-energy-muon physics program at Fermilab that\,---\,even after PSI's High-Intensity Muon Beam (HIMB) upgrade is completed\,---\,would be the world's most sensitive by more than an order of magnitude.

To prepare for a future, world-leading, PIP-II muon facility, it is highly desirable to conduct R\&D now at the MTA to develop and gain experience with the techniques that will be needed at PIP-II, such as those discussed in the next subsection. Such a program at the MTA will also allow  (for the first time) a U.S.-based program of $\mu$SR measurements, obviating the current need for Fermilab superconducting RF-cavity developers (and other U.S. researchers) to take their samples to Canada for $\mu$SR surface studies at TRIUMF.

\subsubsection{Muonium Production in Superfluid Helium}

\begin{figure}
    \centering
    \includegraphics[height=2.25in]{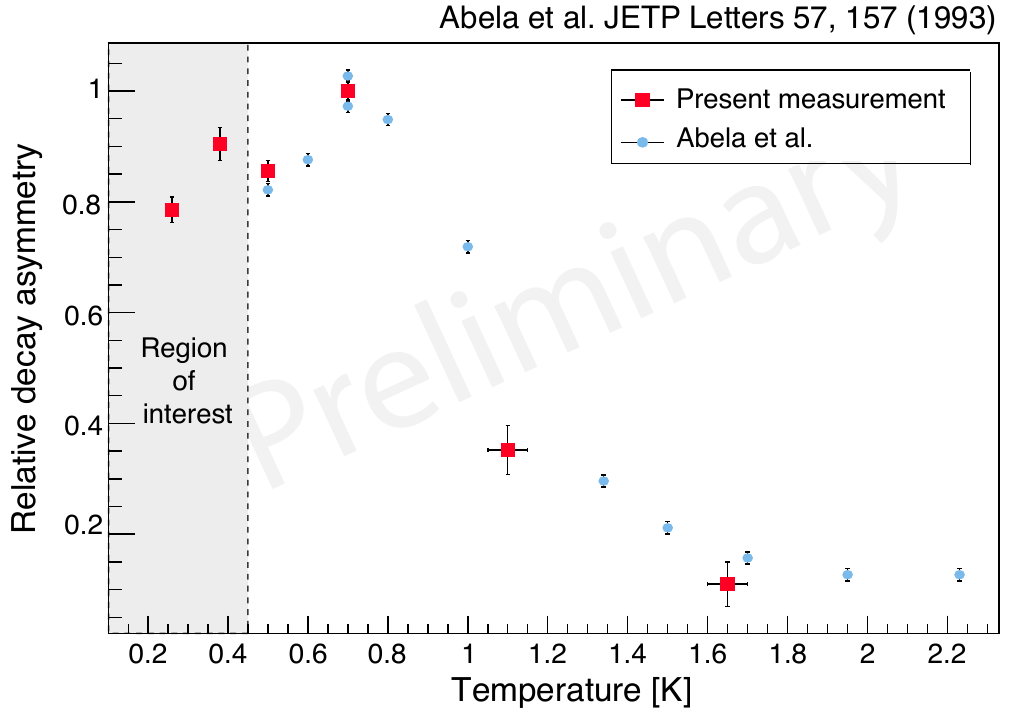}
    \caption{Relative production vs.\ temperature of muonium in superfluid helium (blue points from~\cite{Abela:1993} and recent measurements in red by a PSI--ETH-Z\"{u}rich collaboration~\cite{Soter-private}).}
    \label{fig:SFHe}
\end{figure}

Measurements~\cite{Abela:1993} (Fig.~\ref{fig:SFHe}) have shown that below 1\,K, superfluid helium (SFHe) is an efficient converter of stopped $\mu^+$ to muonium. The signature of M formation is rapid precession of the muon spin; the resulting time evolution of the decay asymmetry in a magnetic field has been used to explore the production of muonium when $\mu^+$ are incident on pure superfluid $^4$He or a solution containing 0.2\% $^3$He~\cite{Abela:1993,Soter-private}. The high conversion efficiency and the immiscibility of hydrogen isotopes in superfluid helium form the basis of a technique proposed by PSI's David Taqqu~\cite{Taqqu:2011} that can make a monochromatic, parallel beam of muonium atoms propagating in vacuum by stopping slow $\mu^+$ in a thin, horizontal SFHe layer. Such a beam would for the first time enable an interferometric muonium gravity experiment (see next subsection) and could also be a game-changer for other high-sensitivity muonium measurements.

Their immiscibility implies that any muonium atoms formed in the SFHe that reach its upper surface will be expelled perpendicular to that surface at a velocity determined by the chemical potential of the atoms in SFHe, predicted to be 270\,K for muonium~\cite{Taqqu:2011}, implying an expelled-atom velocity of 6.3\,mm/$\mu$s. (The prediction depends on isotopic mass and has been verified experimentally for the case of deuterium~\cite{Saarela:1993}; this immiscibility has also been exploited to make a functioning SFHe-coated focusing mirror for hydrogen atoms~\cite{Luppov:1993}.) 
To maximize the efficiency of beam creation, it is desirable to suppress muonium formation in the SFHe bulk by applying an electric field to drift the muons away from their ionization electrons; the required field can be created by a surface-electron pool of $10^8$\,$e$/cm$^2$, maintained e.g.\ by use of a W tip. Such an electron pool is stable below 1\,K~\cite{Barenghi:1986}. Calculations show that for a layer thickness of $\approx$\,300\,$\mu$m, up to $\approx$\,10\% of muons entering the SFHe drift to the surface before decaying and yield escaping M atoms. 

An alternative method of achieving high M escape efficiency is under development at PSI by our PSI and ETH-Z\"{u}rich colleagues (K. Kirch {\it et al.}~\cite{Soter-private}), in which an extremely thin ($\sim$\,1\,$\mu$m) SFHe layer is employed, such that about half of the muonium atoms formed in the bulk diffuse to the upper surface and are expelled. To achieve a substantial $\mu^+$ stopping probability in such a thin layer, the $\mu^+$ beam must be precooled. This ``muCool" apparatus (also proposed by Taqqu~\cite{Taqqu:2006}) has been developed and tested successfully, compressing the $\mu^+$ beam phase space by a factor $10^{10}$ at the expense of $10^{-3}$ efficiency due to muon decay~\cite{Bao:2014,Antognini:2020}. 

\subsubsection{Muonium Antimatter Gravity Experiment (MAGE)}

The gravitational force between antimatter and matter has yet to be directly measured. So far only a crude experimental limit has been established: $-65 \leq \overline{g}/g \leq 110$~\cite{Amole:2013}, where $\overline{g}$ is the gravitational acceleration of antimatter at the earth's surface (indirect limits on $\overline{g}$, such as those derived from torsion-pendulum experiments and lunar radar ranging~\cite{Alves:2009}, are based on the assumed gravitational effects of virtual antiparticles in nuclear binding energy and are not applicable to antimuons). Any deviation of $\overline{g}$ from $g$ would be evidence for new physics\,---\,possibly a ``fifth force," 
as has been invoked to explain multiple muonic experimental anomalies~\cite{Aaij:2022}. From another perspective, antimatter gravity can be viewed as a test of the weak equivalence principle, which some authors suggest may be violated in the case of antimatter~\cite{Nieto:1991,BenoitLevy:2012}. Muonium gravity is thus worth measuring as sensitively as possible.

Measurement of muonium gravity has not heretofore been feasible. The predicted parallel beam produced by superfluid helium is a game changer that will enable it to be measured for the first time. This has led to the approval of the LEMING experiment at PSI, based on the muCool apparatus and use of a micron-thick thin film of SFHe~\cite{LEMING}; however, LEMING is still in an R\&D stage, and a parallel M beam has yet to be demonstrated. Furthermore, the inefficiency of muCool is likely to limit LEMING in its currently designed form to measuring the sign of $\overline{g}$ and determining its magnitude only crudely. The thick-film approach described in the previous subsection, combined with the proton-beam power available from PIP-II and the use of heavy targets, should allow MAGE at Fermilab to measure $\overline{g}$ to 1\% of $g$ or better. Depending on the progress of our R\&D, a $\overline{g}$ measurement at the MTA may also be competitive with LEMING. 

MAGE and LEMING are based on the same idea~\cite{Kirch:2007wa}: 
create a horizontal, parallel, monoenergetic beam of muonium propagating in vacuum and impinging on a three-grating Mach--Zehnder-type interferometer, detecting the products of muonium decay in time--space coincidence downstream of the third grating. The fast positron from $\mu^+$ decay is easily reconstructed using scintillating fibers. The slow atomic electron  left behind when the muon decays can be accelerated electrostatically for detection by a 2D scintillating hodoscope array. The interferometer measures the acceleration of the beam in the direction perpendicular to the grating slits, which (to measure gravitational deflection) should therefore be oriented horizontally\,---\,in contrast to the more usual atom-interferometry applications, in which gravity is an unwanted influence and vertical slits are employed. The de Broglie waves corresponding to the atom beam diffract in each grating, and interference of the diffracted waves creates a sinusoidal intensity distribution at the third grating. The details of the pattern (its phase and period) exceed the position-reconstruction capability of particle detectors but can be measured by translating the third grating up and down periodically over time. The phase is proportional to the gravitational deflection, and its sign gives the direction (deflection up or down).

The beam is created by stopping surface or ``subsurface'' $\mu^+$ in superfluid helium,  deflecting the resulting vertically propagating M beam into the horizontal by means of a SFHe-coated surface angled at 45$^\circ$. (Subsurface muons are surface muons created slightly deeper into the target that lose a fraction of their energy to $dE/dx$ on the way out of the target.) Given the predicted 6.3\,mm/$\mu$s M velocity discussed above, the 2.2\,$\mu$s muon lifetime corresponds to 1.4\,cm of travel, and with the statistically optimal 2-lifetime grating separation (for measurement of an acceleration $\propto t^2$), the apparatus easily fits on a tabletop, or within the sample volume of a dilution refrigerator. Since the measurement resolution is inversely proportional to the grating pitch, the interferometer should have as fine a grating pitch as possible, but the gratings must also be fairly large and as open as possible (i.e., with minimal supporting struts, which potentially could block the passage of a significant fraction of M atoms). The current state of the art is judged to be $\approx$\,1\,cm$^2$ grating area and 100\,nm grating pitch. With $\approx$\,50\%-open (approximately optimal~\cite{Antognini:2020}) gratings, and an approximately parallel beam, the geometric acceptance is close to 100\%, but half of the beam is absorbed at each grating, in addition to decay losses. 

The measurement resolution depends on the pitch $d$, inter-grating travel time $t$, and interferometric contrast $C$ according to~\cite{Oberthaler:2002}
\begin{equation}
    \delta g = \frac{1}{C\sqrt{N}}\frac{d}{2\pi}\frac{1}{t^2}\,,
\end{equation}
where $N$ is the number of detected events. With the (conservative) fringe contrast $C$ = 0.1, and 100,000 M/s detected, the resolution is then $\delta g/g \approx (0.35/\sqrt{\rm \#~days})$ for gratings placed 1 muon lifetime apart, or $\approx (0.1/\sqrt{\rm \#~days})$ with 2-lifetime grating separation. Thus the sign of $\overline{g}$ is determined to 5 standard deviations in less than a day of running, and the magnitude determined to 1\% (if systematic effects can be sufficiently well controlled) in about a year. To these estimates some margin must of course be added for calibration runs and accelerator downtime. Thus precision measurements of $\overline{g}$ (to 1\% or better) will require the highest possible M intensity.

\subsubsection{R\&D Projects / Potential Synergies / Open Questions}
Here is a list of R\&D projects with some comments about potential synergies and some open questions.

\vspace{0.5cm}
\textbf{High priority}
\begin{itemize}
    \item{Pursue SFHe production of muonium}
    \begin{itemize}
        \item{this 
        could greatly improve the efficiency of muonium production for all muonium experiments, and is required for  muonium antimatter-gravity experiments}
    \end{itemize}

    \item{Pursue low-energy muon production at MTA}
    \begin{itemize}
        \item{could be used for R\&D of other experiments (e.g., measure muon production at energies similar to those expected from PIP-II and study pion and muon production-target materials and configurations)}
    \end{itemize}
    
    \item{Make first measurement of $\bar{g}$ with muonium}
    \begin{itemize}
        \item{this would help establish what the limiting systematics of such an experiment are}
    \end{itemize}
 
    \item{How does AMF host all of $\mu^{-}$, $\mu^{+}$, and muonium experiments?}
    \begin{itemize}
        \item{low-energy surface $\mu^{+}$ are needed for both decay and muonium experiments -- can both be supported at the same time?}
    \end{itemize}
\end{itemize}

\vspace{0.5cm}
\textbf{Low priority}

\begin{itemize}
    \item{Determine what improved muon production from SFHe means for all muonium experiments}
    \begin{itemize}
        \item{can we keep the same design for muonium oscillation experiments?}
        \item{can muonium spectroscopy be done in a vacuum?}
    \end{itemize}
    
    \item{Determine feasibility of other muonium experiments (e.g., M\,$\rightarrow \nu \bar{\nu}$)}
\end{itemize}

\newpage
\section{Conclusion}
Charged lepton flavor violating processes open a unique window on the physics of generation and flavor, complementing studies pursued at collider and neutrino experiments. Muons offer a promising avenue to search for CLFV, and a global experimental program is underway to explore these possibilities, promising impressive sensitivity gains in the coming decade. 

A staged program of next-generation experiments and facilities exploiting the full potential of the PIP-II accelerator has been outlined in this document. This program comprises Mu2e-II as a near term evolution of the Mu2e experiment, with the goal of improving the sensitivity to muon-to-electron conversion by an order of magnitude, followed by the Advanced Muon Facility. This new facility would provide the world's most intense positive and negative muon beams, enabling a suite of experiments with unprecedented sensitivity (probing mass scales of the order of $10^4 - 10^5 \TeV$), and offering {\it unique capabilities} to explore in depth the underlying New Physics in case of an observation. This program includes many synergies with the development of a muon collider and a beam dump dark matter experiment at FNAL. 

These experiments and facilities would provide the foundation for a comprehensive muon physics program in the next decade and beyond, an essential component of a global program to search for New Physics. The R\&D roadmap outlined in this document should begin immediately to ensure a timely realization of these possibilities.

\begin{acknowledgments}
This work was supported by the U.S.\ Department of Energy, Office of Science, Office of High Energy Physics, under Award Number DE-SC0019095, DE-SC0007884, DE-SC0011632 and DE-AC02-06CH11357 and by the Walter Burke Institute for Theoretical Physics. This work was partly authored by Fermi Research Alliance, LLC under Contract No. DE-AC02-07CH11359 with the U.S. Department of Energy, Office of Science, Office of High Energy Physics. This work was supported by the EU Horizon 2020 Research and Innovation Program under the Marie Sklodowska-Curie Grant Agreement No. 101006726. J.~Zettlemoyer acknowledges support in part by the DOE grant DE-SC0011784 and the NSF grant OAC-2103889. R.~Plestid is supported by the Neutrino Theory Network Program Grant Award under Number DEAC02-07CHI11359 and the US DOE under Award Number DE-SC0020250. J.~Tang acknowledges support in part by National Natural Science Foundation of China under Grant No. 12075326 and Fundamental Research Funds for the Central Universities (23xkjc017) in Sun Yat-sen University.
\vspace{0.5cm}
\end{acknowledgments}

\bibliography{Workshop_summary}

\begin{thebibliography}{142}%
\makeatletter
\providecommand \@ifxundefined [1]{%
 \@ifx{#1\undefined}
}%
\providecommand \@ifnum [1]{%
 \ifnum #1\expandafter \@firstoftwo
 \else \expandafter \@secondoftwo
 \fi
}%
\providecommand \@ifx [1]{%
 \ifx #1\expandafter \@firstoftwo
 \else \expandafter \@secondoftwo
 \fi
}%
\providecommand \natexlab [1]{#1}%
\providecommand \enquote  [1]{``#1''}%
\providecommand \bibnamefont  [1]{#1}%
\providecommand \bibfnamefont [1]{#1}%
\providecommand \citenamefont [1]{#1}%
\providecommand \href@noop [0]{\@secondoftwo}%
\providecommand \href [0]{\begingroup \@sanitize@url \@href}%
\providecommand \@href[1]{\@@startlink{#1}\@@href}%
\providecommand \@@href[1]{\endgroup#1\@@endlink}%
\providecommand \@sanitize@url [0]{\catcode `\\12\catcode `\$12\catcode
  `\&12\catcode `\#12\catcode `\^12\catcode `\_12\catcode `\%12\relax}%
\providecommand \@@startlink[1]{}%
\providecommand \@@endlink[0]{}%
\providecommand \url  [0]{\begingroup\@sanitize@url \@url }%
\providecommand \@url [1]{\endgroup\@href {#1}{\urlprefix }}%
\providecommand \urlprefix  [0]{URL }%
\providecommand \Eprint [0]{\href }%
\providecommand \doibase [0]{http://dx.doi.org/}%
\providecommand \selectlanguage [0]{\@gobble}%
\providecommand \bibinfo  [0]{\@secondoftwo}%
\providecommand \bibfield  [0]{\@secondoftwo}%
\providecommand \translation [1]{[#1]}%
\providecommand \BibitemOpen [0]{}%
\providecommand \bibitemStop [0]{}%
\providecommand \bibitemNoStop [0]{.\EOS\space}%
\providecommand \EOS [0]{\spacefactor3000\relax}%
\providecommand \BibitemShut  [1]{\csname bibitem#1\endcsname}%
\let\auto@bib@innerbib\@empty
\bibitem [{\citenamefont {Artuso}\ \emph {et~al.}(2022)\citenamefont {Artuso}
  \emph {et~al.}}]{Artuso:2022ouk}%
  \BibitemOpen
  \bibfield  {author} {\bibinfo {author} {\bibfnamefont {M.}~\bibnamefont
  {Artuso}} \emph {et~al.},\ }\href@noop {} {\  (\bibinfo {year} {2022})},\
  \Eprint {http://arxiv.org/abs/2210.04765} {arXiv:2210.04765 [hep-ex]}
  \BibitemShut {NoStop}%
\bibitem [{\citenamefont {Petcov}(1977)}]{Petcov:1976ff}%
  \BibitemOpen
  \bibfield  {author} {\bibinfo {author} {\bibfnamefont {S.~T.}\ \bibnamefont
  {Petcov}},\ }\href@noop {} {\bibfield  {journal} {\bibinfo  {journal} {Sov.
  J. Nucl. Phys.}\ }\textbf {\bibinfo {volume} {25}},\ \bibinfo {pages} {340}
  (\bibinfo {year} {1977})},\ \bibinfo {note} {[Erratum: Sov.J.Nucl.Phys. 25,
  698 (1977), Erratum: Yad.Fiz. 25, 1336 (1977)]}\BibitemShut {NoStop}%
\bibitem [{\citenamefont {Calibbi}\ and\ \citenamefont
  {Signorelli}(2018)}]{Calibbi:2017uvl}%
  \BibitemOpen
  \bibfield  {author} {\bibinfo {author} {\bibfnamefont {L.}~\bibnamefont
  {Calibbi}}\ and\ \bibinfo {author} {\bibfnamefont {G.}~\bibnamefont
  {Signorelli}},\ }\href {\doibase 10.1393/ncr/i2018-10144-0} {\bibfield
  {journal} {\bibinfo  {journal} {Riv. Nuovo Cim.}\ }\textbf {\bibinfo {volume}
  {41}},\ \bibinfo {pages} {71} (\bibinfo {year} {2018})},\ \Eprint
  {http://arxiv.org/abs/1709.00294} {arXiv:1709.00294 [hep-ph]} \BibitemShut
  {NoStop}%
\bibitem [{\citenamefont {Cei}\ and\ \citenamefont
  {Nicolo}(2014)}]{Cei:2014jtm}%
  \BibitemOpen
  \bibfield  {author} {\bibinfo {author} {\bibfnamefont {F.}~\bibnamefont
  {Cei}}\ and\ \bibinfo {author} {\bibfnamefont {D.}~\bibnamefont {Nicolo}},\
  }\href {\doibase 10.1155/2014/282915} {\bibfield  {journal} {\bibinfo
  {journal} {Adv. High Energy Phys.}\ }\textbf {\bibinfo {volume} {2014}},\
  \bibinfo {pages} {282915} (\bibinfo {year} {2014})}\BibitemShut {NoStop}%
\bibitem [{\citenamefont {Cirigliano}\ \emph {et~al.}(2017)\citenamefont
  {Cirigliano}, \citenamefont {Davidson},\ and\ \citenamefont
  {Kuno}}]{Cirigliano:2017azj}%
  \BibitemOpen
  \bibfield  {author} {\bibinfo {author} {\bibfnamefont {V.}~\bibnamefont
  {Cirigliano}}, \bibinfo {author} {\bibfnamefont {S.}~\bibnamefont
  {Davidson}}, \ and\ \bibinfo {author} {\bibfnamefont {Y.}~\bibnamefont
  {Kuno}},\ }\href {\doibase 10.1016/j.physletb.2017.05.053} {\bibfield
  {journal} {\bibinfo  {journal} {Phys. Lett. B}\ }\textbf {\bibinfo {volume}
  {771}},\ \bibinfo {pages} {242} (\bibinfo {year} {2017})},\ \Eprint
  {http://arxiv.org/abs/1703.02057} {arXiv:1703.02057 [hep-ph]} \BibitemShut
  {NoStop}%
\bibitem [{\citenamefont {Davidson}\ \emph {et~al.}(2018)\citenamefont
  {Davidson}, \citenamefont {Kuno},\ and\ \citenamefont
  {Saporta}}]{Davidson:2017nrp}%
  \BibitemOpen
  \bibfield  {author} {\bibinfo {author} {\bibfnamefont {S.}~\bibnamefont
  {Davidson}}, \bibinfo {author} {\bibfnamefont {Y.}~\bibnamefont {Kuno}}, \
  and\ \bibinfo {author} {\bibfnamefont {A.}~\bibnamefont {Saporta}},\ }\href
  {\doibase 10.1140/epjc/s10052-018-5584-8} {\bibfield  {journal} {\bibinfo
  {journal} {Eur. Phys. J. C}\ }\textbf {\bibinfo {volume} {78}},\ \bibinfo
  {pages} {109} (\bibinfo {year} {2018})},\ \Eprint
  {http://arxiv.org/abs/1710.06787} {arXiv:1710.06787 [hep-ph]} \BibitemShut
  {NoStop}%
\bibitem [{\citenamefont {Hoferichter}\ \emph {et~al.}(2023)\citenamefont
  {Hoferichter}, \citenamefont {Men\'endez},\ and\ \citenamefont
  {No\"el}}]{Hoferichter:2022mna}%
  \BibitemOpen
  \bibfield  {author} {\bibinfo {author} {\bibfnamefont {M.}~\bibnamefont
  {Hoferichter}}, \bibinfo {author} {\bibfnamefont {J.}~\bibnamefont
  {Men\'endez}}, \ and\ \bibinfo {author} {\bibfnamefont {F.}~\bibnamefont
  {No\"el}},\ }\href {\doibase 10.1103/PhysRevLett.130.131902} {\bibfield
  {journal} {\bibinfo  {journal} {Phys. Rev. Lett.}\ }\textbf {\bibinfo
  {volume} {130}},\ \bibinfo {pages} {131902} (\bibinfo {year} {2023})},\
  \Eprint {http://arxiv.org/abs/2204.06005} {arXiv:2204.06005 [hep-ph]}
  \BibitemShut {NoStop}%
\bibitem [{\citenamefont {Davidson}(2021)}]{Davidson:2020hkf}%
  \BibitemOpen
  \bibfield  {author} {\bibinfo {author} {\bibfnamefont {S.}~\bibnamefont
  {Davidson}},\ }\href {\doibase 10.1007/JHEP02(2021)172} {\bibfield  {journal}
  {\bibinfo  {journal} {JHEP}\ }\textbf {\bibinfo {volume} {02}},\ \bibinfo
  {pages} {172} (\bibinfo {year} {2021})},\ \Eprint
  {http://arxiv.org/abs/2010.00317} {arXiv:2010.00317 [hep-ph]} \BibitemShut
  {NoStop}%
\bibitem [{\citenamefont {Davidson}\ and\ \citenamefont
  {Echenard}(2022)}]{Davidson:2022nnl}%
  \BibitemOpen
  \bibfield  {author} {\bibinfo {author} {\bibfnamefont {S.}~\bibnamefont
  {Davidson}}\ and\ \bibinfo {author} {\bibfnamefont {B.}~\bibnamefont
  {Echenard}},\ }\href {\doibase 10.1140/epjc/s10052-022-10773-4} {\bibfield
  {journal} {\bibinfo  {journal} {Eur. Phys. J. C}\ }\textbf {\bibinfo {volume}
  {82}},\ \bibinfo {pages} {836} (\bibinfo {year} {2022})},\ \Eprint
  {http://arxiv.org/abs/2204.00564} {arXiv:2204.00564 [hep-ph]} \BibitemShut
  {NoStop}%
\bibitem [{\citenamefont {Okada}\ \emph {et~al.}(2000)\citenamefont {Okada},
  \citenamefont {Okumura},\ and\ \citenamefont {Shimizu}}]{Okada:1999zk}%
  \BibitemOpen
  \bibfield  {author} {\bibinfo {author} {\bibfnamefont {Y.}~\bibnamefont
  {Okada}}, \bibinfo {author} {\bibfnamefont {K.-i.}\ \bibnamefont {Okumura}},
  \ and\ \bibinfo {author} {\bibfnamefont {Y.}~\bibnamefont {Shimizu}},\ }\href
  {\doibase 10.1103/PhysRevD.61.094001} {\bibfield  {journal} {\bibinfo
  {journal} {Phys. Rev. D}\ }\textbf {\bibinfo {volume} {61}},\ \bibinfo
  {pages} {094001} (\bibinfo {year} {2000})},\ \Eprint
  {http://arxiv.org/abs/hep-ph/9906446} {arXiv:hep-ph/9906446} \BibitemShut
  {NoStop}%
\bibitem [{\citenamefont {Kitano}\ \emph
  {et~al.}(2002{\natexlab{a}})\citenamefont {Kitano}, \citenamefont {Koike},\
  and\ \citenamefont {Okada}}]{Kitano:2002mt}%
  \BibitemOpen
  \bibfield  {author} {\bibinfo {author} {\bibfnamefont {R.}~\bibnamefont
  {Kitano}}, \bibinfo {author} {\bibfnamefont {M.}~\bibnamefont {Koike}}, \
  and\ \bibinfo {author} {\bibfnamefont {Y.}~\bibnamefont {Okada}},\ }\href
  {\doibase 10.1103/PhysRevD.76.059902} {\bibfield  {journal} {\bibinfo
  {journal} {Phys. Rev. D}\ }\textbf {\bibinfo {volume} {66}},\ \bibinfo
  {pages} {096002} (\bibinfo {year} {2002}{\natexlab{a}})},\ \bibinfo {note}
  {[Erratum: Phys.Rev.D 76, 059902 (2007)]},\ \Eprint
  {http://arxiv.org/abs/hep-ph/0203110} {arXiv:hep-ph/0203110} \BibitemShut
  {NoStop}%
\bibitem [{\citenamefont {Heeck}\ \emph {et~al.}(2022)\citenamefont {Heeck},
  \citenamefont {Szafron},\ and\ \citenamefont {Uesaka}}]{Heeck:2022wer}%
  \BibitemOpen
  \bibfield  {author} {\bibinfo {author} {\bibfnamefont {J.}~\bibnamefont
  {Heeck}}, \bibinfo {author} {\bibfnamefont {R.}~\bibnamefont {Szafron}}, \
  and\ \bibinfo {author} {\bibfnamefont {Y.}~\bibnamefont {Uesaka}},\ }\href
  {\doibase 10.1016/j.nuclphysb.2022.115833} {\bibfield  {journal} {\bibinfo
  {journal} {Nucl. Phys. B}\ }\textbf {\bibinfo {volume} {980}},\ \bibinfo
  {pages} {115833} (\bibinfo {year} {2022})},\ \Eprint
  {http://arxiv.org/abs/2203.00702} {arXiv:2203.00702 [hep-ph]} \BibitemShut
  {NoStop}%
\bibitem [{\citenamefont {Davidson}\ \emph {et~al.}(2019)\citenamefont
  {Davidson}, \citenamefont {Kuno},\ and\ \citenamefont
  {Yamanaka}}]{Davidson:2018kud}%
  \BibitemOpen
  \bibfield  {author} {\bibinfo {author} {\bibfnamefont {S.}~\bibnamefont
  {Davidson}}, \bibinfo {author} {\bibfnamefont {Y.}~\bibnamefont {Kuno}}, \
  and\ \bibinfo {author} {\bibfnamefont {M.}~\bibnamefont {Yamanaka}},\ }\href
  {\doibase 10.1016/j.physletb.2019.01.042} {\bibfield  {journal} {\bibinfo
  {journal} {Phys. Lett. B}\ }\textbf {\bibinfo {volume} {790}},\ \bibinfo
  {pages} {380} (\bibinfo {year} {2019})},\ \Eprint
  {http://arxiv.org/abs/1810.01884} {arXiv:1810.01884 [hep-ph]} \BibitemShut
  {NoStop}%
\bibitem [{\citenamefont {Byrum}\ \emph {et~al.}(2022)\citenamefont {Byrum}
  \emph {et~al.}}]{Mu2e-II:2022blh}%
  \BibitemOpen
  \bibfield  {author} {\bibinfo {author} {\bibfnamefont {K.}~\bibnamefont
  {Byrum}} \emph {et~al.} (\bibinfo {collaboration} {Mu2e-II}),\ }in\
  \href@noop {} {\emph {\bibinfo {booktitle} {{Snowmass 2021}}}}\ (\bibinfo
  {year} {2022})\ \Eprint {http://arxiv.org/abs/2203.07569} {arXiv:2203.07569
  [hep-ex]} \BibitemShut {NoStop}%
\bibitem [{\citenamefont {Cirigliano}\ \emph {et~al.}(2009)\citenamefont
  {Cirigliano}, \citenamefont {Kitano}, \citenamefont {Okada},\ and\
  \citenamefont {Tuzon}}]{Cirigliano:2009bz}%
  \BibitemOpen
  \bibfield  {author} {\bibinfo {author} {\bibfnamefont {V.}~\bibnamefont
  {Cirigliano}}, \bibinfo {author} {\bibfnamefont {R.}~\bibnamefont {Kitano}},
  \bibinfo {author} {\bibfnamefont {Y.}~\bibnamefont {Okada}}, \ and\ \bibinfo
  {author} {\bibfnamefont {P.}~\bibnamefont {Tuzon}},\ }\href {\doibase
  10.1103/PhysRevD.80.013002} {\bibfield  {journal} {\bibinfo  {journal} {Phys.
  Rev. D}\ }\textbf {\bibinfo {volume} {80}},\ \bibinfo {pages} {013002}
  (\bibinfo {year} {2009})},\ \Eprint {http://arxiv.org/abs/0904.0957}
  {arXiv:0904.0957 [hep-ph]} \BibitemShut {NoStop}%
\bibitem [{\citenamefont {Minkowski}(1977)}]{Minkowski:1977sc}%
  \BibitemOpen
  \bibfield  {author} {\bibinfo {author} {\bibfnamefont {P.}~\bibnamefont
  {Minkowski}},\ }\href {\doibase 10.1016/0370-2693(77)90435-X} {\bibfield
  {journal} {\bibinfo  {journal} {Phys. Lett. B}\ }\textbf {\bibinfo {volume}
  {67}},\ \bibinfo {pages} {421} (\bibinfo {year} {1977})}\BibitemShut
  {NoStop}%
\bibitem [{\citenamefont {Yanagida}(1979)}]{Yanagida:1979as}%
  \BibitemOpen
  \bibfield  {author} {\bibinfo {author} {\bibfnamefont {T.}~\bibnamefont
  {Yanagida}},\ }\href@noop {} {\bibfield  {journal} {\bibinfo  {journal}
  {Conf. Proc. C}\ }\textbf {\bibinfo {volume} {7902131}},\ \bibinfo {pages}
  {95} (\bibinfo {year} {1979})}\BibitemShut {NoStop}%
\bibitem [{\citenamefont {Gell-Mann}\ \emph {et~al.}(1979)\citenamefont
  {Gell-Mann}, \citenamefont {Ramond},\ and\ \citenamefont
  {Slansky}}]{Gell-Mann:1979vob}%
  \BibitemOpen
  \bibfield  {author} {\bibinfo {author} {\bibfnamefont {M.}~\bibnamefont
  {Gell-Mann}}, \bibinfo {author} {\bibfnamefont {P.}~\bibnamefont {Ramond}}, \
  and\ \bibinfo {author} {\bibfnamefont {R.}~\bibnamefont {Slansky}},\
  }\href@noop {} {\bibfield  {journal} {\bibinfo  {journal} {Conf. Proc. C}\
  }\textbf {\bibinfo {volume} {790927}},\ \bibinfo {pages} {315} (\bibinfo
  {year} {1979})},\ \Eprint {http://arxiv.org/abs/1306.4669} {arXiv:1306.4669
  [hep-th]} \BibitemShut {NoStop}%
\bibitem [{\citenamefont {Mohapatra}\ and\ \citenamefont
  {Senjanovic}(1980)}]{Mohapatra:1979ia}%
  \BibitemOpen
  \bibfield  {author} {\bibinfo {author} {\bibfnamefont {R.~N.}\ \bibnamefont
  {Mohapatra}}\ and\ \bibinfo {author} {\bibfnamefont {G.}~\bibnamefont
  {Senjanovic}},\ }\href {\doibase 10.1103/PhysRevLett.44.912} {\bibfield
  {journal} {\bibinfo  {journal} {Phys. Rev. Lett.}\ }\textbf {\bibinfo
  {volume} {44}},\ \bibinfo {pages} {912} (\bibinfo {year} {1980})}\BibitemShut
  {NoStop}%
\bibitem [{\citenamefont {Antusch}\ \emph {et~al.}(2006)\citenamefont
  {Antusch}, \citenamefont {Biggio}, \citenamefont {Fernandez-Martinez},
  \citenamefont {Gavela},\ and\ \citenamefont {Lopez-Pavon}}]{Antusch:2006vwa}%
  \BibitemOpen
  \bibfield  {author} {\bibinfo {author} {\bibfnamefont {S.}~\bibnamefont
  {Antusch}}, \bibinfo {author} {\bibfnamefont {C.}~\bibnamefont {Biggio}},
  \bibinfo {author} {\bibfnamefont {E.}~\bibnamefont {Fernandez-Martinez}},
  \bibinfo {author} {\bibfnamefont {M.~B.}\ \bibnamefont {Gavela}}, \ and\
  \bibinfo {author} {\bibfnamefont {J.}~\bibnamefont {Lopez-Pavon}},\ }\href
  {\doibase 10.1088/1126-6708/2006/10/084} {\bibfield  {journal} {\bibinfo
  {journal} {JHEP}\ }\textbf {\bibinfo {volume} {10}},\ \bibinfo {pages} {084}
  (\bibinfo {year} {2006})},\ \Eprint {http://arxiv.org/abs/hep-ph/0607020}
  {arXiv:hep-ph/0607020} \BibitemShut {NoStop}%
\bibitem [{\citenamefont {Abada}\ \emph {et~al.}(2007)\citenamefont {Abada},
  \citenamefont {Biggio}, \citenamefont {Bonnet}, \citenamefont {Gavela},\ and\
  \citenamefont {Hambye}}]{Abada:2007ux}%
  \BibitemOpen
  \bibfield  {author} {\bibinfo {author} {\bibfnamefont {A.}~\bibnamefont
  {Abada}}, \bibinfo {author} {\bibfnamefont {C.}~\bibnamefont {Biggio}},
  \bibinfo {author} {\bibfnamefont {F.}~\bibnamefont {Bonnet}}, \bibinfo
  {author} {\bibfnamefont {M.~B.}\ \bibnamefont {Gavela}}, \ and\ \bibinfo
  {author} {\bibfnamefont {T.}~\bibnamefont {Hambye}},\ }\href {\doibase
  10.1088/1126-6708/2007/12/061} {\bibfield  {journal} {\bibinfo  {journal}
  {JHEP}\ }\textbf {\bibinfo {volume} {12}},\ \bibinfo {pages} {061} (\bibinfo
  {year} {2007})},\ \Eprint {http://arxiv.org/abs/0707.4058} {arXiv:0707.4058
  [hep-ph]} \BibitemShut {NoStop}%
\bibitem [{\citenamefont {Coy}\ and\ \citenamefont
  {Frigerio}(2019)}]{Coy:2018bxr}%
  \BibitemOpen
  \bibfield  {author} {\bibinfo {author} {\bibfnamefont {R.}~\bibnamefont
  {Coy}}\ and\ \bibinfo {author} {\bibfnamefont {M.}~\bibnamefont {Frigerio}},\
  }\href {\doibase 10.1103/PhysRevD.99.095040} {\bibfield  {journal} {\bibinfo
  {journal} {Phys. Rev. D}\ }\textbf {\bibinfo {volume} {99}},\ \bibinfo
  {pages} {095040} (\bibinfo {year} {2019})},\ \Eprint
  {http://arxiv.org/abs/1812.03165} {arXiv:1812.03165 [hep-ph]} \BibitemShut
  {NoStop}%
\bibitem [{\citenamefont {Broncano}\ \emph {et~al.}(2003)\citenamefont
  {Broncano}, \citenamefont {Gavela},\ and\ \citenamefont
  {Jenkins}}]{Broncano:2003fq}%
  \BibitemOpen
  \bibfield  {author} {\bibinfo {author} {\bibfnamefont {A.}~\bibnamefont
  {Broncano}}, \bibinfo {author} {\bibfnamefont {M.~B.}\ \bibnamefont
  {Gavela}}, \ and\ \bibinfo {author} {\bibfnamefont {E.~E.}\ \bibnamefont
  {Jenkins}},\ }\href {\doibase 10.1016/j.nuclphysb.2003.09.011} {\bibfield
  {journal} {\bibinfo  {journal} {Nucl. Phys. B}\ }\textbf {\bibinfo {volume}
  {672}},\ \bibinfo {pages} {163} (\bibinfo {year} {2003})},\ \Eprint
  {http://arxiv.org/abs/hep-ph/0307058} {arXiv:hep-ph/0307058} \BibitemShut
  {NoStop}%
\bibitem [{\citenamefont {Konetschny}\ and\ \citenamefont
  {Kummer}(1977)}]{Konetschny:1977bn}%
  \BibitemOpen
  \bibfield  {author} {\bibinfo {author} {\bibfnamefont {W.}~\bibnamefont
  {Konetschny}}\ and\ \bibinfo {author} {\bibfnamefont {W.}~\bibnamefont
  {Kummer}},\ }\href {\doibase 10.1016/0370-2693(77)90407-5} {\bibfield
  {journal} {\bibinfo  {journal} {Phys. Lett. B}\ }\textbf {\bibinfo {volume}
  {70}},\ \bibinfo {pages} {433} (\bibinfo {year} {1977})}\BibitemShut
  {NoStop}%
\bibitem [{\citenamefont {Magg}\ and\ \citenamefont
  {Wetterich}(1980)}]{Magg:1980ut}%
  \BibitemOpen
  \bibfield  {author} {\bibinfo {author} {\bibfnamefont {M.}~\bibnamefont
  {Magg}}\ and\ \bibinfo {author} {\bibfnamefont {C.}~\bibnamefont
  {Wetterich}},\ }\href {\doibase 10.1016/0370-2693(80)90825-4} {\bibfield
  {journal} {\bibinfo  {journal} {Phys. Lett. B}\ }\textbf {\bibinfo {volume}
  {94}},\ \bibinfo {pages} {61} (\bibinfo {year} {1980})}\BibitemShut {NoStop}%
\bibitem [{\citenamefont {Schechter}\ and\ \citenamefont
  {Valle}(1980)}]{Schechter:1980gr}%
  \BibitemOpen
  \bibfield  {author} {\bibinfo {author} {\bibfnamefont {J.}~\bibnamefont
  {Schechter}}\ and\ \bibinfo {author} {\bibfnamefont {J.~W.~F.}\ \bibnamefont
  {Valle}},\ }\href {\doibase 10.1103/PhysRevD.22.2227} {\bibfield  {journal}
  {\bibinfo  {journal} {Phys. Rev. D}\ }\textbf {\bibinfo {volume} {22}},\
  \bibinfo {pages} {2227} (\bibinfo {year} {1980})}\BibitemShut {NoStop}%
\bibitem [{\citenamefont {Cheng}\ and\ \citenamefont
  {Li}(1980)}]{Cheng:1980qt}%
  \BibitemOpen
  \bibfield  {author} {\bibinfo {author} {\bibfnamefont {T.~P.}\ \bibnamefont
  {Cheng}}\ and\ \bibinfo {author} {\bibfnamefont {L.-F.}\ \bibnamefont {Li}},\
  }\href {\doibase 10.1103/PhysRevD.22.2860} {\bibfield  {journal} {\bibinfo
  {journal} {Phys. Rev. D}\ }\textbf {\bibinfo {volume} {22}},\ \bibinfo
  {pages} {2860} (\bibinfo {year} {1980})}\BibitemShut {NoStop}%
\bibitem [{\citenamefont {Mohapatra}\ and\ \citenamefont
  {Senjanovic}(1981)}]{Mohapatra:1980yp}%
  \BibitemOpen
  \bibfield  {author} {\bibinfo {author} {\bibfnamefont {R.~N.}\ \bibnamefont
  {Mohapatra}}\ and\ \bibinfo {author} {\bibfnamefont {G.}~\bibnamefont
  {Senjanovic}},\ }\href {\doibase 10.1103/PhysRevD.23.165} {\bibfield
  {journal} {\bibinfo  {journal} {Phys. Rev.}\ }\textbf {\bibinfo {volume}
  {D23}},\ \bibinfo {pages} {165} (\bibinfo {year} {1981})}\BibitemShut
  {NoStop}%
\bibitem [{\citenamefont {Cai}\ \emph {et~al.}(2017)\citenamefont {Cai},
  \citenamefont {Herrero-Garc\'\i{}a}, \citenamefont {Schmidt}, \citenamefont
  {Vicente},\ and\ \citenamefont {Volkas}}]{Cai:2017jrq}%
  \BibitemOpen
  \bibfield  {author} {\bibinfo {author} {\bibfnamefont {Y.}~\bibnamefont
  {Cai}}, \bibinfo {author} {\bibfnamefont {J.}~\bibnamefont
  {Herrero-Garc\'\i{}a}}, \bibinfo {author} {\bibfnamefont {M.~A.}\
  \bibnamefont {Schmidt}}, \bibinfo {author} {\bibfnamefont {A.}~\bibnamefont
  {Vicente}}, \ and\ \bibinfo {author} {\bibfnamefont {R.~R.}\ \bibnamefont
  {Volkas}},\ }\href {\doibase 10.3389/fphy.2017.00063} {\bibfield  {journal}
  {\bibinfo  {journal} {Front. in Phys.}\ }\textbf {\bibinfo {volume} {5}},\
  \bibinfo {pages} {63} (\bibinfo {year} {2017})},\ \Eprint
  {http://arxiv.org/abs/1706.08524} {arXiv:1706.08524 [hep-ph]} \BibitemShut
  {NoStop}%
\bibitem [{\citenamefont {Calibbi}\ \emph
  {et~al.}(2021{\natexlab{a}})\citenamefont {Calibbi}, \citenamefont {Essig},
  \citenamefont {Papa}, \citenamefont {Redigolo}, \citenamefont {Zhong},
  \citenamefont {Ziegler},\ and\ \citenamefont {Zupan}}]{LoI080}%
  \BibitemOpen
  \bibfield  {author} {\bibinfo {author} {\bibfnamefont {L.}~\bibnamefont
  {Calibbi}}, \bibinfo {author} {\bibfnamefont {R.}~\bibnamefont {Essig}},
  \bibinfo {author} {\bibfnamefont {A.}~\bibnamefont {Papa}}, \bibinfo {author}
  {\bibfnamefont {D.}~\bibnamefont {Redigolo}}, \bibinfo {author}
  {\bibfnamefont {Y.-M.}\ \bibnamefont {Zhong}}, \bibinfo {author}
  {\bibfnamefont {R.}~\bibnamefont {Ziegler}}, \ and\ \bibinfo {author}
  {\bibfnamefont {J.}~\bibnamefont {Zupan}},\ }\href@noop {} {\enquote
  {\bibinfo {title} {{“Letter of Interest: rare muon decays and light new
  physics,” https://www.snowmass21.org/docs/files/
  summaries/RF/SNOWMASS21-RF5 RF0-080.pdf}},}\ } (\bibinfo {year}
  {2021}{\natexlab{a}})\BibitemShut {NoStop}%
\bibitem [{\citenamefont {Chikashige}\ \emph {et~al.}(1981)\citenamefont
  {Chikashige}, \citenamefont {Mohapatra},\ and\ \citenamefont
  {Peccei}}]{Chikashige:1980ui}%
  \BibitemOpen
  \bibfield  {author} {\bibinfo {author} {\bibfnamefont {Y.}~\bibnamefont
  {Chikashige}}, \bibinfo {author} {\bibfnamefont {R.~N.}\ \bibnamefont
  {Mohapatra}}, \ and\ \bibinfo {author} {\bibfnamefont {R.~D.}\ \bibnamefont
  {Peccei}},\ }\href {\doibase 10.1016/0370-2693(81)90011-3} {\bibfield
  {journal} {\bibinfo  {journal} {Phys. Lett. B}\ }\textbf {\bibinfo {volume}
  {98}},\ \bibinfo {pages} {265} (\bibinfo {year} {1981})}\BibitemShut
  {NoStop}%
\bibitem [{\citenamefont {Schechter}\ and\ \citenamefont
  {Valle}(1982)}]{Schechter:1981cv}%
  \BibitemOpen
  \bibfield  {author} {\bibinfo {author} {\bibfnamefont {J.}~\bibnamefont
  {Schechter}}\ and\ \bibinfo {author} {\bibfnamefont {J.~W.~F.}\ \bibnamefont
  {Valle}},\ }\href {\doibase 10.1103/PhysRevD.25.774} {\bibfield  {journal}
  {\bibinfo  {journal} {Phys. Rev. D}\ }\textbf {\bibinfo {volume} {25}},\
  \bibinfo {pages} {774} (\bibinfo {year} {1982})}\BibitemShut {NoStop}%
\bibitem [{\citenamefont {Kim}(1987)}]{Kim:1986ax}%
  \BibitemOpen
  \bibfield  {author} {\bibinfo {author} {\bibfnamefont {J.~E.}\ \bibnamefont
  {Kim}},\ }\href {\doibase 10.1016/0370-1573(87)90017-2} {\bibfield  {journal}
  {\bibinfo  {journal} {Phys. Rept.}\ }\textbf {\bibinfo {volume} {150}},\
  \bibinfo {pages} {1} (\bibinfo {year} {1987})}\BibitemShut {NoStop}%
\bibitem [{\citenamefont {Calibbi}\ \emph
  {et~al.}(2021{\natexlab{b}})\citenamefont {Calibbi}, \citenamefont
  {Redigolo}, \citenamefont {Ziegler},\ and\ \citenamefont
  {Zupan}}]{Calibbi:2020jvd}%
  \BibitemOpen
  \bibfield  {author} {\bibinfo {author} {\bibfnamefont {L.}~\bibnamefont
  {Calibbi}}, \bibinfo {author} {\bibfnamefont {D.}~\bibnamefont {Redigolo}},
  \bibinfo {author} {\bibfnamefont {R.}~\bibnamefont {Ziegler}}, \ and\
  \bibinfo {author} {\bibfnamefont {J.}~\bibnamefont {Zupan}},\ }\href
  {\doibase 10.1007/JHEP09(2021)173} {\bibfield  {journal} {\bibinfo  {journal}
  {JHEP}\ }\textbf {\bibinfo {volume} {09}},\ \bibinfo {pages} {173} (\bibinfo
  {year} {2021}{\natexlab{b}})},\ \Eprint {http://arxiv.org/abs/2006.04795}
  {arXiv:2006.04795} \BibitemShut {NoStop}%
\bibitem [{\citenamefont {Foot}\ \emph {et~al.}(1994)\citenamefont {Foot},
  \citenamefont {He}, \citenamefont {Lew},\ and\ \citenamefont
  {Volkas}}]{Foot:1994vd}%
  \BibitemOpen
  \bibfield  {author} {\bibinfo {author} {\bibfnamefont {R.}~\bibnamefont
  {Foot}}, \bibinfo {author} {\bibfnamefont {X.~G.}\ \bibnamefont {He}},
  \bibinfo {author} {\bibfnamefont {H.}~\bibnamefont {Lew}}, \ and\ \bibinfo
  {author} {\bibfnamefont {R.~R.}\ \bibnamefont {Volkas}},\ }\href {\doibase
  10.1103/PhysRevD.50.4571} {\bibfield  {journal} {\bibinfo  {journal} {Phys.
  Rev. D}\ }\textbf {\bibinfo {volume} {50}},\ \bibinfo {pages} {4571}
  (\bibinfo {year} {1994})},\ \Eprint {http://arxiv.org/abs/hep-ph/9401250}
  {arXiv:hep-ph/9401250} \BibitemShut {NoStop}%
\bibitem [{\citenamefont {Heeck}(2016)}]{Heeck:2016xkh}%
  \BibitemOpen
  \bibfield  {author} {\bibinfo {author} {\bibfnamefont {J.}~\bibnamefont
  {Heeck}},\ }\href {\doibase 10.1016/j.physletb.2016.05.007} {\bibfield
  {journal} {\bibinfo  {journal} {Phys. Lett. B}\ }\textbf {\bibinfo {volume}
  {758}},\ \bibinfo {pages} {101} (\bibinfo {year} {2016})},\ \Eprint
  {http://arxiv.org/abs/1602.03810} {arXiv:1602.03810 [hep-ph]} \BibitemShut
  {NoStop}%
\bibitem [{\citenamefont {Knecht}\ \emph {et~al.}(2020)\citenamefont {Knecht},
  \citenamefont {Skawran},\ and\ \citenamefont {Vogiatzi}}]{Knecht:2020}%
  \BibitemOpen
  \bibfield  {author} {\bibinfo {author} {\bibfnamefont {A.}~\bibnamefont
  {Knecht}}, \bibinfo {author} {\bibfnamefont {A.}~\bibnamefont {Skawran}}, \
  and\ \bibinfo {author} {\bibfnamefont {S.}~\bibnamefont {Vogiatzi}},\ }\href
  {\doibase 10.1140/epjp/s13360-020-00777-y} {\bibfield  {journal} {\bibinfo
  {journal} {Eur. Phys. J. Plus}\ }\textbf {\bibinfo {volume} {135}},\ \bibinfo
  {pages} {777} (\bibinfo {year} {2020})}\BibitemShut {NoStop}%
\bibitem [{\citenamefont {Y.J.Lee}\ \emph {et~al.}(2023)\citenamefont {Y.J.Lee}
  \emph {et~al.}}]{YJL1}%
  \BibitemOpen
  \bibfield  {author} {\bibinfo {author} {\bibnamefont {Y.J.Lee}} \emph
  {et~al.},\ }\href@noop {} {\enquote {\bibinfo {title} {{Spallation Materials
  Research for Second Target Station Project at ORNL}},}\ } (\bibinfo {year}
  {2023}),\ \bibinfo {note} {{The 15th International Workshop on Spallation
  Materials Technology}}\BibitemShut {NoStop}%
\bibitem [{\citenamefont {Y.J.Lee}(2023)}]{YJL2}%
  \BibitemOpen
  \bibfield  {author} {\bibinfo {author} {\bibnamefont {Y.J.Lee}},\ }\href@noop
  {} {\enquote {\bibinfo {title} {{Tungsten in accelerator environments}},}\ }
  (\bibinfo {year} {2023}),\ \bibinfo {note} {{Muon program at Fermilab
  workshop, Caltech, Pasadena, CA}}\BibitemShut {NoStop}%
\bibitem [{\citenamefont {Wilcox}(2023)}]{Dan}%
  \BibitemOpen
  \bibfield  {author} {\bibinfo {author} {\bibfnamefont {D.}~\bibnamefont
  {Wilcox}},\ }\href@noop {} {\enquote {\bibinfo {title} {{Fluidized Tungsten
  Powder as a Muon Production Target}},}\ } (\bibinfo {year} {2023}),\ \bibinfo
  {note} {{Muon program at Fermilab workshop, Caltech, Pasadena,
  CA}}\BibitemShut {NoStop}%
\bibitem [{\citenamefont {Carelli}(2023)}]{Carelli}%
  \BibitemOpen
  \bibfield  {author} {\bibinfo {author} {\bibfnamefont {C.}~\bibnamefont
  {Carelli}},\ }\href@noop {} {\enquote {\bibinfo {title} {{Liquid Heavy Metal
  applications for particle accelerators}},}\ } (\bibinfo {year} {2023}),\
  \bibinfo {note} {{Muon program at Fermilab workshop, Caltech, Pasadena,
  CA}}\BibitemShut {NoStop}%
\bibitem [{\citenamefont {Hedges}(2023)}]{Hedges}%
  \BibitemOpen
  \bibfield  {author} {\bibinfo {author} {\bibfnamefont {M.}~\bibnamefont
  {Hedges}},\ }\href@noop {} {\enquote {\bibinfo {title} {{The Mu2e target}},}\
  } (\bibinfo {year} {2023}),\ \bibinfo {note} {{Muon program at Fermilab
  workshop, Caltech, Pasadena, CA}}\BibitemShut {NoStop}%
\bibitem [{\citenamefont {Pellemoine}(2023)}]{Frederique}%
  \BibitemOpen
  \bibfield  {author} {\bibinfo {author} {\bibfnamefont {F.}~\bibnamefont
  {Pellemoine}},\ }\href@noop {} {\enquote {\bibinfo {title} {{HPT R\&D, FNAL
  plans and needs}},}\ } (\bibinfo {year} {2023}),\ \bibinfo {note} {{Muon
  program at Fermilab workshop, Caltech, Pasadena, CA}}\BibitemShut {NoStop}%
\bibitem [{\citenamefont {Agostinelli}\ \emph {et~al.}(2003)\citenamefont
  {Agostinelli} \emph {et~al.}}]{GEANT4_2003}%
  \BibitemOpen
  \bibfield  {author} {\bibinfo {author} {\bibfnamefont {S.}~\bibnamefont
  {Agostinelli}} \emph {et~al.},\ }\href {\doibase
  https://doi.org/10.1016/S0168-9002(03)01368-8} {\bibfield  {journal}
  {\bibinfo  {journal} {Nuclear Instruments and Methods in Physics Research
  Section A: Accelerators, Spectrometers, Detectors and Associated Equipment}\
  }\textbf {\bibinfo {volume} {506}},\ \bibinfo {pages} {250} (\bibinfo {year}
  {2003})}\BibitemShut {NoStop}%
\bibitem [{\citenamefont {Allison}\ \emph {et~al.}(2006)\citenamefont {Allison}
  \emph {et~al.}}]{GEANT4_2006}%
  \BibitemOpen
  \bibfield  {author} {\bibinfo {author} {\bibfnamefont {J.}~\bibnamefont
  {Allison}} \emph {et~al.},\ }\href {\doibase 10.1109/TNS.2006.869826}
  {\bibfield  {journal} {\bibinfo  {journal} {IEEE Transactions on Nuclear
  Science}\ }\textbf {\bibinfo {volume} {53}},\ \bibinfo {pages} {270}
  (\bibinfo {year} {2006})}\BibitemShut {NoStop}%
\bibitem [{\citenamefont {Allison}\ \emph {et~al.}(2016)\citenamefont {Allison}
  \emph {et~al.}}]{GEANT4_2016}%
  \BibitemOpen
  \bibfield  {author} {\bibinfo {author} {\bibfnamefont {J.}~\bibnamefont
  {Allison}} \emph {et~al.},\ }\href {\doibase
  https://doi.org/10.1016/j.nima.2016.06.125} {\bibfield  {journal} {\bibinfo
  {journal} {Nuclear Instruments and Methods in Physics Research Section A:
  Accelerators, Spectrometers, Detectors and Associated Equipment}\ }\textbf
  {\bibinfo {volume} {835}},\ \bibinfo {pages} {186} (\bibinfo {year}
  {2016})}\BibitemShut {NoStop}%
\bibitem [{\citenamefont {Volkov}\ \emph {et~al.}(2021)\citenamefont {Volkov},
  \citenamefont {Evtoukhovich}, \citenamefont {Kravchenko}, \citenamefont
  {Kuno}, \citenamefont {Mihara}, \citenamefont {Nishiguchi}, \citenamefont
  {Pavlov},\ and\ \citenamefont {Tsamalaidze}}]{VOLKOV2021165242}%
  \BibitemOpen
  \bibfield  {author} {\bibinfo {author} {\bibfnamefont {A.}~\bibnamefont
  {Volkov}}, \bibinfo {author} {\bibfnamefont {P.}~\bibnamefont
  {Evtoukhovich}}, \bibinfo {author} {\bibfnamefont {M.}~\bibnamefont
  {Kravchenko}}, \bibinfo {author} {\bibfnamefont {Y.}~\bibnamefont {Kuno}},
  \bibinfo {author} {\bibfnamefont {S.}~\bibnamefont {Mihara}}, \bibinfo
  {author} {\bibfnamefont {H.}~\bibnamefont {Nishiguchi}}, \bibinfo {author}
  {\bibfnamefont {A.}~\bibnamefont {Pavlov}}, \ and\ \bibinfo {author}
  {\bibfnamefont {Z.}~\bibnamefont {Tsamalaidze}},\ }\href {\doibase
  https://doi.org/10.1016/j.nima.2021.165242} {\bibfield  {journal} {\bibinfo
  {journal} {Nuclear Instruments and Methods in Physics Research Section A:
  Accelerators, Spectrometers, Detectors and Associated Equipment}\ }\textbf
  {\bibinfo {volume} {1004}},\ \bibinfo {pages} {165242} (\bibinfo {year}
  {2021})}\BibitemShut {NoStop}%
\bibitem [{\citenamefont {Meng}\ \emph {et~al.}(2022)\citenamefont {Meng},
  \citenamefont {Zhao},\ and\ \citenamefont {Wan}}]{MENG2022353}%
  \BibitemOpen
  \bibfield  {author} {\bibinfo {author} {\bibfnamefont {B.}~\bibnamefont
  {Meng}}, \bibinfo {author} {\bibfnamefont {R.}~\bibnamefont {Zhao}}, \ and\
  \bibinfo {author} {\bibfnamefont {M.}~\bibnamefont {Wan}},\ }in\ \href
  {\doibase https://doi.org/10.1016/B978-0-12-819726-4.00014-4} {\emph
  {\bibinfo {booktitle} {Encyclopedia of Materials: Metals and Alloys}}},\
  \bibinfo {editor} {edited by\ \bibinfo {editor} {\bibfnamefont {F.~G.}\
  \bibnamefont {Caballero}}}\ (\bibinfo  {publisher} {Elsevier},\ \bibinfo
  {address} {Oxford},\ \bibinfo {year} {2022})\ pp.\ \bibinfo {pages}
  {353--370}\BibitemShut {NoStop}%
\bibitem [{\citenamefont {Murtas}(2020)}]{MURTAS2020106421}%
  \BibitemOpen
  \bibfield  {author} {\bibinfo {author} {\bibfnamefont {F.}~\bibnamefont
  {Murtas}},\ }\href {\doibase https://doi.org/10.1016/j.radmeas.2020.106421}
  {\bibfield  {journal} {\bibinfo  {journal} {Radiation Measurements}\ }\textbf
  {\bibinfo {volume} {138}},\ \bibinfo {pages} {106421} (\bibinfo {year}
  {2020})}\BibitemShut {NoStop}%
\bibitem [{\citenamefont {Wu}\ \emph {et~al.}(2021)\citenamefont {Wu},
  \citenamefont {Wong}, \citenamefont {Kuno}, \citenamefont {Moritsu},
  \citenamefont {Nakazawa}, \citenamefont {Sato}, \citenamefont {Sakamoto},
  \citenamefont {Tran}, \citenamefont {Wong}, \citenamefont {Yoshida},
  \citenamefont {Yamane},\ and\ \citenamefont {Zhang}}]{WU2021165756}%
  \BibitemOpen
  \bibfield  {author} {\bibinfo {author} {\bibfnamefont {C.}~\bibnamefont
  {Wu}}, \bibinfo {author} {\bibfnamefont {T.}~\bibnamefont {Wong}}, \bibinfo
  {author} {\bibfnamefont {Y.}~\bibnamefont {Kuno}}, \bibinfo {author}
  {\bibfnamefont {M.}~\bibnamefont {Moritsu}}, \bibinfo {author} {\bibfnamefont
  {Y.}~\bibnamefont {Nakazawa}}, \bibinfo {author} {\bibfnamefont
  {A.}~\bibnamefont {Sato}}, \bibinfo {author} {\bibfnamefont {H.}~\bibnamefont
  {Sakamoto}}, \bibinfo {author} {\bibfnamefont {N.}~\bibnamefont {Tran}},
  \bibinfo {author} {\bibfnamefont {M.}~\bibnamefont {Wong}}, \bibinfo {author}
  {\bibfnamefont {H.}~\bibnamefont {Yoshida}}, \bibinfo {author} {\bibfnamefont
  {T.}~\bibnamefont {Yamane}}, \ and\ \bibinfo {author} {\bibfnamefont
  {J.}~\bibnamefont {Zhang}},\ }\href {\doibase
  https://doi.org/10.1016/j.nima.2021.165756} {\bibfield  {journal} {\bibinfo
  {journal} {Nuclear Instruments and Methods in Physics Research Section A:
  Accelerators, Spectrometers, Detectors and Associated Equipment}\ }\textbf
  {\bibinfo {volume} {1015}},\ \bibinfo {pages} {165756} (\bibinfo {year}
  {2021})}\BibitemShut {NoStop}%
\bibitem [{\citenamefont {Chiarello}\ \emph {et~al.}(2017)\citenamefont
  {Chiarello}, \citenamefont {Chiri}, \citenamefont {Cocciolo}, \citenamefont
  {Corvaglia}, \citenamefont {Grancagnolo}, \citenamefont {Panareo},
  \citenamefont {Pepino}, \citenamefont {Renga}, \citenamefont {Tassielli},\
  and\ \citenamefont {Voena}}]{Chiarello_2017}%
  \BibitemOpen
  \bibfield  {author} {\bibinfo {author} {\bibfnamefont {G.}~\bibnamefont
  {Chiarello}}, \bibinfo {author} {\bibfnamefont {C.}~\bibnamefont {Chiri}},
  \bibinfo {author} {\bibfnamefont {G.}~\bibnamefont {Cocciolo}}, \bibinfo
  {author} {\bibfnamefont {A.}~\bibnamefont {Corvaglia}}, \bibinfo {author}
  {\bibfnamefont {F.}~\bibnamefont {Grancagnolo}}, \bibinfo {author}
  {\bibfnamefont {M.}~\bibnamefont {Panareo}}, \bibinfo {author} {\bibfnamefont
  {A.}~\bibnamefont {Pepino}}, \bibinfo {author} {\bibfnamefont
  {F.}~\bibnamefont {Renga}}, \bibinfo {author} {\bibfnamefont
  {G.}~\bibnamefont {Tassielli}}, \ and\ \bibinfo {author} {\bibfnamefont
  {C.}~\bibnamefont {Voena}},\ }\href {\doibase 10.1088/1748-0221/12/07/C07021}
  {\bibfield  {journal} {\bibinfo  {journal} {Journal of Instrumentation}\
  }\textbf {\bibinfo {volume} {12}},\ \bibinfo {pages} {C07021} (\bibinfo
  {year} {2017})}\BibitemShut {NoStop}%
\bibitem [{\citenamefont {Grancagnolo}(2012)}]{i-tracker}%
  \BibitemOpen
  \bibfield  {author} {\bibinfo {author} {\bibfnamefont {F.}~\bibnamefont
  {Grancagnolo}},\ }\href@noop {} {\enquote {\bibinfo {title} {The i-tracker
  hardware},}\ } (\bibinfo {year} {2012}),\ \bibinfo {note} {mu2e Document
  2450}\BibitemShut {NoStop}%
\bibitem [{\citenamefont {Tassielli}(2020)}]{altTracker}%
  \BibitemOpen
  \bibfield  {author} {\bibinfo {author} {\bibfnamefont {G.}~\bibnamefont
  {Tassielli}},\ }\href@noop {} {\enquote {\bibinfo {title} {Conceptual idea
  for a potential mu2e-ii tracker option},}\ } (\bibinfo {year} {2020}),\
  \bibinfo {note} {mu2e Document 33929}\BibitemShut {NoStop}%
\bibitem [{\citenamefont {Atanov}\ \emph {et~al.}(2018)\citenamefont {Atanov}
  \emph {et~al.}}]{Calo:Atanov2018}%
  \BibitemOpen
  \bibfield  {author} {\bibinfo {author} {\bibfnamefont {N.}~\bibnamefont
  {Atanov}} \emph {et~al.},\ }\href@noop {} {\enquote {\bibinfo {title} {{The
  Mu2e Calorimeter Final Technical Design Report}},}\ } (\bibinfo {year}
  {2018}),\ \Eprint {http://arxiv.org/abs/1802.06341} {arXiv:1802.06341
  [physics.ins-det]} \BibitemShut {NoStop}%
\bibitem [{\citenamefont {Zhu}(2017)}]{Zhu2017}%
  \BibitemOpen
  \bibfield  {author} {\bibinfo {author} {\bibfnamefont {R.}~\bibnamefont
  {Zhu}},\ }\href {\doibase 10.1117/12.2274617} {\bibfield  {journal} {\bibinfo
   {journal} {Proc. SPIE Int. Soc. Opt. Eng.}\ }\textbf {\bibinfo {volume}
  {10392}},\ \bibinfo {pages} {103920G} (\bibinfo {year} {2017})}\BibitemShut
  {NoStop}%
\bibitem [{\citenamefont {Hitlin}\ and\ \citenamefont
  {Zhu}(2020)}]{Hitlin2020a}%
  \BibitemOpen
  \bibfield  {author} {\bibinfo {author} {\bibfnamefont {D.}~\bibnamefont
  {Hitlin}}\ and\ \bibinfo {author} {\bibfnamefont {R.}~\bibnamefont {Zhu}},\
  }\href@noop {} {\enquote {\bibinfo {title} {Crystal and photosensor
  development for a {BaF$_2$} electromagnetic calorimeter},}\ } (\bibinfo
  {year} {2020}),\ \bibinfo {note} {lOI for Snowmass 21}\BibitemShut {NoStop}%
\bibitem [{\citenamefont {Ceravolo}\ \emph {et~al.}(2022)\citenamefont
  {Ceravolo}, \citenamefont {Colao}, \citenamefont {Curatolo}, \citenamefont
  {Di~Meco}, \citenamefont {Diociaiuti}, \citenamefont {Lucchesi},
  \citenamefont {Paesani}, \citenamefont {Pastrone}, \citenamefont {Saputi},
  \citenamefont {Sarra} \emph {et~al.}}]{2022crilin}%
  \BibitemOpen
  \bibfield  {author} {\bibinfo {author} {\bibfnamefont {S.}~\bibnamefont
  {Ceravolo}}, \bibinfo {author} {\bibfnamefont {F.}~\bibnamefont {Colao}},
  \bibinfo {author} {\bibfnamefont {C.}~\bibnamefont {Curatolo}}, \bibinfo
  {author} {\bibfnamefont {E.}~\bibnamefont {Di~Meco}}, \bibinfo {author}
  {\bibfnamefont {E.}~\bibnamefont {Diociaiuti}}, \bibinfo {author}
  {\bibfnamefont {D.}~\bibnamefont {Lucchesi}}, \bibinfo {author}
  {\bibfnamefont {D.}~\bibnamefont {Paesani}}, \bibinfo {author} {\bibfnamefont
  {N.}~\bibnamefont {Pastrone}}, \bibinfo {author} {\bibfnamefont
  {A.}~\bibnamefont {Saputi}}, \bibinfo {author} {\bibfnamefont
  {I.}~\bibnamefont {Sarra}},  \emph {et~al.},\ }\href@noop {} {\bibfield
  {journal} {\bibinfo  {journal} {Journal of Instrumentation}\ }\textbf
  {\bibinfo {volume} {17}},\ \bibinfo {pages} {P09033} (\bibinfo {year}
  {2022})}\BibitemShut {NoStop}%
\bibitem [{\citenamefont {Artikov}\ \emph {et~al.}(2018)\citenamefont {Artikov}
  \emph {et~al.}}]{Mu2e:2017lae}%
  \BibitemOpen
  \bibfield  {author} {\bibinfo {author} {\bibfnamefont {A.}~\bibnamefont
  {Artikov}} \emph {et~al.} (\bibinfo {collaboration} {Mu2e}),\ }\href
  {\doibase 10.1016/j.nima.2018.02.023} {\bibfield  {journal} {\bibinfo
  {journal} {Nucl. Instrum. Meth. A}\ }\textbf {\bibinfo {volume} {890}},\
  \bibinfo {pages} {84} (\bibinfo {year} {2018})},\ \Eprint
  {http://arxiv.org/abs/1709.06587} {arXiv:1709.06587 [physics.ins-det]}
  \BibitemShut {NoStop}%
\bibitem [{\citenamefont {Dukes}\ \emph {et~al.}(2018)\citenamefont {Dukes},
  \citenamefont {Farris}, \citenamefont {Group}, \citenamefont {Lam},
  \citenamefont {Shooltz},\ and\ \citenamefont {Oksuzian}}]{Dukes:2018scs}%
  \BibitemOpen
  \bibfield  {author} {\bibinfo {author} {\bibfnamefont {E.~C.}\ \bibnamefont
  {Dukes}}, \bibinfo {author} {\bibfnamefont {P.~J.}\ \bibnamefont {Farris}},
  \bibinfo {author} {\bibfnamefont {R.~C.}\ \bibnamefont {Group}}, \bibinfo
  {author} {\bibfnamefont {T.}~\bibnamefont {Lam}}, \bibinfo {author}
  {\bibfnamefont {D.}~\bibnamefont {Shooltz}}, \ and\ \bibinfo {author}
  {\bibfnamefont {Y.}~\bibnamefont {Oksuzian}},\ }\href {\doibase
  10.1088/1748-0221/13/12/P12028} {\bibfield  {journal} {\bibinfo  {journal}
  {JINST}\ }\textbf {\bibinfo {volume} {13}},\ \bibinfo {pages} {P12028}
  (\bibinfo {year} {2018})},\ \Eprint {http://arxiv.org/abs/1811.04874}
  {arXiv:1811.04874 [physics.ins-det]} \BibitemShut {NoStop}%
\bibitem [{\citenamefont {Blazey}\ \emph {et~al.}(2019)\citenamefont {Blazey}
  \emph {et~al.}}]{Blazey:2019vdr}%
  \BibitemOpen
  \bibfield  {author} {\bibinfo {author} {\bibfnamefont {G.}~\bibnamefont
  {Blazey}} \emph {et~al.},\ }\href {\doibase 10.1016/j.nima.2018.12.050}
  {\bibfield  {journal} {\bibinfo  {journal} {Nucl. Instrum. Meth. A}\ }\textbf
  {\bibinfo {volume} {927}},\ \bibinfo {pages} {463} (\bibinfo {year}
  {2019})},\ \Eprint {http://arxiv.org/abs/1906.07237} {arXiv:1906.07237
  [physics.ins-det]} \BibitemShut {NoStop}%
\bibitem [{\citenamefont {Solt}\ \emph {et~al.}(2023)\citenamefont {Solt},
  \citenamefont {Coveyou}, \citenamefont {Dukes}, \citenamefont {Oksuzian},\
  and\ \citenamefont {Roberts}}]{Solt:2023uxp}%
  \BibitemOpen
  \bibfield  {author} {\bibinfo {author} {\bibfnamefont {M.}~\bibnamefont
  {Solt}}, \bibinfo {author} {\bibfnamefont {D.}~\bibnamefont {Coveyou}},
  \bibinfo {author} {\bibfnamefont {E.~C.}\ \bibnamefont {Dukes}}, \bibinfo
  {author} {\bibfnamefont {Y.}~\bibnamefont {Oksuzian}}, \ and\ \bibinfo
  {author} {\bibfnamefont {S.}~\bibnamefont {Roberts}} (\bibinfo
  {collaboration} {R. C. Group}),\ }\href@noop {} {\enquote {\bibinfo {title}
  {{Performance of the wavelength-shifting fiber upgrade for the Mu2e
  cosmic-ray veto detector}},}\ } (\bibinfo {year} {2023}),\ \Eprint
  {http://arxiv.org/abs/2302.09172} {arXiv:2302.09172 [hep-ex]} \BibitemShut
  {NoStop}%
\bibitem [{\citenamefont {Artikov}\ \emph {et~al.}(2019)\citenamefont {Artikov}
  \emph {et~al.}}]{Artikov:2017fmr}%
  \BibitemOpen
  \bibfield  {author} {\bibinfo {author} {\bibfnamefont {A.}~\bibnamefont
  {Artikov}} \emph {et~al.},\ }\href {\doibase 10.1016/j.nima.2019.03.087}
  {\bibfield  {journal} {\bibinfo  {journal} {Nucl. Instrum. Meth. A}\ }\textbf
  {\bibinfo {volume} {930}},\ \bibinfo {pages} {87} (\bibinfo {year} {2019})},\
  \Eprint {http://arxiv.org/abs/1711.11393} {arXiv:1711.11393
  [physics.ins-det]} \BibitemShut {NoStop}%
\bibitem [{\citenamefont {Kitano}\ \emph
  {et~al.}(2002{\natexlab{b}})\citenamefont {Kitano}, \citenamefont {Koike},\
  and\ \citenamefont {Okada}}]{Kitano2002}%
  \BibitemOpen
  \bibfield  {author} {\bibinfo {author} {\bibfnamefont {R.}~\bibnamefont
  {Kitano}}, \bibinfo {author} {\bibfnamefont {M.}~\bibnamefont {Koike}}, \
  and\ \bibinfo {author} {\bibfnamefont {Y.}~\bibnamefont {Okada}},\ }\href
  {\doibase 10.1103/PhysRevD.66.096002} {\bibfield  {journal} {\bibinfo
  {journal} {Phys. Rev. D}\ }\textbf {\bibinfo {volume} {66}},\ \bibinfo
  {pages} {096002} (\bibinfo {year} {2002}{\natexlab{b}})}\BibitemShut
  {NoStop}%
\bibitem [{\citenamefont {Barrett}(1970)}]{Barrett1970}%
  \BibitemOpen
  \bibfield  {author} {\bibinfo {author} {\bibfnamefont {R.}~\bibnamefont
  {Barrett}},\ }\href {\doibase https://doi.org/10.1016/0370-2693(70)90611-8}
  {\bibfield  {journal} {\bibinfo  {journal} {Phys. Lett. B}\ }\textbf
  {\bibinfo {volume} {33}},\ \bibinfo {pages} {388} (\bibinfo {year}
  {1970})}\BibitemShut {NoStop}%
\bibitem [{\citenamefont {Weinberg}\ and\ \citenamefont
  {Feinberg}(1959)}]{Weinberg1959}%
  \BibitemOpen
  \bibfield  {author} {\bibinfo {author} {\bibfnamefont {S.}~\bibnamefont
  {Weinberg}}\ and\ \bibinfo {author} {\bibfnamefont {G.}~\bibnamefont
  {Feinberg}},\ }\href {\doibase 10.1103/PhysRevLett.3.111} {\bibfield
  {journal} {\bibinfo  {journal} {Phys. Rev. Lett.}\ }\textbf {\bibinfo
  {volume} {3}},\ \bibinfo {pages} {111} (\bibinfo {year} {1959})}\BibitemShut
  {NoStop}%
\bibitem [{\citenamefont {Suzuki}\ \emph {et~al.}(1987)\citenamefont {Suzuki},
  \citenamefont {Measday},\ and\ \citenamefont {Roalsvig}}]{suzuki1987}%
  \BibitemOpen
  \bibfield  {author} {\bibinfo {author} {\bibfnamefont {T.}~\bibnamefont
  {Suzuki}}, \bibinfo {author} {\bibfnamefont {D.}~\bibnamefont {Measday}}, \
  and\ \bibinfo {author} {\bibfnamefont {J.}~\bibnamefont {Roalsvig}},\
  }\href@noop {} {\bibfield  {journal} {\bibinfo  {journal} {Phys. Rev.}\
  }\textbf {\bibinfo {volume} {C35}},\ \bibinfo {pages} {2212} (\bibinfo {year}
  {1987})}\BibitemShut {NoStop}%
\bibitem [{\citenamefont {Wheeler}(1949)}]{Wheeler1949}%
  \BibitemOpen
  \bibfield  {author} {\bibinfo {author} {\bibfnamefont {J.}~\bibnamefont
  {Wheeler}},\ }\href@noop {} {\bibfield  {journal} {\bibinfo  {journal} {Rev.
  Mod. Phys}\ }\textbf {\bibinfo {volume} {21}},\ \bibinfo {pages} {133}
  (\bibinfo {year} {1949})}\BibitemShut {NoStop}%
\bibitem [{\citenamefont {Borrel}\ \emph {et~al.}(2023)\citenamefont {Borrel},
  \citenamefont {Hitlin},\ and\ \citenamefont {Middleton}}]{Borrel2023}%
  \BibitemOpen
  \bibfield  {author} {\bibinfo {author} {\bibfnamefont {L.}~\bibnamefont
  {Borrel}}, \bibinfo {author} {\bibfnamefont {D.}~\bibnamefont {Hitlin}}, \
  and\ \bibinfo {author} {\bibfnamefont {S.}~\bibnamefont {Middleton}},\
  }\href@noop {} {\enquote {\bibinfo {title} {{Mu to e matters}},}\ } (\bibinfo
  {year} {2023})\BibitemShut {NoStop}%
\bibitem [{\citenamefont {Lee}(2018)}]{Lee:2018wcx}%
  \BibitemOpen
  \bibfield  {author} {\bibinfo {author} {\bibfnamefont {M.}~\bibnamefont
  {Lee}},\ }\href {\doibase 10.3389/fphy.2018.00133} {\bibfield  {journal}
  {\bibinfo  {journal} {Front.\ in Phys.}\ }\textbf {\bibinfo {volume} {6}}
  (\bibinfo {year} {2018}),\ 10.3389/fphy.2018.00133}\BibitemShut {NoStop}%
\bibitem [{\citenamefont {Bernstein}(2019)}]{Bernstein:2019fyh}%
  \BibitemOpen
  \bibfield  {author} {\bibinfo {author} {\bibfnamefont {R.~H.}\ \bibnamefont
  {Bernstein}} (\bibinfo {collaboration} {Mu2e}),\ }\href {\doibase
  10.3389/fphy.2019.00001} {\bibfield  {journal} {\bibinfo  {journal} {Front.
  in Phys.}\ }\textbf {\bibinfo {volume} {7}},\ \bibinfo {pages} {1} (\bibinfo
  {year} {2019})},\ \Eprint {http://arxiv.org/abs/1901.11099} {arXiv:1901.11099
  [physics.ins-det]} \BibitemShut {NoStop}%
\bibitem [{\citenamefont {Baldini}\ \emph {et~al.}(2021)\citenamefont {Baldini}
  \emph {et~al.}}]{MEGII:2021fah}%
  \BibitemOpen
  \bibfield  {author} {\bibinfo {author} {\bibfnamefont {A.~M.}\ \bibnamefont
  {Baldini}} \emph {et~al.} (\bibinfo {collaboration} {MEG II}),\ }\href
  {\doibase 10.3390/sym13091591} {\bibfield  {journal} {\bibinfo  {journal}
  {Symmetry}\ }\textbf {\bibinfo {volume} {13}},\ \bibinfo {pages} {1591}
  (\bibinfo {year} {2021})},\ \Eprint {http://arxiv.org/abs/2107.10767}
  {arXiv:2107.10767 [hep-ex]} \BibitemShut {NoStop}%
\bibitem [{\citenamefont {Hesketh}\ \emph {et~al.}(2022)\citenamefont {Hesketh}
  \emph {et~al.}}]{Hesketh:2022wgw}%
  \BibitemOpen
  \bibfield  {author} {\bibinfo {author} {\bibfnamefont {G.}~\bibnamefont
  {Hesketh}} \emph {et~al.} (\bibinfo {collaboration} {Mu3e}),\ }in\ \href@noop
  {} {\emph {\bibinfo {booktitle} {{Snowmass 2021}}}}\ (\bibinfo {year}
  {2022})\ \Eprint {http://arxiv.org/abs/2204.00001} {arXiv:2204.00001
  [hep-ex]} \BibitemShut {NoStop}%
\bibitem [{\citenamefont {Aguilar-Arevalo}\ \emph {et~al.}(2018)\citenamefont
  {Aguilar-Arevalo} \emph {et~al.}}]{MiniBooNEDM:2018cxm}%
  \BibitemOpen
  \bibfield  {author} {\bibinfo {author} {\bibfnamefont {A.~A.}\ \bibnamefont
  {Aguilar-Arevalo}} \emph {et~al.} (\bibinfo {collaboration} {MiniBooNE DM}),\
  }\href {\doibase 10.1103/PhysRevD.98.112004} {\bibfield  {journal} {\bibinfo
  {journal} {Phys. Rev. D}\ }\textbf {\bibinfo {volume} {98}},\ \bibinfo
  {pages} {112004} (\bibinfo {year} {2018})},\ \Eprint
  {http://arxiv.org/abs/1807.06137} {arXiv:1807.06137 [hep-ex]} \BibitemShut
  {NoStop}%
\bibitem [{\citenamefont {Abi}\ \emph {et~al.}(2021)\citenamefont {Abi} \emph
  {et~al.}}]{DUNE:2020zfm}%
  \BibitemOpen
  \bibfield  {author} {\bibinfo {author} {\bibfnamefont {B.}~\bibnamefont
  {Abi}} \emph {et~al.} (\bibinfo {collaboration} {DUNE}),\ }\href {\doibase
  10.1140/epjc/s10052-021-09166-w} {\bibfield  {journal} {\bibinfo  {journal}
  {Eur. Phys. J. C}\ }\textbf {\bibinfo {volume} {81}},\ \bibinfo {pages} {423}
  (\bibinfo {year} {2021})},\ \Eprint {http://arxiv.org/abs/2008.06647}
  {arXiv:2008.06647 [hep-ex]} \BibitemShut {NoStop}%
\bibitem [{\citenamefont {Takenaka}\ \emph {et~al.}(2020)\citenamefont
  {Takenaka} \emph {et~al.}}]{Super-Kamiokande:2020wjk}%
  \BibitemOpen
  \bibfield  {author} {\bibinfo {author} {\bibfnamefont {A.}~\bibnamefont
  {Takenaka}} \emph {et~al.} (\bibinfo {collaboration} {Super-Kamiokande}),\
  }\href {\doibase 10.1103/PhysRevD.102.112011} {\bibfield  {journal} {\bibinfo
   {journal} {Phys. Rev. D}\ }\textbf {\bibinfo {volume} {102}},\ \bibinfo
  {pages} {112011} (\bibinfo {year} {2020})},\ \Eprint
  {http://arxiv.org/abs/2010.16098} {arXiv:2010.16098 [hep-ex]} \BibitemShut
  {NoStop}%
\bibitem [{\citenamefont {Abi}\ \emph {et~al.}(2020)\citenamefont {Abi} \emph
  {et~al.}}]{DUNE:2020ypp}%
  \BibitemOpen
  \bibfield  {author} {\bibinfo {author} {\bibfnamefont {B.}~\bibnamefont
  {Abi}} \emph {et~al.} (\bibinfo {collaboration} {DUNE}),\ }\href@noop {}
  {\enquote {\bibinfo {title} {{Deep Underground Neutrino Experiment (DUNE),
  Far Detector Technical Design Report, Volume II: DUNE Physics}},}\ }
  (\bibinfo {year} {2020}),\ \Eprint {http://arxiv.org/abs/2002.03005}
  {arXiv:2002.03005 [hep-ex]} \BibitemShut {NoStop}%
\bibitem [{\citenamefont {Abe}\ \emph {et~al.}(2021)\citenamefont {Abe} \emph
  {et~al.}}]{Super-Kamiokande:2021jaq}%
  \BibitemOpen
  \bibfield  {author} {\bibinfo {author} {\bibfnamefont {K.}~\bibnamefont
  {Abe}} \emph {et~al.} (\bibinfo {collaboration} {Super-Kamiokande}),\ }\href
  {\doibase 10.1103/PhysRevD.104.122002} {\bibfield  {journal} {\bibinfo
  {journal} {Phys. Rev. D}\ }\textbf {\bibinfo {volume} {104}},\ \bibinfo
  {pages} {122002} (\bibinfo {year} {2021})},\ \Eprint
  {http://arxiv.org/abs/2109.11174} {arXiv:2109.11174 [astro-ph.HE]}
  \BibitemShut {NoStop}%
\bibitem [{\citenamefont {Abusleme}\ \emph {et~al.}(2022)\citenamefont
  {Abusleme} \emph {et~al.}}]{JUNO:2022qgr}%
  \BibitemOpen
  \bibfield  {author} {\bibinfo {author} {\bibfnamefont {A.}~\bibnamefont
  {Abusleme}} \emph {et~al.} (\bibinfo {collaboration} {JUNO}),\ }\href@noop {}
  {\enquote {\bibinfo {title} {{JUNO Sensitivity on Proton Decay $p\to \bar\nu
  K^+$ Searches}},}\ } (\bibinfo {year} {2022}),\ \Eprint
  {http://arxiv.org/abs/2212.08502} {arXiv:2212.08502 [hep-ex]} \BibitemShut
  {NoStop}%
\bibitem [{\citenamefont {Hill}\ \emph {et~al.}(2023)\citenamefont {Hill},
  \citenamefont {Plestid},\ and\ \citenamefont {Zupan}}]{Plestid_forthcoming}%
  \BibitemOpen
  \bibfield  {author} {\bibinfo {author} {\bibfnamefont {R.}~\bibnamefont
  {Hill}}, \bibinfo {author} {\bibfnamefont {R.}~\bibnamefont {Plestid}}, \
  and\ \bibinfo {author} {\bibfnamefont {J.}~\bibnamefont {Zupan}},\
  }\href@noop {} {\enquote {\bibinfo {title} {{In preparation}},}\ } (\bibinfo
  {year} {2023})\BibitemShut {NoStop}%
\bibitem [{Ple()}]{Plestid_Ack}%
  \BibitemOpen
  \href@noop {} {}\bibinfo {note} {{This work grew out of detailed discussions
  with Pavel Murat, David Koltick, and Shihua Huang. Their understanding of
  experimental details has been crucial in our modelling of projected
  sensitivity.}}\BibitemShut {Stop}%
\bibitem [{\citenamefont {Bayes}\ \emph {et~al.}(2015)\citenamefont {Bayes}
  \emph {et~al.}}]{TWIST:2014ymv}%
  \BibitemOpen
  \bibfield  {author} {\bibinfo {author} {\bibfnamefont {R.}~\bibnamefont
  {Bayes}} \emph {et~al.} (\bibinfo {collaboration} {TWIST}),\ }\href {\doibase
  10.1103/PhysRevD.91.052020} {\bibfield  {journal} {\bibinfo  {journal} {Phys.
  Rev. D}\ }\textbf {\bibinfo {volume} {91}},\ \bibinfo {pages} {052020}
  (\bibinfo {year} {2015})},\ \Eprint {http://arxiv.org/abs/1409.0638}
  {arXiv:1409.0638 [hep-ex]} \BibitemShut {NoStop}%
\bibitem [{\citenamefont {Athanassopoulos}\ \emph {et~al.}(1997)\citenamefont
  {Athanassopoulos} \emph {et~al.}}]{LSND:1996jxj}%
  \BibitemOpen
  \bibfield  {author} {\bibinfo {author} {\bibfnamefont {C.}~\bibnamefont
  {Athanassopoulos}} \emph {et~al.} (\bibinfo {collaboration} {LSND}),\ }\href
  {\doibase 10.1016/S0168-9002(96)01155-2} {\bibfield  {journal} {\bibinfo
  {journal} {Nucl. Instrum. Meth. A}\ }\textbf {\bibinfo {volume} {388}},\
  \bibinfo {pages} {149} (\bibinfo {year} {1997})},\ \Eprint
  {http://arxiv.org/abs/nucl-ex/9605002} {arXiv:nucl-ex/9605002} \BibitemShut
  {NoStop}%
\bibitem [{\citenamefont {Akimov}\ \emph {et~al.}(2015)\citenamefont {Akimov}
  \emph {et~al.}}]{COHERENT:2015mry}%
  \BibitemOpen
  \bibfield  {author} {\bibinfo {author} {\bibfnamefont {D.}~\bibnamefont
  {Akimov}} \emph {et~al.} (\bibinfo {collaboration} {COHERENT}),\ }\href@noop
  {} {\enquote {\bibinfo {title} {{The COHERENT Experiment at the Spallation
  Neutron Source}},}\ } (\bibinfo {year} {2015}),\ \Eprint
  {http://arxiv.org/abs/1509.08702} {arXiv:1509.08702 [physics.ins-det]}
  \BibitemShut {NoStop}%
\bibitem [{\citenamefont {Ajimura}\ \emph {et~al.}(2021)\citenamefont {Ajimura}
  \emph {et~al.}}]{JSNS2:2021hyk}%
  \BibitemOpen
  \bibfield  {author} {\bibinfo {author} {\bibfnamefont {S.}~\bibnamefont
  {Ajimura}} \emph {et~al.} (\bibinfo {collaboration} {JSNS2}),\ }\href
  {\doibase 10.1016/j.nima.2021.165742} {\bibfield  {journal} {\bibinfo
  {journal} {Nucl. Instrum. Meth. A}\ }\textbf {\bibinfo {volume} {1014}},\
  \bibinfo {pages} {165742} (\bibinfo {year} {2021})},\ \Eprint
  {http://arxiv.org/abs/2104.13169} {arXiv:2104.13169 [physics.ins-det]}
  \BibitemShut {NoStop}%
\bibitem [{\citenamefont {Altmannshofer}\ \emph {et~al.}(2022)\citenamefont
  {Altmannshofer} \emph {et~al.}}]{PIONEER:2022yag}%
  \BibitemOpen
  \bibfield  {author} {\bibinfo {author} {\bibfnamefont {W.}~\bibnamefont
  {Altmannshofer}} \emph {et~al.} (\bibinfo {collaboration} {PIONEER}),\
  }\href@noop {} {\enquote {\bibinfo {title} {{PIONEER: Studies of Rare Pion
  Decays}},}\ } (\bibinfo {year} {2022}),\ \Eprint
  {http://arxiv.org/abs/2203.01981} {arXiv:2203.01981 [hep-ex]} \BibitemShut
  {NoStop}%
\bibitem [{\citenamefont {Toups}\ \emph {et~al.}(2022)\citenamefont {Toups},
  \citenamefont {Van~de Water}, \citenamefont {Batell}, \citenamefont {Brice},
  \citenamefont {deNiverville}, \citenamefont {Eldred}, \citenamefont {Harnik},
  \citenamefont {Kelly}, \citenamefont {Kim}, \citenamefont {Kobilarcik},
  \citenamefont {Krnjaic}, \citenamefont {Littlejohn}, \citenamefont {Louis},
  \citenamefont {Machado}, \citenamefont {Pavlovic}, \citenamefont {Pellico},
  \citenamefont {Shaevitz}, \citenamefont {Snopok}, \citenamefont {Tayloe},
  \citenamefont {Thornton}, \citenamefont {Zettlemoyer},\ and\ \citenamefont
  {Zwaska}}]{toupsDM}%
  \BibitemOpen
  \bibfield  {author} {\bibinfo {author} {\bibfnamefont {M.}~\bibnamefont
  {Toups}}, \bibinfo {author} {\bibfnamefont {R.~G.}\ \bibnamefont {Van~de
  Water}}, \bibinfo {author} {\bibfnamefont {B.}~\bibnamefont {Batell}},
  \bibinfo {author} {\bibfnamefont {S.~J.}\ \bibnamefont {Brice}}, \bibinfo
  {author} {\bibfnamefont {P.}~\bibnamefont {deNiverville}}, \bibinfo {author}
  {\bibfnamefont {J.}~\bibnamefont {Eldred}}, \bibinfo {author} {\bibfnamefont
  {R.}~\bibnamefont {Harnik}}, \bibinfo {author} {\bibfnamefont {K.~J.}\
  \bibnamefont {Kelly}}, \bibinfo {author} {\bibfnamefont {D.}~\bibnamefont
  {Kim}}, \bibinfo {author} {\bibfnamefont {T.}~\bibnamefont {Kobilarcik}},
  \bibinfo {author} {\bibfnamefont {G.}~\bibnamefont {Krnjaic}}, \bibinfo
  {author} {\bibfnamefont {B.~R.}\ \bibnamefont {Littlejohn}}, \bibinfo
  {author} {\bibfnamefont {B.}~\bibnamefont {Louis}}, \bibinfo {author}
  {\bibfnamefont {P.~A.~N.}\ \bibnamefont {Machado}}, \bibinfo {author}
  {\bibfnamefont {Z.}~\bibnamefont {Pavlovic}}, \bibinfo {author}
  {\bibfnamefont {W.}~\bibnamefont {Pellico}}, \bibinfo {author} {\bibfnamefont
  {M.}~\bibnamefont {Shaevitz}}, \bibinfo {author} {\bibfnamefont
  {P.}~\bibnamefont {Snopok}}, \bibinfo {author} {\bibfnamefont
  {R.}~\bibnamefont {Tayloe}}, \bibinfo {author} {\bibfnamefont {R.~T.}\
  \bibnamefont {Thornton}}, \bibinfo {author} {\bibfnamefont {J.}~\bibnamefont
  {Zettlemoyer}}, \ and\ \bibinfo {author} {\bibfnamefont {B.}~\bibnamefont
  {Zwaska}},\ }\href {\doibase 10.48550/ARXIV.2203.08079} {\enquote {\bibinfo
  {title} {{PIP2-BD: GeV Proton Beam Dump at Fermilab's PIP-II Linac}},}\ }
  (\bibinfo {year} {2022})\BibitemShut {NoStop}%
\bibitem [{\citenamefont {Pellico}(2023)}]{ACE_Workshop}%
  \BibitemOpen
  \bibfield  {author} {\bibinfo {author} {\bibfnamefont {B.}~\bibnamefont
  {Pellico}},\ }\href@noop {} {\enquote {\bibinfo {title} {{PAR and ACE}},}\ }
  (\bibinfo {year} {2023}),\ \bibinfo {note} {{Accelerators Capabilities
  Enhancement Workshop, Batavia, IL}}\BibitemShut {NoStop}%
\bibitem [{\citenamefont {Echenard}\ \emph {et~al.}(2022)\citenamefont
  {Echenard} \emph {et~al.}}]{CLFV:Snowmass}%
  \BibitemOpen
  \bibfield  {author} {\bibinfo {author} {\bibfnamefont {B.}~\bibnamefont
  {Echenard}} \emph {et~al.} (\bibinfo {collaboration} {AMF}),\ }in\ \href
  {\doibase 0.48550/arXiv.2203.08278} {\emph {\bibinfo {booktitle} {{Snowmass
  2021}}}}\ (\bibinfo {year} {2022})\ \Eprint {http://arxiv.org/abs/2203.08278}
  {arXiv:2203.08278 [hep-ex]} \BibitemShut {NoStop}%
\bibitem [{\citenamefont {Cousineau}\ \emph {et~al.}(2017)\citenamefont
  {Cousineau} \emph {et~al.}}]{Cousinau:PRAB}%
  \BibitemOpen
  \bibfield  {author} {\bibinfo {author} {\bibfnamefont {S.}~\bibnamefont
  {Cousineau}} \emph {et~al.},\ }\href {\doibase
  10.1103/PhysRevAccelBeams.20.120402} {\bibfield  {journal} {\bibinfo
  {journal} {Phys. Rev. D}\ }\textbf {\bibinfo {volume} {20}},\ \bibinfo
  {pages} {120402} (\bibinfo {year} {2017})}\BibitemShut {NoStop}%
\bibitem [{\citenamefont {Abadzhev}\ \emph {et~al.}(1992)\citenamefont
  {Abadzhev} \emph {et~al.}}]{Abadzhev:1992nx}%
  \BibitemOpen
  \bibfield  {author} {\bibinfo {author} {\bibfnamefont {V.~S.}\ \bibnamefont
  {Abadzhev}} \emph {et~al.},\ }\href@noop {} {\  (\bibinfo {year}
  {1992})}\BibitemShut {NoStop}%
\bibitem [{\citenamefont {Symon}\ \emph {et~al.}(1956)\citenamefont {Symon},
  \citenamefont {Kerst}, \citenamefont {Jones}, \citenamefont {Laslett},\ and\
  \citenamefont {Terwilliger}}]{Symon:1956pr}%
  \BibitemOpen
  \bibfield  {author} {\bibinfo {author} {\bibfnamefont {K.~R.}\ \bibnamefont
  {Symon}}, \bibinfo {author} {\bibfnamefont {D.~W.}\ \bibnamefont {Kerst}},
  \bibinfo {author} {\bibfnamefont {L.~W.}\ \bibnamefont {Jones}}, \bibinfo
  {author} {\bibfnamefont {L.~J.}\ \bibnamefont {Laslett}}, \ and\ \bibinfo
  {author} {\bibfnamefont {K.~M.}\ \bibnamefont {Terwilliger}},\ }\href
  {\doibase 10.1103/PhysRev.103.1837} {\bibfield  {journal} {\bibinfo
  {journal} {Phys. Rev.}\ }\textbf {\bibinfo {volume} {103}},\ \bibinfo {pages}
  {1837} (\bibinfo {year} {1956})}\BibitemShut {NoStop}%
\bibitem [{\citenamefont {Collot}\ \emph {et~al.}(2008)\citenamefont {Collot},
  \citenamefont {Mori}, \citenamefont {Mandrillon}, \citenamefont {Meot},\ and\
  \citenamefont {Edgecock}}]{Collot:2008zz}%
  \BibitemOpen
  \bibfield  {author} {\bibinfo {author} {\bibfnamefont {J.}~\bibnamefont
  {Collot}}, \bibinfo {author} {\bibfnamefont {Y.}~\bibnamefont {Mori}},
  \bibinfo {author} {\bibfnamefont {P.}~\bibnamefont {Mandrillon}}, \bibinfo
  {author} {\bibfnamefont {F.}~\bibnamefont {Meot}}, \ and\ \bibinfo {author}
  {\bibfnamefont {R.}~\bibnamefont {Edgecock}},\ }\href@noop {} {\bibfield
  {journal} {\bibinfo  {journal} {CERN Cour.}\ }\textbf {\bibinfo {volume}
  {48N7}},\ \bibinfo {pages} {21} (\bibinfo {year} {2008})}\BibitemShut
  {NoStop}%
\bibitem [{\citenamefont {et~al.}(2000)}]{Aiba:2000}%
  \BibitemOpen
  \bibfield  {author} {\bibinfo {author} {\bibfnamefont {A.~M.}\ \bibnamefont
  {et~al.}},\ }\href@noop {} {\bibfield  {journal} {\bibinfo  {journal}
  {JACoW}\ }\textbf {\bibinfo {volume} {EPAC2000}} (\bibinfo {year}
  {2000})}\BibitemShut {NoStop}%
\bibitem [{\citenamefont {Kuno}(2005)}]{KUNO2005376}%
  \BibitemOpen
  \bibfield  {author} {\bibinfo {author} {\bibfnamefont {Y.}~\bibnamefont
  {Kuno}},\ }\href {\doibase https://doi.org/10.1016/j.nuclphysbps.2005.05.073}
  {\bibfield  {journal} {\bibinfo  {journal} {Nuclear Physics B - Proceedings
  Supplements}\ }\textbf {\bibinfo {volume} {149}},\ \bibinfo {pages} {376}
  (\bibinfo {year} {2005})},\ \bibinfo {note} {nuFact04}\BibitemShut {NoStop}%
\bibitem [{\citenamefont {Ohmori}\ \emph {et~al.}(2008)\citenamefont {Ohmori},
  \citenamefont {Aoki}, \citenamefont {Arimoto}, \citenamefont {Itahashi},
  \citenamefont {Kuno}, \citenamefont {Kuriyama}, \citenamefont {Sato},
  \citenamefont {Yoshida}, \citenamefont {Iwashita},\ and\ \citenamefont
  {Mori}}]{Ohmori:2008zza}%
  \BibitemOpen
  \bibfield  {author} {\bibinfo {author} {\bibfnamefont {C.}~\bibnamefont
  {Ohmori}}, \bibinfo {author} {\bibfnamefont {M.}~\bibnamefont {Aoki}},
  \bibinfo {author} {\bibfnamefont {Y.}~\bibnamefont {Arimoto}}, \bibinfo
  {author} {\bibfnamefont {I.}~\bibnamefont {Itahashi}}, \bibinfo {author}
  {\bibfnamefont {Y.}~\bibnamefont {Kuno}}, \bibinfo {author} {\bibfnamefont
  {Y.}~\bibnamefont {Kuriyama}}, \bibinfo {author} {\bibfnamefont
  {A.}~\bibnamefont {Sato}}, \bibinfo {author} {\bibfnamefont {M.~Y.}\
  \bibnamefont {Yoshida}}, \bibinfo {author} {\bibfnamefont {Y.}~\bibnamefont
  {Iwashita}}, \ and\ \bibinfo {author} {\bibfnamefont {Y.}~\bibnamefont
  {Mori}},\ }\href@noop {} {\bibfield  {journal} {\bibinfo  {journal} {Conf.
  Proc. C}\ }\textbf {\bibinfo {volume} {0806233}},\ \bibinfo {pages} {MOPP103}
  (\bibinfo {year} {2008})}\BibitemShut {NoStop}%
\bibitem [{\citenamefont {Alekou}\ \emph {et~al.}(2013)\citenamefont {Alekou}
  \emph {et~al.}}]{Alekou:2013eta}%
  \BibitemOpen
  \bibfield  {author} {\bibinfo {author} {\bibfnamefont {A.}~\bibnamefont
  {Alekou}} \emph {et~al.},\ }in\ \href@noop {} {\emph {\bibinfo {booktitle}
  {{Community Summer Study 2013}: {Snowmass on the Mississippi}}}}\ (\bibinfo
  {year} {2013})\ \Eprint {http://arxiv.org/abs/1310.0804} {arXiv:1310.0804
  [physics.acc-ph]} \BibitemShut {NoStop}%
\bibitem [{\citenamefont {Sato}\ \emph {et~al.}(2008)\citenamefont {Sato} \emph
  {et~al.}}]{Sato:2008zze}%
  \BibitemOpen
  \bibfield  {author} {\bibinfo {author} {\bibfnamefont {A.}~\bibnamefont
  {Sato}} \emph {et~al.},\ }\href@noop {} {\bibfield  {journal} {\bibinfo
  {journal} {Conf. Proc. C}\ }\textbf {\bibinfo {volume} {0806233}},\ \bibinfo
  {pages} {THPP007} (\bibinfo {year} {2008})}\BibitemShut {NoStop}%
\bibitem [{\citenamefont {Lagrange}\ \emph {et~al.}(2018)\citenamefont
  {Lagrange}, \citenamefont {Appleby}, \citenamefont {Garland}, \citenamefont
  {Pasternak},\ and\ \citenamefont {Tygier}}]{Lagrange:2018triplet}%
  \BibitemOpen
  \bibfield  {author} {\bibinfo {author} {\bibfnamefont {J.-B.}\ \bibnamefont
  {Lagrange}}, \bibinfo {author} {\bibfnamefont {R.}~\bibnamefont {Appleby}},
  \bibinfo {author} {\bibfnamefont {J.}~\bibnamefont {Garland}}, \bibinfo
  {author} {\bibfnamefont {J.}~\bibnamefont {Pasternak}}, \ and\ \bibinfo
  {author} {\bibfnamefont {S.}~\bibnamefont {Tygier}},\ }\href {\doibase
  10.1088/1748-0221/13/09/p09013} {\bibfield  {journal} {\bibinfo  {journal}
  {Journal of Instrumentation}\ }\textbf {\bibinfo {volume} {13}},\ \bibinfo
  {pages} {P09013} (\bibinfo {year} {2018})}\BibitemShut {NoStop}%
\bibitem [{\citenamefont {Paraliev}(2018)}]{Paraliev:2018ski}%
  \BibitemOpen
  \bibfield  {author} {\bibinfo {author} {\bibfnamefont {M.}~\bibnamefont
  {Paraliev}},\ }\href {\doibase 10.23730/CYRSP-2018-005.363} {\bibfield
  {journal} {\bibinfo  {journal} {CERN Yellow Rep. School Proc.}\ }\textbf
  {\bibinfo {volume} {5}},\ \bibinfo {pages} {363} (\bibinfo {year}
  {2018})}\BibitemShut {NoStop}%
\bibitem [{\citenamefont {{MURA~staff}}(1994)}]{39289}%
  \BibitemOpen
  \bibfield  {author} {\bibinfo {author} {\bibnamefont {{MURA~staff}}},\
  }\href@noop {} {\enquote {\bibinfo {title} {{The MURA two way electron
  accelerator}},}\ } (\bibinfo {year} {1994})\BibitemShut {NoStop}%
\bibitem [{\citenamefont {Bartoszek}\ \emph {et~al.}(2015)\citenamefont
  {Bartoszek} \emph {et~al.}}]{bartoszek2015mu2e}%
  \BibitemOpen
  \bibfield  {author} {\bibinfo {author} {\bibfnamefont {L.}~\bibnamefont
  {Bartoszek}} \emph {et~al.},\ }\href@noop {} {\enquote {\bibinfo {title}
  {{Mu2e Technical Design Report}},}\ } (\bibinfo {year} {2015}),\ \Eprint
  {http://arxiv.org/abs/1501.05241} {arXiv:1501.05241 [physics.ins-det]}
  \BibitemShut {NoStop}%
\bibitem [{\citenamefont {Abramishvili}\ \emph {et~al.}(2020)\citenamefont
  {Abramishvili}, \citenamefont {Adamov}, \citenamefont {Akhmetshin},
  \citenamefont {Allin}, \citenamefont {Angélique}, \citenamefont {Anishchik},
  \citenamefont {Aoki}, \citenamefont {Aznabayev}, \citenamefont {Bagaturia},
  \citenamefont {Ban} \emph {et~al.}}]{comet-tdr}%
  \BibitemOpen
  \bibfield  {author} {\bibinfo {author} {\bibfnamefont {R.}~\bibnamefont
  {Abramishvili}}, \bibinfo {author} {\bibfnamefont {G.}~\bibnamefont
  {Adamov}}, \bibinfo {author} {\bibfnamefont {R.~R.}\ \bibnamefont
  {Akhmetshin}}, \bibinfo {author} {\bibfnamefont {A.}~\bibnamefont {Allin}},
  \bibinfo {author} {\bibfnamefont {J.~C.}\ \bibnamefont {Angélique}},
  \bibinfo {author} {\bibfnamefont {V.}~\bibnamefont {Anishchik}}, \bibinfo
  {author} {\bibfnamefont {M.}~\bibnamefont {Aoki}}, \bibinfo {author}
  {\bibfnamefont {D.}~\bibnamefont {Aznabayev}}, \bibinfo {author}
  {\bibfnamefont {I.}~\bibnamefont {Bagaturia}}, \bibinfo {author}
  {\bibfnamefont {G.}~\bibnamefont {Ban}},  \emph {et~al.},\ }\href {\doibase
  10.1093/ptep/ptz125} {\bibfield  {journal} {\bibinfo  {journal} {Progress of
  Theoretical and Experimental Physics}\ }\textbf {\bibinfo {volume} {2020}}
  (\bibinfo {year} {2020}),\ 10.1093/ptep/ptz125}\BibitemShut {NoStop}%
\bibitem [{\citenamefont {Gaponenko}(2022)}]{sensitivity-scaling}%
  \BibitemOpen
  \bibfield  {author} {\bibinfo {author} {\bibfnamefont {A.}~\bibnamefont
  {Gaponenko}},\ }\href@noop {} {\enquote {\bibinfo {title} {{Momentum
  resolution requirement for muon-to-electron conversion searches}},}\
  }\bibinfo {howpublished} {FERMILAB-PUB-22-117-PPD} (\bibinfo {year}
  {2022})\BibitemShut {NoStop}%
\bibitem [{\citenamefont {Abdi}\ \emph {et~al.}(2023)\citenamefont {Abdi} \emph
  {et~al.}}]{Mu2e:2022ggl}%
  \BibitemOpen
  \bibfield  {author} {\bibinfo {author} {\bibfnamefont {F.}~\bibnamefont
  {Abdi}} \emph {et~al.} (\bibinfo {collaboration} {Mu2e}),\ }\href {\doibase
  10.3390/universe9010054} {\bibfield  {journal} {\bibinfo  {journal}
  {Universe}\ }\textbf {\bibinfo {volume} {9}},\ \bibinfo {pages} {54}
  (\bibinfo {year} {2023})},\ \Eprint {http://arxiv.org/abs/2210.11380}
  {arXiv:2210.11380 [hep-ex]} \BibitemShut {NoStop}%
\bibitem [{\citenamefont {Aiba}\ \emph {et~al.}(2021)\citenamefont {Aiba} \emph
  {et~al.}}]{Aiba:2021bxe}%
  \BibitemOpen
  \bibfield  {author} {\bibinfo {author} {\bibfnamefont {M.}~\bibnamefont
  {Aiba}} \emph {et~al.},\ }\href@noop {} {\enquote {\bibinfo {title} {{Science
  Case for the new High-Intensity Muon Beams HIMB at PSI}},}\ } (\bibinfo
  {year} {2021}),\ \Eprint {http://arxiv.org/abs/2111.05788} {arXiv:2111.05788
  [hep-ex]} \BibitemShut {NoStop}%
\bibitem [{\citenamefont {Renga}(2019)}]{Renga:2019mpg}%
  \BibitemOpen
  \bibfield  {author} {\bibinfo {author} {\bibfnamefont {F.}~\bibnamefont
  {Renga}},\ }\href {\doibase 10.1016/j.revip.2019.100029} {\bibfield
  {journal} {\bibinfo  {journal} {Rev. Phys.}\ }\textbf {\bibinfo {volume}
  {4}},\ \bibinfo {pages} {100029} (\bibinfo {year} {2019})},\ \Eprint
  {http://arxiv.org/abs/1902.06291} {arXiv:1902.06291 [hep-ex]} \BibitemShut
  {NoStop}%
\bibitem [{\citenamefont {Papa}\ \emph {et~al.}(2023)\citenamefont {Papa} \emph
  {et~al.}}]{Papa:2023voa}%
  \BibitemOpen
  \bibfield  {author} {\bibinfo {author} {\bibfnamefont {A.}~\bibnamefont
  {Papa}} \emph {et~al.},\ }\href {\doibase 10.1016/j.nima.2022.167997}
  {\bibfield  {journal} {\bibinfo  {journal} {Nucl. Instrum. Meth. A}\ }\textbf
  {\bibinfo {volume} {1049}},\ \bibinfo {pages} {167997} (\bibinfo {year}
  {2023})}\BibitemShut {NoStop}%
\bibitem [{\citenamefont {Bellgardt}\ \emph {et~al.}(1988)\citenamefont
  {Bellgardt} \emph {et~al.}}]{SINDRUM:1987nra}%
  \BibitemOpen
  \bibfield  {author} {\bibinfo {author} {\bibfnamefont {U.}~\bibnamefont
  {Bellgardt}} \emph {et~al.} (\bibinfo {collaboration} {SINDRUM}),\ }\href
  {\doibase 10.1016/0550-3213(88)90462-2} {\bibfield  {journal} {\bibinfo
  {journal} {Nucl. Phys. B}\ }\textbf {\bibinfo {volume} {299}},\ \bibinfo
  {pages} {1} (\bibinfo {year} {1988})}\BibitemShut {NoStop}%
\bibitem [{\citenamefont {Arndt}\ \emph {et~al.}(2021)\citenamefont {Arndt}
  \emph {et~al.}}]{Mu3e:2020gyw}%
  \BibitemOpen
  \bibfield  {author} {\bibinfo {author} {\bibfnamefont {K.}~\bibnamefont
  {Arndt}} \emph {et~al.} (\bibinfo {collaboration} {Mu3e}),\ }\href {\doibase
  10.1016/j.nima.2021.165679} {\bibfield  {journal} {\bibinfo  {journal} {Nucl.
  Instrum. Meth. A}\ }\textbf {\bibinfo {volume} {1014}},\ \bibinfo {pages}
  {165679} (\bibinfo {year} {2021})},\ \Eprint
  {http://arxiv.org/abs/2009.11690} {arXiv:2009.11690 [physics.ins-det]}
  \BibitemShut {NoStop}%
\bibitem [{\citenamefont {Baldini}\ \emph {et~al.}(2016)\citenamefont {Baldini}
  \emph {et~al.}}]{MEG:2016leq}%
  \BibitemOpen
  \bibfield  {author} {\bibinfo {author} {\bibfnamefont {A.~M.}\ \bibnamefont
  {Baldini}} \emph {et~al.} (\bibinfo {collaboration} {MEG}),\ }\href {\doibase
  10.1140/epjc/s10052-016-4271-x} {\bibfield  {journal} {\bibinfo  {journal}
  {Eur. Phys. J. C}\ }\textbf {\bibinfo {volume} {76}},\ \bibinfo {pages} {434}
  (\bibinfo {year} {2016})},\ \Eprint {http://arxiv.org/abs/1605.05081}
  {arXiv:1605.05081 [hep-ex]} \BibitemShut {NoStop}%
\bibitem [{\citenamefont {Baldini}\ \emph {et~al.}(2018)\citenamefont
  {Baldini}, \citenamefont {Baracchini}, \citenamefont {Bemporad},
  \citenamefont {Berg}, \citenamefont {Biasotti}, \citenamefont {Boca},
  \citenamefont {Cattaneo}, \citenamefont {Cavoto}, \citenamefont {Cei},
  \citenamefont {Chiappini}, \citenamefont {Chiarello}, \citenamefont {Chiri},
  \citenamefont {Cocciolo} \emph {et~al.}}]{MEGII2018}%
  \BibitemOpen
  \bibfield  {author} {\bibinfo {author} {\bibfnamefont {A.~M.}\ \bibnamefont
  {Baldini}}, \bibinfo {author} {\bibfnamefont {E.}~\bibnamefont {Baracchini}},
  \bibinfo {author} {\bibfnamefont {C.}~\bibnamefont {Bemporad}}, \bibinfo
  {author} {\bibfnamefont {F.}~\bibnamefont {Berg}}, \bibinfo {author}
  {\bibfnamefont {M.}~\bibnamefont {Biasotti}}, \bibinfo {author}
  {\bibfnamefont {G.}~\bibnamefont {Boca}}, \bibinfo {author} {\bibfnamefont
  {P.~W.}\ \bibnamefont {Cattaneo}}, \bibinfo {author} {\bibfnamefont
  {G.}~\bibnamefont {Cavoto}}, \bibinfo {author} {\bibfnamefont
  {F.}~\bibnamefont {Cei}}, \bibinfo {author} {\bibfnamefont {M.}~\bibnamefont
  {Chiappini}}, \bibinfo {author} {\bibfnamefont {G.}~\bibnamefont
  {Chiarello}}, \bibinfo {author} {\bibfnamefont {C.}~\bibnamefont {Chiri}},
  \bibinfo {author} {\bibfnamefont {G.}~\bibnamefont {Cocciolo}},  \emph
  {et~al.},\ }\href {\doibase 10.1140/epjc/s10052-018-5845-6} {\bibfield
  {journal} {\bibinfo  {journal} {The European Physical Journal C}\ }\textbf
  {\bibinfo {volume} {78}} (\bibinfo {year} {2018}),\
  10.1140/epjc/s10052-018-5845-6}\BibitemShut {NoStop}%
\bibitem [{\citenamefont {Cavoto}\ \emph {et~al.}(2018)\citenamefont {Cavoto},
  \citenamefont {Papa}, \citenamefont {Renga}, \citenamefont {Ripiccini},\ and\
  \citenamefont {Voena}}]{next_meg}%
  \BibitemOpen
  \bibfield  {author} {\bibinfo {author} {\bibfnamefont {G.}~\bibnamefont
  {Cavoto}}, \bibinfo {author} {\bibfnamefont {A.}~\bibnamefont {Papa}},
  \bibinfo {author} {\bibfnamefont {F.}~\bibnamefont {Renga}}, \bibinfo
  {author} {\bibfnamefont {E.}~\bibnamefont {Ripiccini}}, \ and\ \bibinfo
  {author} {\bibfnamefont {C.}~\bibnamefont {Voena}},\ }\href {\doibase
  10.1140/epjc/s10052-017-5444-y} {\bibfield  {journal} {\bibinfo  {journal}
  {Eur. Phys. J. C}\ }\textbf {\bibinfo {volume} {78}},\ \bibinfo {pages} {37}
  (\bibinfo {year} {2018})},\ \Eprint {http://arxiv.org/abs/1707.01805}
  {arXiv:1707.01805 [hep-ex]} \BibitemShut {NoStop}%
\bibitem [{\citenamefont {Jho}\ \emph {et~al.}(2022)\citenamefont {Jho},
  \citenamefont {Knapen},\ and\ \citenamefont {Redigolo}}]{Jho:2022snj}%
  \BibitemOpen
  \bibfield  {author} {\bibinfo {author} {\bibfnamefont {Y.}~\bibnamefont
  {Jho}}, \bibinfo {author} {\bibfnamefont {S.}~\bibnamefont {Knapen}}, \ and\
  \bibinfo {author} {\bibfnamefont {D.}~\bibnamefont {Redigolo}},\ }\href
  {\doibase 10.1007/JHEP10(2022)029} {\bibfield  {journal} {\bibinfo  {journal}
  {JHEP}\ }\textbf {\bibinfo {volume} {10}},\ \bibinfo {pages} {029} (\bibinfo
  {year} {2022})},\ \Eprint {http://arxiv.org/abs/2203.11222} {arXiv:2203.11222
  [hep-ph]} \BibitemShut {NoStop}%
\bibitem [{\citenamefont {Willmann}\ \emph {et~al.}(1999)\citenamefont
  {Willmann} \emph {et~al.}}]{MACS:1999}%
  \BibitemOpen
  \bibfield  {author} {\bibinfo {author} {\bibfnamefont {L.}~\bibnamefont
  {Willmann}} \emph {et~al.} (\bibinfo {collaboration} {MACS}),\ }\href@noop {}
  {\bibfield  {journal} {\bibinfo  {journal} {Phys. Rev. Lett.}\ }\textbf
  {\bibinfo {volume} {82}},\ \bibinfo {pages} {49} (\bibinfo {year}
  {1999})}\BibitemShut {NoStop}%
\bibitem [{\citenamefont {Oram}\ \emph {et~al.}(1984)\citenamefont {Oram} \emph
  {et~al.}}]{LambShift:1984}%
  \BibitemOpen
  \bibfield  {author} {\bibinfo {author} {\bibfnamefont {C.~J.}\ \bibnamefont
  {Oram}} \emph {et~al.},\ }\href@noop {} {\bibfield  {journal} {\bibinfo
  {journal} {Phys. Rev. Lett.}\ }\textbf {\bibinfo {volume} {52}},\ \bibinfo
  {pages} {910} (\bibinfo {year} {1984})}\BibitemShut {NoStop}%
\bibitem [{\citenamefont {Meyer}\ \emph {et~al.}(2000)\citenamefont {Meyer}
  \emph {et~al.}}]{1s-2s:2000}%
  \BibitemOpen
  \bibfield  {author} {\bibinfo {author} {\bibfnamefont {V.}~\bibnamefont
  {Meyer}} \emph {et~al.},\ }\href@noop {} {\bibfield  {journal} {\bibinfo
  {journal} {Phys. Rev. Lett.}\ }\textbf {\bibinfo {volume} {84}},\ \bibinfo
  {pages} {1136} (\bibinfo {year} {2000})}\BibitemShut {NoStop}%
\bibitem [{\citenamefont {Liu}\ \emph {et~al.}(1999)\citenamefont {Liu} \emph
  {et~al.}}]{hyperfine:1999}%
  \BibitemOpen
  \bibfield  {author} {\bibinfo {author} {\bibfnamefont {W.}~\bibnamefont
  {Liu}} \emph {et~al.},\ }\href@noop {} {\bibfield  {journal} {\bibinfo
  {journal} {Phys. Rev. Lett.}\ }\textbf {\bibinfo {volume} {82}},\ \bibinfo
  {pages} {711} (\bibinfo {year} {1999})}\BibitemShut {NoStop}%
\bibitem [{\citenamefont {Kanda}\ \emph {et~al.}(2021)\citenamefont {Kanda}
  \emph {et~al.}}]{MuSEUM:2021}%
  \BibitemOpen
  \bibfield  {author} {\bibinfo {author} {\bibfnamefont {S.}~\bibnamefont
  {Kanda}} \emph {et~al.} (\bibinfo {collaboration} {MuSEUM}),\ }\href@noop {}
  {\bibfield  {journal} {\bibinfo  {journal} {Phys. Lett. B}\ }\textbf
  {\bibinfo {volume} {815}},\ \bibinfo {pages} {136154} (\bibinfo {year}
  {2021})}\BibitemShut {NoStop}%
\bibitem [{\citenamefont {Nishimura}\ \emph {et~al.}(2021)\citenamefont
  {Nishimura} \emph {et~al.}}]{MuSEUM-Nishimura:2021}%
  \BibitemOpen
  \bibfield  {author} {\bibinfo {author} {\bibfnamefont {S.}~\bibnamefont
  {Nishimura}} \emph {et~al.} (\bibinfo {collaboration} {MuSEUM}),\ }\href@noop
  {} {\bibfield  {journal} {\bibinfo  {journal} {Phys. Rev. A}\ }\textbf
  {\bibinfo {volume} {104}},\ \bibinfo {pages} {L020801} (\bibinfo {year}
  {2021})}\BibitemShut {NoStop}%
\bibitem [{\citenamefont {{P. Crivelli}}(2018)}]{MuMass-Crivelli:2018}%
  \BibitemOpen
  \bibfield  {author} {\bibinfo {author} {\bibnamefont {{P. Crivelli}}},\
  }\href@noop {} {\bibfield  {journal} {\bibinfo  {journal} {Hyp. Int.}\
  }\textbf {\bibinfo {volume} {239}},\ \bibinfo {pages} {49} (\bibinfo {year}
  {2018})}\BibitemShut {NoStop}%
\bibitem [{\citenamefont {{B. Ohayon, Z. Burkley, P.
  Crivelli}}(2021)}]{MuMass-Ohayon:2021}%
  \BibitemOpen
  \bibfield  {author} {\bibinfo {author} {\bibnamefont {{B. Ohayon, Z. Burkley,
  P. Crivelli}}},\ }\href@noop {} {\bibfield  {journal} {\bibinfo  {journal}
  {SciPost Phys. Proc.}\ }\textbf {\bibinfo {volume} {5}},\ \bibinfo {pages}
  {029} (\bibinfo {year} {2021})}\BibitemShut {NoStop}%
\bibitem [{LEM(2022)}]{LEMING}%
  \BibitemOpen
  \href@noop {} {}\bibinfo {howpublished}
  {\url{https://lepp.ethz.ch/research/leming.html}} (\bibinfo {year}
  {2022})\BibitemShut {NoStop}%
\bibitem [{\citenamefont {Beer}\ \emph {et~al.}(2014)\citenamefont {Beer} \emph
  {et~al.}}]{Beer:2014}%
  \BibitemOpen
  \bibfield  {author} {\bibinfo {author} {\bibfnamefont {G.}~\bibnamefont
  {Beer}} \emph {et~al.},\ }\href {https://doi.org/10.1093/ptep/ptu116}
  {\bibfield  {journal} {\bibinfo  {journal} {Prog. Theor. Exp. Phys.}\
  }\textbf {\bibinfo {volume} {2014}},\ \bibinfo {pages} {091C01} (\bibinfo
  {year} {2014})}\BibitemShut {NoStop}%
\bibitem [{\citenamefont {Bai}\ \emph {et~al.}(2022)\citenamefont {Bai} \emph
  {et~al.}}]{MACE-Snowmass:2022}%
  \BibitemOpen
  \bibfield  {author} {\bibinfo {author} {\bibfnamefont {A.-Y.}\ \bibnamefont
  {Bai}} \emph {et~al.} (\bibinfo {collaboration} {MACE}),\ }\href@noop {}
  {\enquote {\bibinfo {title} {{Muonium to antimuonium conversion: Contributed
  paper for Snowmass 21}},}\ } (\bibinfo {year} {2022}),\ \Eprint
  {http://arxiv.org/abs/2203.11406} {arXiv:2203.11406 [hep-ph]} \BibitemShut
  {NoStop}%
\bibitem [{\citenamefont {Cochran}\ \emph {et~al.}(1972)\citenamefont {Cochran}
  \emph {et~al.}}]{Cochran:1972}%
  \BibitemOpen
  \bibfield  {author} {\bibinfo {author} {\bibfnamefont {D.}~\bibnamefont
  {Cochran}} \emph {et~al.},\ }\href@noop {} {\bibfield  {journal} {\bibinfo
  {journal} {Phys. Rev. D}\ }\textbf {\bibinfo {volume} {6}},\ \bibinfo {pages}
  {3085} (\bibinfo {year} {1972})}\BibitemShut {NoStop}%
\bibitem [{\citenamefont {{A. Bungau, R. Cywinski, C.
  Bungau}}(2013)}]{Bungau:2013}%
  \BibitemOpen
  \bibfield  {author} {\bibinfo {author} {\bibnamefont {{A. Bungau, R.
  Cywinski, C. Bungau}}},\ }\href@noop {} {\bibfield  {journal} {\bibinfo
  {journal} {Phys. Rev. ST Accel. Beams}\ }\textbf {\bibinfo {volume} {16}},\
  \bibinfo {pages} {014701} (\bibinfo {year} {2013})}\BibitemShut {NoStop}%
\bibitem [{\citenamefont {Abela}\ \emph {et~al.}(1993)\citenamefont {Abela}
  \emph {et~al.}}]{Abela:1993}%
  \BibitemOpen
  \bibfield  {author} {\bibinfo {author} {\bibfnamefont {R.}~\bibnamefont
  {Abela}} \emph {et~al.},\ }\href@noop {} {\bibfield  {journal} {\bibinfo
  {journal} {JETP Lett.}\ }\textbf {\bibinfo {volume} {57}},\ \bibinfo {pages}
  {157} (\bibinfo {year} {1993})}\BibitemShut {NoStop}%
\bibitem [{\citenamefont {Soter}\ \emph {et~al.}(2022)\citenamefont {Soter}
  \emph {et~al.}}]{Soter-private}%
  \BibitemOpen
  \bibfield  {author} {\bibinfo {author} {\bibfnamefont {A.}~\bibnamefont
  {Soter}} \emph {et~al.},\ }\href@noop {} {}\bibinfo {howpublished} {private
  communication} (\bibinfo {year} {2022})\BibitemShut {NoStop}%
\bibitem [{\citenamefont {Taqqu}(2011)}]{Taqqu:2011}%
  \BibitemOpen
  \bibfield  {author} {\bibinfo {author} {\bibfnamefont {D.}~\bibnamefont
  {Taqqu}},\ }\href@noop {} {\bibfield  {journal} {\bibinfo  {journal} {Phys.
  Procedia}\ }\textbf {\bibinfo {volume} {17}},\ \bibinfo {pages} {216}
  (\bibinfo {year} {2011})}\BibitemShut {NoStop}%
\bibitem [{\citenamefont {{M. Saarela, E. Krotscheck}}(1993)}]{Saarela:1993}%
  \BibitemOpen
  \bibfield  {author} {\bibinfo {author} {\bibnamefont {{M. Saarela, E.
  Krotscheck}}},\ }\href@noop {} {\bibfield  {journal} {\bibinfo  {journal} {J.
  Low Temp. Phys.}\ }\textbf {\bibinfo {volume} {90}},\ \bibinfo {pages} {415}
  (\bibinfo {year} {1993})}\BibitemShut {NoStop}%
\bibitem [{\citenamefont {Luppov}\ \emph {et~al.}(1993)\citenamefont {Luppov}
  \emph {et~al.}}]{Luppov:1993}%
  \BibitemOpen
  \bibfield  {author} {\bibinfo {author} {\bibfnamefont {V.~G.}\ \bibnamefont
  {Luppov}} \emph {et~al.},\ }\href@noop {} {\bibfield  {journal} {\bibinfo
  {journal} {Phys. Rev. Lett}\ }\textbf {\bibinfo {volume} {71}},\ \bibinfo
  {pages} {2405} (\bibinfo {year} {1993})}\BibitemShut {NoStop}%
\bibitem [{\citenamefont {{C. Barenghi, C. Mellor, C. Muirhead, W.
  Vinen}}(1986)}]{Barenghi:1986}%
  \BibitemOpen
  \bibfield  {author} {\bibinfo {author} {\bibnamefont {{C. Barenghi, C.
  Mellor, C. Muirhead, W. Vinen}}},\ }\href@noop {} {\bibfield  {journal}
  {\bibinfo  {journal} {J. Phys. C: Solid State Phys.}\ }\textbf {\bibinfo
  {volume} {19}},\ \bibinfo {pages} {1135} (\bibinfo {year}
  {1986})}\BibitemShut {NoStop}%
\bibitem [{\citenamefont {{D. Taqqu}}(2006)}]{Taqqu:2006}%
  \BibitemOpen
  \bibfield  {author} {\bibinfo {author} {\bibnamefont {{D. Taqqu}}},\
  }\href@noop {} {\bibfield  {journal} {\bibinfo  {journal} {Phys. Rev. Lett.}\
  }\textbf {\bibinfo {volume} {97}},\ \bibinfo {pages} {194801} (\bibinfo
  {year} {2006})}\BibitemShut {NoStop}%
\bibitem [{\citenamefont {Bao}\ \emph {et~al.}(2014)\citenamefont {Bao} \emph
  {et~al.}}]{Bao:2014}%
  \BibitemOpen
  \bibfield  {author} {\bibinfo {author} {\bibfnamefont {Y.}~\bibnamefont
  {Bao}} \emph {et~al.},\ }\href@noop {} {\bibfield  {journal} {\bibinfo
  {journal} {Phys. Rev. Lett.}\ }\textbf {\bibinfo {volume} {112}},\ \bibinfo
  {pages} {224801} (\bibinfo {year} {2014})}\BibitemShut {NoStop}%
\bibitem [{\citenamefont {Antognini}\ \emph {et~al.}(2020)\citenamefont
  {Antognini} \emph {et~al.}}]{Antognini:2020}%
  \BibitemOpen
  \bibfield  {author} {\bibinfo {author} {\bibfnamefont {A.}~\bibnamefont
  {Antognini}} \emph {et~al.},\ }\href@noop {} {\bibfield  {journal} {\bibinfo
  {journal} {Phys. Rev. Lett.}\ }\textbf {\bibinfo {volume} {125}},\ \bibinfo
  {pages} {164802} (\bibinfo {year} {2020})}\BibitemShut {NoStop}%
\bibitem [{\citenamefont {Amole}\ \emph {et~al.}(2013)\citenamefont {Amole}
  \emph {et~al.}}]{Amole:2013}%
  \BibitemOpen
  \bibfield  {author} {\bibinfo {author} {\bibfnamefont {C.}~\bibnamefont
  {Amole}} \emph {et~al.},\ }\href@noop {} {\bibfield  {journal} {\bibinfo
  {journal} {Nature Comm.}\ }\textbf {\bibinfo {volume} {4}},\ \bibinfo {pages}
  {1785} (\bibinfo {year} {2013})}\BibitemShut {NoStop}%
\bibitem [{\citenamefont {{D. S. M. Alves, M. Jankowiak, P.
  Saraswat}}(2009)}]{Alves:2009}%
  \BibitemOpen
  \bibfield  {author} {\bibinfo {author} {\bibnamefont {{D. S. M. Alves, M.
  Jankowiak, P. Saraswat}}},\ }\href@noop {} {\enquote {\bibinfo {title}
  {{Experimental constraints on the free fall acceleration of antimatter}},}\ }
  (\bibinfo {year} {2009}),\ \Eprint {http://arxiv.org/abs/0907.4110}
  {arXiv:0907.4110 [hep-ph]} \BibitemShut {NoStop}%
\bibitem [{\citenamefont {Aaij}\ \emph {et~al.}(2022)\citenamefont {Aaij} \emph
  {et~al.}}]{Aaij:2022}%
  \BibitemOpen
  \bibfield  {author} {\bibinfo {author} {\bibfnamefont {R.}~\bibnamefont
  {Aaij}} \emph {et~al.} (\bibinfo {collaboration} {LHCb}),\ }\href@noop {}
  {\bibfield  {journal} {\bibinfo  {journal} {Nature Phys.}\ }\textbf {\bibinfo
  {volume} {18}},\ \bibinfo {pages} {277} (\bibinfo {year} {2022})},\ \bibinfo
  {note} {and references therein}\BibitemShut {NoStop}%
\bibitem [{\citenamefont {{M. M Nieto, T. Goldman}}(1991)}]{Nieto:1991}%
  \BibitemOpen
  \bibfield  {author} {\bibinfo {author} {\bibnamefont {{M. M Nieto, T.
  Goldman}}},\ }\href@noop {} {\bibfield  {journal} {\bibinfo  {journal} {Phys.
  Rep.}\ }\textbf {\bibinfo {volume} {205}},\ \bibinfo {pages} {221–281}
  (\bibinfo {year} {1991})}\BibitemShut {NoStop}%
\bibitem [{\citenamefont {{A. Benoit-L\'evy, G.
  Chardin}}(2012)}]{BenoitLevy:2012}%
  \BibitemOpen
  \bibfield  {author} {\bibinfo {author} {\bibnamefont {{A. Benoit-L\'evy, G.
  Chardin}}},\ }\href@noop {} {\bibfield  {journal} {\bibinfo  {journal}
  {Astron. \& Astrophys.}\ }\textbf {\bibinfo {volume} {537}},\ \bibinfo
  {pages} {A78} (\bibinfo {year} {2012})},\ \bibinfo {note} {and many
  more}\BibitemShut {NoStop}%
\bibitem [{\citenamefont {Kirch}(2007)}]{Kirch:2007wa}%
  \BibitemOpen
  \bibfield  {author} {\bibinfo {author} {\bibfnamefont {K.}~\bibnamefont
  {Kirch}},\ }\href@noop {} {\enquote {\bibinfo {title} {{Testing gravity with
  muonium}},}\ } (\bibinfo {year} {2007}),\ \Eprint
  {http://arxiv.org/abs/physics/0702143} {arXiv:physics/0702143} \BibitemShut
  {NoStop}%
\bibitem [{\citenamefont {{M. K. Oberthaler}}(2002)}]{Oberthaler:2002}%
  \BibitemOpen
  \bibfield  {author} {\bibinfo {author} {\bibnamefont {{M. K. Oberthaler}}},\
  }\href@noop {} {\bibfield  {journal} {\bibinfo  {journal} {{Nucl. Instr.
  Meth. B}}\ }\textbf {\bibinfo {volume} {192}},\ \bibinfo {pages} {129}
  (\bibinfo {year} {2002})}\BibitemShut {NoStop}%
\end{thebibliography}%

\newpage
\appendix
\section{Workshop program}
\label{program}
\vspace{0.5cm}

{\bf \noindent Introduction (chair: Bertrand Echenard)}\\
\phantom{000}Welcome and workshop goals - Bertrand Echenard\\
\phantom{000}Theory overview - Sacha Davidson\\
\phantom{000}Mu2e-II - status and perspectives - James Miller\\
\phantom{000}AMF - status and perspectives - Robert Bernstein\\

{\bf \noindent Mu2e-II: Magnets (chair: Karie Badgley)}\\
\phantom{000}Muon collider HTS solenoid - Luca Bottura\\
\phantom{000}HTS (high temperature superconductor) magnets - Zach Hartwig\\
\phantom{000}Field calculations - Cole Kampa\\

{\bf \noindent Mu2e-II: Targetry\\
(chair: Kevin Lynch, Michael Mackenzie, Stefan Mueller, Vitaly Pronskikh) }\\
\phantom{000}Introduction - Kevin Lynch\\
\phantom{000}What we know and do not know about tungsten in accelerator environments - Yong Joong Lee\\
\phantom{000}Granular tungsten target R\&D at RAL - Dan Wilcox\\
\phantom{000}HPR R\&D - Frederique Pellemoine\\
\phantom{000}Muon Collider, fluidized targets - Carlo Carelli\\
\phantom{000}Mu2e - Michael Hedges\\
\phantom{000}Mu2e-II LDRD - David Neuffer\\

{\bf \noindent Mu2e II: Tracker (chair: Daniel Ambrose)}\\
\phantom{000}Mu2e-II Tracker Introduction - Daniel Ambrose\\
\phantom{000}Mu2e-II Tracker Intrinsic Resolution, Electronics, Readout - Richard Bonventre\\
\phantom{000}Mu2e-II Research topics and Discussion - James Popp\\
\phantom{000}Alternative Tracker ideas - David Brown\\
\phantom{000}Simulating Vacuum effects on Resolution and Beam Flash Radiation - Andrew Edmonds\\
\phantom{000}Mu2e lesson learned - Daniel Ambrose\\

{\bf \noindent Mu2e II: Calorimeter (chair: David Hitlin, Luca Morescalchi, Ivano Sarra)}\\
\phantom{000}Mu2e-II Calorimeter: Overview \& Requirements - Luca Morescalchi\\
\phantom{000}Crystals 1 - Ren Yuan Zhu\\
\phantom{000}Crystals 2 - Yury Davidov\\
\phantom{000}Photosensors - David Hitlin\\
\phantom{000}Alternatives - Ivano \\

{\bf \noindent Mu2e II: Comsic Ray Veto (chair: Edmond Craig Dukes, Craig Group, Yuri Oksuzian)}\\
\phantom{000}Welcome and goals - Craig Group\\
\phantom{000}CRV at Mu2e II Overview - Simon Corrodi\\
\phantom{000}Studies of Triangular Counters for Mu2eII - Ralf Ehrlich\\
\phantom{000}Discussion and plans - Craig Group\\

{\bf \noindent Mu2e-II: Trigger and DAQ (chair: Antonio Gioiosa, Gianantonio Pezzullo)}\\
\phantom{000}TDAQ summary - Gianantonio Pezzullo\\

\newpage
{\bf \noindent  Mu2e-II: Physics and Sensitivity (chair: Sophie Middleton)}\\
\phantom{000}Introduction - Sophie Middleton\\
\phantom{000}Stopping target - Leo Borrel\\
\phantom{000}Normalization - David Hitlin\\
\phantom{000}Exotic physics signatures - Ryan Plestid\\
\phantom{000}Implication of production target design on sensitivity - Michael Mackenzie\\
\phantom{000}Implication of tracker design on sensitivity - David Brown\\

{\bf\noindent  AMF: Proton compressor ring (chair: Jeffrey Eldred)}\\
\phantom{000}PAR Proposal - Ben Simmons\\
\phantom{000}C-PAR Concept - Jeffrey Eldred\\
\phantom{000}SNS Experience - Vasiliy Morozov\\
\phantom{000}PIP2-BD Experiment - Jacob Zettlemoyer\\

{\bf \noindent AMF: Muon FFA (chair: Jaroslaw Pasternak, Robert Bernstein)}\\
\phantom{000}FFAs for Muons - Francois Meot\\
\phantom{000}Building blocks and requirements of PRISM-like FFA system - Jaroslaw Pasternak\\
\phantom{000}Review of PRISM FFA ring designs - Jean-Baptiste Lagrange\\
\phantom{000}Designing injection system for PRISM FFA ring - Jaroslaw Pasternak\\

{\bf \noindent AMF: Conversion experiments (chair: Cole Kampa, Craig Group) }\\
\phantom{000}Introduction - Craig Group\\
\phantom{000}Overview of AMF Conversion Experiment - Bertrand Echenard\\
\phantom{000}Signal Resolution Requirements - Andrei Gaponenko\\
\phantom{000}Discussing an AMF Tracker - Daniel Ambrose\\
\phantom{000}Tracker Hit Resolution - Richard Bonventre\\
\phantom{000}Tracker Design; Simulations - David Brown\\
\phantom{000}Cosmic Ray Veto Considerations - Craig Group\\

{\bf \noindent AMF: Decay experiments  (chair: Angela Papa, Francesco Renga)}\\
\phantom{000}Status of the MEG II experiment - Dylan Palo\\
\phantom{000}HiMB: status and physics cases (incl. Mu3e) - Angela Papa\\
\phantom{000}Exotic muon decays - Deigo Redigolo\\
\phantom{000}Limiting factors of future mu $\rightarrow$ e gamma searches - Francesco Renga\\
\phantom{000}Conceptual design and R\&D activities for a future mu $\rightarrow$ e gamma search - Wataru Ootani\\

{\bf \noindent AMF: Other experiments (chair: Andrew Edmonds, Daniel Kaplan)}\\
\phantom{000}Muonium overview - Dan Kaplan\\
\phantom{000}M-Mbar mixing - Alexey Petrov\\
\phantom{000}Proposed MACE M-Mbar experiment - Shihan Zhao\\
\phantom{000}Muonium spectroscopy - David Kawall\\
\phantom{000}Low-energy muons at Fermilab - Carol Johnstone\\
\phantom{000}Making muonium with superfluid helium - Thomas Phillips\\
\phantom{000}Muonium gravity - Daniel Kaplan\\

\newpage
{\bf \noindent Summary Mu2e-II (chair: Frank Porter)}\\
\phantom{000}PIP-II - David Neuffer\\
\phantom{000}Calorimeter - Ivano Sarra\\
\phantom{000}Trigger and DAQ - Gianantonio Pezzullo\\
\phantom{000}Magnets - Karie Badgley\\
\phantom{000}Cosmic ray veto - Simon Corrodi\\
\phantom{000}Production target - Vitaly Pronskikh\\
\phantom{000}Tracking - Mete Yucel\\
\phantom{000}Sensitivity - Sophie Middleton\\

{\bf \noindent Summary AMF (chair: Bertrand Echenard)}\\
\phantom{000}Compressor summary - Jeffrey Eldred\\
\phantom{000}FFA accelerator summary - Robert Bernstein\\
\phantom{000}Conversion experiment summary - Cole Kampa\\
\phantom{000}Other experiment summary - Daniel Kaplan\\
\phantom{000}Decay experiments summary - Francesco Renga\\

\end{document}